\documentclass[12pt,a4paper]{article}
\usepackage[OT2,T1]{fontenc}
\usepackage{amssymb}
\usepackage{amsmath}
\usepackage{stmaryrd}
\usepackage{mathrsfs}
\usepackage{mathtools}
\usepackage{bm}
\usepackage{graphicx}
\usepackage[leftcaption]{sidecap}
\usepackage{caption}
\usepackage{subcaption}
\usepackage{here}
\usepackage[margin=1in]{geometry}
\usepackage{cite}
\DeclareMathAlphabet{\mathscrbf}{OMS}{mdugm}{b}{n}
\usepackage{tikz}
\usepackage{needspace}

\usepackage{sectsty}
\allsectionsfont{\sffamily}
\usepackage{titlesec}
\titleformat*{\paragraph}{\sffamily\bfseries\boldmath}

\usepackage{tocloft}

\usepackage{enumerate}

\let\OLDthebibliography\thebibliography
\renewcommand\thebibliography[1]{
  \OLDthebibliography{#1}
  \setlength{\parskip}{0pt}
  \setlength{\itemsep}{0pt plus 0.3ex}
}

\usepackage[colorlinks,allcolors=blue]{hyperref}

\newcommand{\be}{\begin{equation}}
\newcommand{\ee}{\end{equation}}
\newcommand{\ben}{\begin{enumerate}}
\newcommand{\een}{\end{enumerate}}
\newcommand{\bi}{\begin{itemize}}
\newcommand{\ei}{\end{itemize}}
\newcommand{\bmm}{\begin{pmatrix}}
\newcommand{\emm}{\end{pmatrix}}

\newcommand{\Ad}{\text{Ad}}
\newcommand{\ad}{\text{ad}}

\newcommand{\dd}{\text{d}}
\newcommand{\demi}{\frac{1}{2}}
\newcommand{\der}{\partial}
\newcommand{\Diff}{\text{Diff}\,S^1}

\newcommand{\ds}{\displaystyle}
\newcommand{\dt}{\left.\frac{\text{d}}{\text{d}t}\right|_0}
\newcommand{\eg}{{\it e.g.}\ }

\newcommand{\hDiff}{\widehat{\text{Diff}}\,S^1}

\newcommand{\hG}{\widehat{G}}
\newcommand{\hg}{\widehat{\mathfrak{g}}}

\newcommand{\hVect}{\widehat{\text{Vect}}\,S^1}
\newcommand{\ie}{{\it i.e.}\ }

\newcommand{\sn}{\,\text{sn}}

\newcommand{\Vect}{\text{Vect}\,S^1}

\newcommand{\bL}{\textbf L}

\newcommand{\bbomega}{{\boldsymbol\omega}}

\newcommand{\cA}{{\cal A}}

\newcommand{\cE}{{\cal E}}
\newcommand{\cF}{{\cal F}}
\newcommand{\cG}{{\cal G}}
\newcommand{\cH}{{\cal H}}
\newcommand{\cI}{{\cal I}}

\newcommand{\cL}{{\cal L}}
\newcommand{\cM}{{\cal M}}
\newcommand{\cO}{{\cal O}}
\newcommand{\cR}{{\cal R}}

\newcommand{\cV}{{\cal V}}

\newcommand{\mg}{\mathfrak{g}}

\newcommand{\sfC}{\mathsf{C}}

\newcommand{\sfG}{\mathsf{G}}

\newcommand{\sfL}{\mathsf{L}}

\newcommand{\sfR}{\mathsf{R}}
\newcommand{\sfS}{\mathsf{S}}

\newcommand{\II}{\mathbb{I}}

\newcommand{\RR}{\mathbb{R}}
\newcommand{\ZZ}{\mathbb{Z}}

\begin{document}

\hrule
\begin{center}
\Large{\bfseries{\textsf{Berry Phases in the Reconstructed KdV Equation}}}
\end{center}
\hrule
~\\

\begin{center}
\large{\textsf{Blagoje Oblak$\,^{a}$ and Gregory Kozyreff$\,^{b}$}}
\end{center}
%~\\

\begin{center}
\begin{minipage}{.8\textwidth}\small\it
\begin{center}
$^a$
{\tt{boblak@lpthe.jussieu.fr}}\\
Laboratoire de Physique Th\'eorique et Hautes Energies,\\
Sorbonne Universit\'e and CNRS UMR 7589, F-75005 Paris, France;\\
Institut f\"ur Theoretische Physik, ETH Z\"urich,\\
Campus H\"onggerberg, CH-8093 Z\"urich, Switzerland.
~\\
~\\
$^b$
{\tt{gkozyref@ulb.ac.be}}\\
Optique Non-lin\'eaire Th\'eorique, Universit\'e Libre de Bruxelles,\\
Campus Plaine C.P.\ 231, B-1050 Bruxelles, Belgium.
\end{center}
\end{minipage}
\end{center}

\vspace{1cm}

\begin{center}
\begin{minipage}{.92\textwidth}
\begin{center}{\bfseries{\textsf{Abstract}}}\end{center}
We consider the KdV equation on a circle and its Lie-Poisson reconstruction, which is reminiscent of an equation of motion for fluid particles. For periodic waves, the stroboscopic reconstructed motion is governed by an iterated map whose Poincar\'e rotation number yields the drift velocity. We show that this number has a geometric origin: it is the sum of a dynamical phase, a Berry phase, and an `anomalous phase'. The last two quantities are universal: they are solely due to the underlying Virasoro group structure. The Berry phase, in particular, was previously described in \cite{Oblak:2017ect} for two-dimensional conformal field theories, and follows from adiabatic deformations produced by the propagating wave. We illustrate these general results with cnoidal waves, for which all phases can be evaluated in closed form thanks to a uniformizing map that we derive. Along the way, we encounter `orbital bifurcations' occurring when a wave becomes non-uniformizable: there exists a resonance wedge, in the cnoidal parameter space, where particle motion is locked to the wave, while no such locking occurs outside of the wedge.
\end{minipage}
\end{center}

\newpage

\textsf{\tableofcontents}
~\\[-.3cm]%%PARBREAK%%
\hrule

\section{Introduction and summary of results}
\label{secIN}

It is quite generally true that the state vector of a quantum system undergoing cyclic changes of reference frames picks up Berry phases \cite{Berry:1984jv,berry1985classical}. Typical examples of this behaviour include Thomas precession \cite{Thomas:1926dy}, a spin in a slowly rotating magnetic field \cite{Berry:1984jv,Nakahara:2003nw,leek2007observation}, and its non-compact analogue \cite{svensmark1994experimental} which appears in the quantum Hall effect \cite{AvronSeiler}. In \cite{Oblak:2017ect}, such Berry phases were shown to arise in two-dimensional conformal field theories (CFTs) coupled to an environment that produces adiabatic conformal transformations. These phases can be computed exactly despite the infinite-dimensional parameter space, and coincide with `geometric actions' of the Virasoro group \cite{Alekseev:1988ce}. From now on, we refer to them as {\it Virasoro Berry phases}. They are reminiscent of the response of a quantum Hall fluid to metric deformations, where the parameter space is infinite-dimensional as well \cite{Bradlyn:2015wsa}.\\[-.3cm]%%PARBREAK%%

The goal of this paper is to exhibit {\it classical} systems where Virasoro Berry phases are realized dynamically, \ie without any implicit coupling to the `environment'.\footnote{We write `Berry phases' despite the lack of quantization. `Geometric phases' or `Hannay angles' \cite{Hannay} would be more appropriate, but the distinction is unimportant as the classical geometric phase will coincide with its quantum analogue \cite{Oblak:2017ect}.} This notably includes the Korteweg-de Vries (KdV) equation \cite{Korteweg}, or rather its `reconstruction' \cite{Marsden90,Marsden,Holm,HolmTyr}, but it applies more generally to any Lie-Poisson equation based on the Virasoro group \cite{Khesin2003,Khesin}, such as the Hunter-Saxton and Camassa-Holm equations \cite{Hunter}. Indeed, the group structure underlying Lie-Poisson equations provides powerful geometric tools that can be used to predict universal properties of the reconstructed dynamics --- such as Berry phases appearing when the system's motion in momentum space is periodic. For example, the Lie-Poisson system of SO$(3)$ yields the standard Euler equations for the angular momentum of a rigid body. When angular momentum completes one period of its motion, the final orientation of the body in space differs from its initial one by a rotation whose angle is known as a Montgomery phase \cite{Montgomery,Natario}; it is the sum of a dynamical phase and a geometric phase due to adiabatic rotations. The purpose of this paper is thus to describe the Virasoro analogue of Montgomery phases.\\[-.3cm]%%PARBREAK%%

For the record, this is not the first time that geometric phases are found in the KdV equation: such phases were indeed exhibited in \cite{Alber} and reproduced, among other things, the standard phase shift occurring after the collision of two solitons. However, \cite{Alber} crucially used the effective, finite-dimensional phase space description of KdV solitons, and the corresponding geometric phases are Hannay angles in a finite-dimensional parameter space. This is radically different from what we do here, since we, by contrast, explicitly use the infinite-dimensional nature of the Virasoro group and never rely on soliton dynamics {\it per se}. In this sense, there is, to our knowledge, no overlap between \cite{Alber} and the present work, other than general context.\\[-.3cm]%%PARBREAK%%

We now explain how Virasoro Berry phases affect the motion of suitable `fluid particles', and how these phases can be computed. We then expose the plan of the paper.

\paragraph{Summary of results.} This work relies on a fair amount of symplectic geometry and Virasoro group theory, none of which is reviewed in a self-contained manner --- we refer \eg to \cite{Abraham} for an introduction to the former, and to \cite{Guieu,Oblak:2016eij} for the latter. It is nevertheless straightforward to describe our main conclusions with minimal technicalities. Namely, let $p(x,t)$ be a (spatially $2\pi$-periodic) wave profile that solves the KdV equation\footnote{We choose $p$ to satisfy KdV, but virtually identical arguments apply to Camassa-Holm and Hunter-Saxton upon suitably adapting the inertia operator \cite{Khesin2003}.}
\be
\frac{\der p}{\der t}+3p\frac{\der p}{\der x}-\frac{c}{12}\frac{\der^3p}{\der x^3}
=
0,
\label{KADA}
\ee
where $c\neq0$ is a constant parameter (the Virasoro central charge). Suppose, then, that a particle on the line has a position $x(t)$ that satisfies
\be
\frac{\dd x}{\dd t}
=
p\big(x(t),t\big),
\label{sb1}
\ee
with initial position $x(0)=x_0$ say. This particle could be, for example, a small fluid element in a shallow water channel supporting the wave $p$.\footnote{We will return to this interpretation repeatedly below, particularly in section \ref{secSTOKES}. Up to a (crucial!) mismatch in reference frames, the interpretation of (\ref{sb1}) as a fluid transport equation holds.} Our goal is to find, analytically, general properties of the resulting solution $x(t)$, such as the drift velocity
\be
v_{\text{Drift}}
\equiv
\lim_{t\to+\infty}\frac{x(t)-x(0)}{t}.
\label{VIDI}
\ee
To ensure that the latter is well-defined, we add one extra condition: we require the wave $p$ to be {\it periodic in time}, \ie $p(x,t+T)=p(x,t)$ for some $T>0$. Then, there exists a (time-independent) diffeomorphism $x\mapsto F(x)$ of $\RR$ such that, after $N$ periods,
\be
x(NT)
=
F\circ F\circ...\circ F(x_0)
\equiv F^N(x_0).
\label{FEQ}
\ee
One can thus think of the `stroboscopic' motion of particles at integer multiples of the period $T$ as a discrete-time dynamical system governed by iterations of the map $F$. From that perspective, the drift velocity (\ref{VIDI}) reads
\be
v_{\text{Drift}}
=
\frac{\Delta\phi}{T},
\qquad
\Delta\phi
\equiv
\lim_{N\to+\infty}\frac{F^N(x_0)-x_0}{N}
\label{DAFA}
\ee
where $\Delta\phi$ is the {\it Poincar\'e rotation number} of $F$ \cite[sec.\ 4.4.3]{Guieu}. It is easily read off by integrating eq.\ (\ref{sb1}) numerically over many periods. As we now explain, there is in fact a way to predict the value of $\Delta\phi$, analytically, using group theory and symplectic geometry. This value involves, in particular, a Virasoro Berry phase.\\[-.3cm]%%PARBREAK%%

To see where symplectic geometry plays a role, one has to think of the KdV equation (\ref{KADA}) as a Lie-Poisson equation\footnote{Sometimes also known as an `Euler-Poisson' or `Euler-Arnold' equation; see footnote \ref{fonosix}.} for the Virasoro group. The phase space of any such system is the cotangent bundle of a Lie group, where the cotangent part consists of `momenta', while the group manifold is a space of `positions' or `configurations'. In the KdV case, for instance, $p(x,t)$ is a Virasoro momentum (which justifies our notation). By construction, the motion of momenta determines that of configurations through Lie-Poisson reconstruction \cite{Marsden90,Marsden,Holm}. In the KdV case, this reconstruction turns out to precisely take the form of eq.\ (\ref{sb1}), as explained in greater detail in section \ref{secEPP}. Importantly, periodic motion of momenta does {\it not}, in general, imply periodicity of configurations. Instead, when the system performs a loop in momentum space, its configuration typically traces an open path, and the difference between the initial and final positions can be interpreted as an (an)holonomy. The latter involves a Berry phase associated with adiabatic changes of reference frames, exactly as in the aforementioned example of Montgomery phases \cite{Montgomery,Natario}.\\[-.3cm]%%PARBREAK%%

For Lie-Poisson systems based on the Virasoro group, such as KdV, the holonomy in the space of configurations is precisely the rotation angle $\Delta\phi$ of (\ref{DAFA}). It can be written as a sum of three terms, derived in eq.\ (\ref{DPHI}) below, whose schematic form is
\be
\Delta\phi
=
\text{Dynamical phase}+\underbrace{\text{Berry phase}+\text{Anomalous phase}}_{\text{Universal}}.
\label{DBIB}
\ee
In that expression, the first term, proportional to the period $T$, is a dynamical phase, while the second term is a Berry phase associated with adiabatic diffeomorphisms \cite{Oblak:2017ect}. The anomalous term is a contribution due to the Virasoro central extension and may be seen as the integral of a Berry connection along the {\it inverse} of the reconstructed path. Both the Berry phase and the anomalous term are universal: they solely follow from Virasoro group theory and take the same form regardless of dynamics (though the path $p(x,t)$ that determines their value does, of course, depend on dynamics). Furthermore, the dynamical phase and Berry phase are known functionals of $p(x,t)$; the anomalous term, on the other hand, is an implicit integral (\ref{KLO}). All these quantities turn out to simplify greatly for travelling waves $p(x,t)=p(x-vt)$, which eventually yields eq.\ (\ref{SANTO}) for $\Delta\phi$. As a result, for cnoidal waves, the three terms of (\ref{DBIB}) can be evaluated analytically at any point in parameter space; they are displayed in eqs.\ (\ref{VIDY})-(\ref{VIAN}). Their sum coincides with the value (\ref{DAFA}) that can be computed by other means --- thanks to a `uniformizing map' that we derive ---, and leads to the compact formula (\ref{s14}) for the drift velocity.\\[-.3cm]%%PARBREAK%%

It should be noted that our derivation of eq.\ (\ref{DBIB}) for KdV rests on one key technical assumption: the profile $p(x,t)$ must be uniformizable, or {\it amenable}, in the sense that there exists a conformal transformation (\ie a diffeomorphism of the circle) mapping it on some uniform, $x$-independent, profile $k$. This is generally not guaranteed, since a great many Virasoro coadjoint orbits have no uniform representative \cite{Lazutkin,Witten:1987ty}. For any profile that does not satisfy the assumption of amenability, a notion of drift does exist in the sense of eqs.\ (\ref{VIDI})-(\ref{DAFA}), but the corresponding $\Delta\phi$ is an integer multiple of $2\pi$ and cannot be written as a sum of phases (\ref{DBIB}). Following \cite{Oblak:2019llc}, we will show that such a regime occurs for cnoidal waves with sufficient pointedness: there exists a {\it resonance wedge} in the cnoidal parameter space where (\ref{DBIB}) does not apply, and, in that wedge, particle motion is `locked' to the travelling wave: $v_{\text{Drift}}=v_{\text{Wave}}$. The transition along the boundary of the wedge is reminiscent of the {\sc sniper} bifurcation of the Adler equation \cite{Adler}. Outside of the wedge, cnoidal waves are amenable and eq.\ (\ref{DBIB}) applies, leading to a drift velocity $v_{\text{Drift}}\neq v_{\text{Wave}}$.\\[-.3cm]%%PARBREAK%%

An important motivation for this work stems from fluid dynamics, where the KdV equation notoriously describes shallow water waves \cite{Ockendon}. Indeed, in a comoving frame, the leading equation of motion for fluid particles in a two-dimensional channel supporting KdV waves is nearly identical to the reconstruction equation (\ref{sb1}): the only difference is a factor 2 and the presence of a (large) constant term on the right-hand side (see eq.\ (\ref{XDOT}) below and the surrounding discussion). The symplectic formula (\ref{DBIB}) for $\Delta\phi$, along with the drift velocity (\ref{VIDI}), thus suggests that Virasoro Berry phases contribute to the {\it Stokes drift} velocity of particles in shallow water \cite{Longuet53}, similarly to the crest slowdown observed in wave breaking \cite{Banner}. However, the seemingly innocuous change of reference frames that distinguishes eq.\ (\ref{sb1}) from the actual equation of motion for fluid particles turns out to be crucial: it implies that Stokes drift in standard shallow water dynamics differs from the drift velocity introduced here, whose (subleading) effect is entirely washed out by the overwhelming, dominant contribution of the overall velocity $\sqrt{gh}$ of the comoving frame. More on that in section \ref{secSTOKES}. Prospects for actual observations of Virasoro Berry phases are relegated to the \hyperref[sec5]{conclusion} of this paper.

\paragraph{Plan of the paper.} This work is not self-contained: the necessary prerequisites include symplectic geometry \cite{Abraham} and Virasoro group theory \cite{Guieu,Oblak:2016eij}, and the parts concerning cnoidal waves heavily rely on \cite{Oblak:2019llc}. We will not review any of that content here, but we do adopt a logical flow that corresponds to the way one would naturally teach the subject. Accordingly, the structure is as follows.\\[-.3cm]%%PARBREAK%%

First, section \ref{sec2} contains general prerequisites in symplectic geometry. In it, we introduce the notion of reconstruction and derive an abstract formula for the rotation $\Delta\phi$ associated with any periodic solution of a Lie-Poisson system based on a centrally extended group, provided the solution has a $\text{U}(1)$ stabilizer in a suitable sense. This leads to eq.\ (\ref{KAX}), which is critical to the rest of the paper (and is new to our knowledge, as it contains an `anomalous phase' that appears to have been overlooked so far). In section \ref{secEKO} we apply this formula to any Lie-Poisson wave equation based on the Virasoro group, resulting in eq.\ (\ref{DPHI}) for $\Delta\phi$. We also establish the link between reconstruction and the equation of motion (\ref{sb1}), hence between geometric phases and the drift velocity (\ref{VIDI}), and comment on the important difference between the latter notion and that of Stokes drift \cite{Longuet53}. Section \ref{secNUM} is devoted to the application of these arguments to travelling waves, and to a comparison between the geometric prediction (\ref{DBIB}) and the value of $\Delta\phi$ computed analytically. To that end, we actually find a general formula for `uniformizing maps' of travelling waves satisfying KdV, and deduce an exact expression for the solution of the equation of motion (\ref{sb1}), from which the drift velocity (\ref{VIDI}) follows. As we shall see, the drift velocity is indeed perfectly predicted by the symplectic formula (\ref{DBIB}), but it is strongly affected by the values of cnoidal parameters --- in particular, waves located in a certain `resonance wedge' produce particle motion that is locked to the wave, confirming the existence of `orbital bifurcations' anticipated in \cite{Oblak:2019llc}. Finally, we conclude in section \ref{sec5} with a discussion of potential follow-ups of our work. For completeness, appendices \ref{appa} and \ref{appb} collect some details of group theory and symplectic geometry needed in section \ref{sec2}.

\section{Reconstruction and Berry phases}
\label{sec2}

This section is a prelude on geometric phases in symplectic geometry, providing crucial background material for the entire paper. It is organized as follows. We start by briefly reviewing general aspects of Lie groups and Lie-Poisson equations \cite{Khesin}, which describe the time evolution of a vector $p(t)$ living in a suitable `momentum space'. We then turn to the notion of {\it reconstruction}, which associates with $p(t)$ a path $g_t$ in a group manifold.\footnote{We write paths in groups as $g_t$ instead of $g(t)$ for notational convenience: from section \ref{secEKO} onwards, each $g_t$ will be a function $g_t(x)$ of $x\in\RR$ and the subscript will stress the asymmetric roles of $t$ and $x$.} As mentioned in the \hyperref[secIN]{introduction}, one can think of $g_t$ as the time-dependent configuration, or `position', of a system whose momentum $p(t)$ satisfies a Lie-Poisson equation. Finally, we consider periodic momenta (satisfying $p(T)=p(0)$) and ask what this entails for the reconstructed curve $g_t$. As we shall see, despite the periodicity of $p(t)$, the path $g_t$ is generally {\it not} closed: $g_T\neq g_0$. We will measure this inequality by a rotation angle $\Delta\phi$ which will turn out to equal the sum of a dynamical phase and a Berry phase (section \ref{secDYBE}), plus an `anomalous phase' when the group $G$ is centrally extended (section \ref{sec24}).\\[-.3cm]%%PARBREAK%%

Applications of this formalism include the Montgomery phases of rigid bodies \cite{Montgomery,Natario}, and of course Berry phases in the KdV equation, whose detailed treatment is relegated to sections \ref{secEKO} and \ref{secNUM}. We warn the reader that the discussion relies heavily on Lie group theory and symplectic geometry; some technical details are covered in appendices \ref{appa} and \ref{appb}. For a pedagogical presentation, see \eg \cite{Khesin,Abraham}; see also \cite{Marsden90,Marsden} for a detailed account of Lie-Poisson reconstruction in general, including geometric phases.

\subsection{Lie groups and Lie-Poisson equations}
\label{sec21}

Lie-Poisson equations are Hamiltonian systems whose dynamics is almost entirely fixed by a parent Lie group. For instance, the rotation group SO(3) leads to the motion of free-falling rigid bodies, while the Virasoro group is associated with a host of non-linear wave equations that includes the KdV, inviscid Burgers, Hunter-Saxton and Camassa-Holm equations \cite{Khesin2003,Khesin}. Here, as a preparation for KdV and its cousins, we recall the derivation of Lie-Poisson equations in a general group-theoretic setting. We refer to appendix \ref{appa} for the minimal necessary background on Lie groups and symplectic geometry.\\[-.3cm]%%PARBREAK%%

Let $G$ be a Lie group with algebra $\mg$, whose dual space is $\mg^*$. The {\it adjoint representation} of $G$ on $\mg$ is defined, for all $g\in G$ and any $\xi\in\mg$, by $\Ad_g(\xi)\equiv\der_t\big|_0\big(g\,e^{t\xi}\,g^{-1}\big)$, where $e^{\cdots}$ denotes the exponential map. For matrix groups, this boils down to $g\,\xi\,g^{-1}$. The dual of the adjoint is the {\it coadjoint representation} of $G$, given for all $g\in G$, $p\in\mg^*$, $\xi\in\mg$ by
\be
\langle g\cdot p,\xi\rangle
\equiv
\langle p,\Ad_g^{-1}\xi\rangle.
\label{coadef}
\ee
In what follows, the coadjoint representation will play a key role, so we reduce clutter by writing it as $g\cdot p$, instead of the heavier notation $\Ad^*_g(p)$. We will also refer to elements of $\mg^*$ as {\it coadjoint vectors}, or {\it momenta} for brevity. The Lie-algebraic analogue of the coadjoint representation will be denoted as $\ad^*$ and is defined by the derivative of $\Ad^*$, that is, $\ad^*_{\xi}\equiv\der_t|_0\Ad^*_{e^{t\xi}}$. Using (\ref{coadef}), this is equivalent to
\be
\langle\ad^*_{\xi}p,\zeta\rangle
\equiv
-\langle p,[\xi,\zeta]\rangle
\label{coadefa}
\ee
where $p\in\mg^*$ and $\xi,\zeta\in\mg$, with $[\cdot,\cdot]$ the Lie bracket.

\paragraph{Lie-Poisson equations.} As a starting point towards the Lie-Poisson construction, note that $\mg^*$ is essentially a phase space, since it admits a Poisson structure. Indeed, given any real function $F$ on $\mg^*$, its differential $\dd F_p$ at a point $p$ is a linear map from $T_p\mg^*\cong\mg^*$ to $\RR$. Hence $\dd F_p$ may be seen as an element of the Lie algebra $\mg\cong(\mg^*)^*$, and one defines the {\it Kirillov-Kostant bracket} on $\mg^*$ by\footnote{The bracket (\ref{ss18}) is degenerate, so $\mg^*$ is a Poisson manifold but {\it not} a symplectic manifold. Its symplectic leaves are coadjoint orbits of $G$, to which we will turn shortly.}
\be
\{F,G\}(p)
\equiv
\langle p,[\dd F_p,\dd G_p]\rangle
\stackrel{\text{(\ref{coadefa})}}{=}
-\langle\ad^*_{\dd F_p}(p),\dd G_p\rangle.
\label{ss18}
\ee
Thinking of $\mg^*$ as a phase space, it is immediate to write down evolution equations: any Hamiltonian $H$ on $\mg^*$ determines the time-dependence of a function $F$ according to
\be
\dot F(p)
=
\{F,H\}(p)
\stackrel{\text{(\ref{ss18})}}{=}
\langle\ad^*_{\dd H_p}(p),\dd F_p\rangle
\equiv
\dd F_p\big(\ad^*_{\dd H_p}(p)\big)
\label{THEQ}
\ee
where $\dot F\equiv\dd F/\dd t$. The left-hand side can be written as $\dot F(p)=\dd F_p(\dot p)$, where $\dot p$ is the vector field given by the Hamiltonian flow. Thus, removing the differential $\dd F_p$ from both sides of (\ref{THEQ}), one reads off the equation of motion
\be
\dot p(t)
=
\ad^*_{\dd H_{p(t)}}\big(p(t)\big).
\label{s20}
\ee
This is the {\it Lie-Poisson equation} of $G$ associated with the Hamiltonian $H$.\footnote{\label{fonosix}Lie-Poisson equations have a long history and go under various names in the literature; they are sometimes known as `Euler-Poisson' or `Euler-Arnold' equations, and their Lagrangian analogues are called Euler-Poincar\'e equations \cite{Marsden90,Marsden,Holm,HolmTyr,Khesin2003,Khesin}. Here, following \cite{Marsden}, we will only use the `Lie-Poisson' terminology, since all our considerations are Hamiltonian (as opposed to Lagrangian).} It determines a unique curve $p(t)$ in phase space $\mg^*$ for any initial condition $p(0)$.\\[-.3cm]%%PARBREAK%%

The Lie-Poisson equations that are of interest for fluid mechanics \cite{Arnold}, KdV and its cousins \cite{Khesin2003,Khesin,Hunter} have quadratic Hamiltonians. This requires an extra bit of terminology: by definition, an {\it inertia operator} is an invertible linear map
\be
\cI:
\mg\rightarrow\mg^*:
\xi\mapsto\cI(\xi)
\label{t20}
\ee
which is self-adjoint in the sense that $\langle\cI(\xi),\zeta\rangle=\langle\cI(\zeta),\xi\rangle$, and positive-definite in the sense that $\langle\cI(\xi),\xi\rangle>0$ for any non-zero $\xi\in\mg$. The ensuing quadratic Hamiltonian
\be
H(p)
=
\demi\big<p,\cI^{-1}(p)\big>
\label{tt20}
\ee
may be seen as `kinetic energy' and the corresponding Lie-Poisson equation (\ref{s20}) reads
\be
\dot p
=
\ad^*_{\cI^{-1}(p)}(p).
\label{ss20}
\ee
One can show that this is equivalent to a geodesic equation in $G$ for the right-invariant metric induced by the inertia operator $\cI$ \cite[sec.\ 4.3]{Khesin}. The point of `reconstruction' will precisely be to recover a geodesic $g(t)\equiv g_t$ in $G$ from a solution $p(t)$ of (\ref{ss20}).\\[-.3cm]%%PARBREAK%%

From now on, we shall always restrict attention to Lie-Poisson systems of the form (\ref{ss20}), associated with an inertia operator $\cI$ and a quadratic Hamiltonian (\ref{tt20}). More general Hamiltonians are only revisited briefly at the very end of section \ref{secDYBE}, and in appendix \ref{appb}.

\paragraph{Remarks on coadjoint orbits.} The name `inertia operator' for the map (\ref{t20}) stresses that Lie-Poisson equations generalize the Euler equations of motion for free-falling rigid bodies. The latter have a configuration space $G=\text{SO}(3)$ and an inertia tensor specified by their mass distribution. The Lie algebra $\mathfrak{so}(3)$ and its dual $\mathfrak{so}(3)^*$ respectively consist of angular velocities and angular momenta, the two being related through the inertia tensor $\cI$. As seen from a (non-inertial) reference frame attached to the body at its centre of mass, the time evolution of angular momentum is given by eq.\ (\ref{ss20}). By contrast, in any {\it inertial} frame, the angular momentum vector is constant.\\[-.3cm]%%PARBREAK%%

This example illustrates an important property of Lie-Poisson equations. Namely, since eqs.\ (\ref{s20}) and (\ref{ss20}) have the structure $\dot p=\ad^*_{\cdots}(p)$, any solution $p(t)$ is such that $p(t)=f_t\cdot p(0)$ for some path $f_t$ in the group manifold. As a result, once an initial condition has been fixed, the motion of $p(t)$ takes place on a single {\it coadjoint orbit} of $G$,
\be
\cO_{p(0)}
\equiv
\big\{
f\cdot p(0)
\big|
f\in G
\big\},
\label{s23}
\ee
so that $p(t)$ and $p(t')$ are related to one another by a change of reference frames, for all $t,t'$. In particular, there always exists a frame where the motion of $p(t)$ is trivial, namely $p(0)=f_t^{-1}\cdot p(t)=\text{const}$. For the Euler top, this is achieved in any inertial frame.\\[-.3cm]%%PARBREAK%%

The orbit (\ref{s23}) is a submanifold of $\mg^*$, so it is specified by a certain number of continuous parameters whose value remains constant in time. In that sense, the statement $p(t)\in\cO_{p(0)}$ is a conservation law. This will allow us to fix a particular orbit representative, say $k\in\mg^*$, and write time evolution as $p(t)=g_t\cdot k$ for some `reconstructed' path $g_t$ in $G$. Suitable choices of $k$ will then greatly simplify the analysis of reconstructed motion.

\subsection{Lie-Poisson reconstruction}
\label{sec22}

\begin{figure}[t]
\centering
\includegraphics[width=0.40\textwidth]{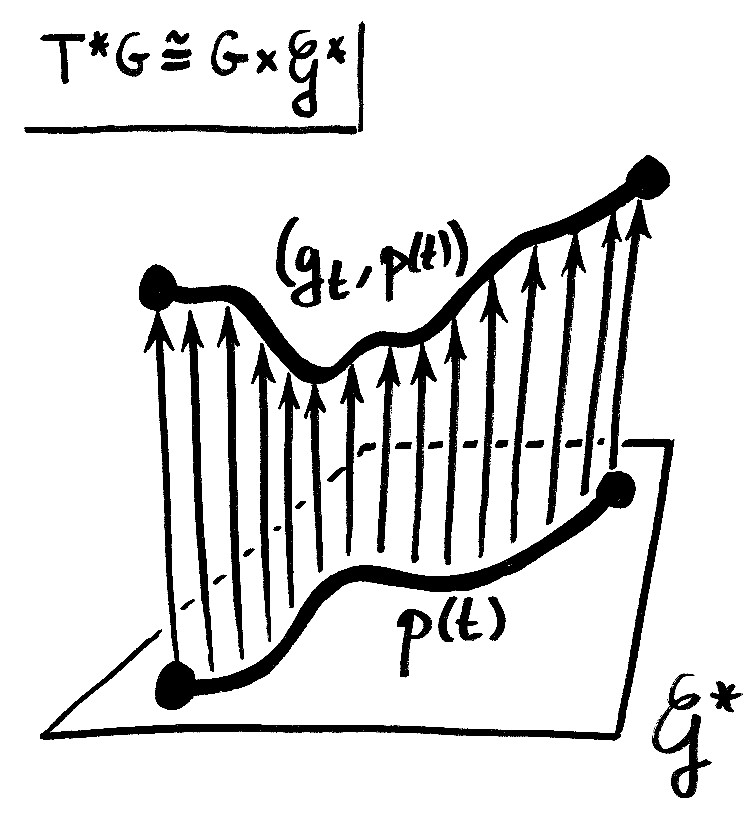}
\caption{A schematic picture of Lie-Poisson reconstruction: a path $p(t)$ in $\mg^*$, solving eq.\ (\ref{ss20}), is lifted to a pair of paths $(g_t,p(t))$ in $G\times\mg^*\cong T^*G$.\label{s7b}}
\end{figure}

As just reviewed, the dual $\mg^*$ of an algebra $\mg$ is a space of momenta endowed with a bracket (\ref{ss18}); the Lie-Poisson equations (\ref{s20})-(\ref{ss20}) describe Hamiltonian systems in that space. We now extend this picture by thinking of the group $G$ as the configuration manifold of the system, with a larger phase space given by the cotangent bundle $T^*G\cong G\times\mg^*$. From that perspective, the momentum dynamics described so far is a `reduction' of more complete, parent dynamics in $T^*G$. In the opposite direction, {\it reconstruction} will lift the motion $p(t)$ in $\mg^*$ to a curve $\big(g_t,p(t)\big)$ in $G\times\mg^*$, with $g_t$ determined by $p(t)$ (see fig.\ \ref{s7b}).\\[-.3cm]%%PARBREAK%%

Since these concepts are crucial for our purposes, we now define Lie-Poisson reconstruction and state some of its elementary properties. The presentation will be self-contained but somewhat heuristic. Some background is relegated to appendix \ref{appa}, and a more systematic derivation, based on group actions on $T^*G$, is exposed in appendix \ref{appb}.

\paragraph{Defining reconstruction.} We show in appendix \ref{appa} that the cotangent bundle $T^*G$, \ie the phase space of the reconstructed system, is a trivial bundle: it is equivalent (symplectomorphic) to the product $G\times\mg^*$. In particular, the symplectic form $\omega=-\dd\cA$ of $G\times\mg^*$ is obtained by pulling back the standard symplectic structure of $T^*G$, with\footnote{See eq.\ (\ref{ss3030}) at the end of appendix \ref{appa}.}
\be
\cA_{(g,p)}
=
\big(\langle p,\dd g\,g^{-1}\rangle,0\big)
\label{ss30}
\ee
where $\dd g\,g^{-1}\equiv\dd(R_{g^{-1}})_g$ is the right Maurer-Cartan form (and $R_{g^{-1}}$ denotes right multiplication by $g^{-1}$). The one-form (\ref{ss30}) is the group-theoretic version of what is commonly written in mechanics as $p\,\dd q$, with $\dd q$ the Maurer-Cartan form of the group $\RR$. It is an essential object for all that follows, since it is also the Berry connection that eventually gives rise to geometric phases in reconstructed dynamics --- which is why we call it $`\cA$'.\\[-.3cm]%%PARBREAK%%

Now consider the Lie-Poisson equation (\ref{ss20}) from the point of view of the full phase space $T^*G\cong G\times\mg^*$. The key to reconstruction is the fact, emphasized in section \ref{sec21}, that any curve $p(t)$ solving (\ref{ss20}) lies entirely on a single coadjoint orbit of $G$. Thus, the path in momentum space can be written as
\be
p(t)
=
g_t\cdot k
\label{gkk}
\ee
for some fixed coadjoint vector $k$ and some curve $g_t$ in $G$. Note that $k$ need not coincide with $p(0)=g_0\cdot k$, as $g_0$ may well differ from the identity --- this will indeed occur below.\footnote{\label{tobemade}From the perspective of reconstructed dynamics, $k$ is the value of the momentum map associated with the right action of $G$ on $T^*G$, evaluated on the solution of eqs.\ (\ref{ss20})-(\ref{ss31b}). See appendix \ref{appb}.} It is therefore tempting to consider paths of the form $\big(g_t,g_t\cdot k\big)$ in $G\times\mg^*$, and declare that any such path is a reconstruction of $g_t\cdot k$. However, this na\"ive definition suffers from a `gauge redundancy': for any curve $h_t$ in $G$ such that $h_t\cdot k=k$, one has $p(t)=g_t\cdot k=g_th_t\cdot k$ even though $g_t\neq g_th_t$. To fix this, one constrains the time-dependence of $g_t$ by defining the Hamiltonian on $G\times\mg^*$ to be $\cH(g,p)=H(p)$, where $H$ is the kinetic energy (\ref{tt20}). Using the symplectic form that follows from the potential (\ref{ss30}), the ensuing equations of motion\footnote{For details, see appendix \ref{appb} and the derivation leading to eqs.\ (\ref{eoma}).} are given by (\ref{ss20}) and the {\it reconstruction condition}
\be
p(t)
=
\cI\big(\dot g_t\,g_t^{-1}\big).
\label{ss31b}
\ee
This generalizes the rigid body relation $\bL=\cI(\bbomega)$ linking angular momentum $\bL\equiv p$ to angular velocity $\bbomega\equiv\dot gg^{-1}$. Given an initial condition $g_0$ such that $g_0\cdot k=p(0)$, the resulting unique solution $\big(g_t,p(t)\big)=\big(g_t,g_t\cdot k\big)$ in $G\times\mg^*$ is called a {\it Lie-Poisson reconstruction} of $p(t)$. In particular, $g_t$ turns out to be a geodesic with respect to a right-invariant metric determined by the inertia operator \cite[chap.\ 13]{Marsden}.

\paragraph{Remarks.} We show in sections \ref{secDYBE} and \ref{sec24} how reconstruction leads to geometric phases when $p(t)$ is periodic. Two concluding comments are in order first:
\begin{enumerate}[i]
\setlength\itemsep{0em}
\item[(i)] The reconstruction condition (\ref{ss31b}) implies the Lie-Poisson equation (\ref{ss20}). Indeed, one can act with both sides of $\dot g\,g^{-1}=\cI^{-1}(p)$ on a momentum $p(t)=g_t\cdot k$ through the coadjoint action (\ref{coadefa}); this yields (leaving the time dependence implicit)
\be
\ad^*_{\dot g\,g^{-1}}(g\cdot k)
=
\frac{\dd}{\dd t}(g\cdot k)
\stackrel{!}{=}\ad^*_{\cI^{-1}(g\cdot k)}(g\cdot k),
\label{s32b}
\ee
which coincides as announced with eq.\ (\ref{ss20}). Intuitively, this states that the time-dependence of `velocity' $\dot g\,g^{-1}$ determines the time-dependence of momentum. The converse is not true, although eq.\ (\ref{s32b}) does imply that $g^{-1}\dot g$ and $\Ad_{g^{-1}}(\cI^{-1}(\Ad^*_g(k)))$ only differ by an element of the Lie algebra of the stabilizer of $k$.
\item[(ii)] Given a quadratic Hamiltonian (\ref{tt20}), non-trivial dynamics only occurs when the inertia operator breaks $G$ symmetry, \ie when in general
\be
\Ad^*_g\circ\cI\circ\Ad_g^{-1}
\neq
\cI.
\label{b20}
\ee
Indeed, suppose instead that $G$ symmetry were preserved, \ie that the inequality (\ref{b20}) were replaced by an equality for all $g\in G$. Then the adjoint and coadjoint representations of $G$ would be equivalent and the reconstruction condition (\ref{ss31b}) would yield $g^{-1}\dot g=\cI^{-1}(k)$, whose solution is $g_t=g_0\,e^{\,t\,\cI^{-1}(k)}$. This occurs \eg in isotropic rigid bodies, which rotate without precession. Such trivialities do not arise in {\it anisotropic} systems, where the inequality (\ref{b20}) holds for at least some (typically most) choices of $g\in G$. In particular, when the adjoint and coadjoint representations of $G$ are inequivalent (as is the case for Virasoro), any inertia operator is necessarily anisotropic.
\end{enumerate}

\subsection{Dynamical phase and Berry phase}
\label{secDYBE}

\begin{figure}[t]
\centering
\includegraphics[width=0.40\textwidth]{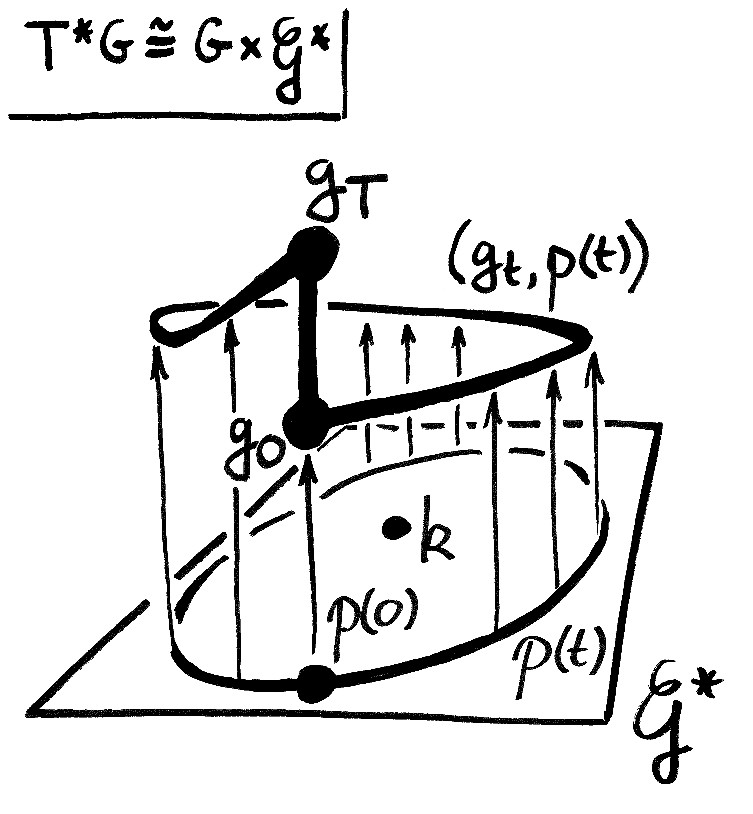}
\caption{The fate of fig.\ \ref{s7b} when the path $p(t)=g_t\cdot k$, in momentum space, is closed. Its reconstruction $(g_t,g_t\cdot k)$ generally contains a curve $g_t$ that does not close, corresponding to a non-trivial group element $g_0^{-1}g_T$. To compensate this effect and close the loop $\bar g_t$ as in (\ref{s32t}), we add a path $g_Th_t$ (depicted here as a vertical segment) where $h_t$ stabilizes $k$.\label{s33}}
\end{figure}

Geometric phases appear when one traces a closed curve in a suitable parameter space \cite{Berry:1984jv,Hannay,Nakahara:2003nw}. In the case at hand, parameter space is momentum space (or rather a coadjoint orbit therein), so we wish to ask the following question: given a solution $p(t)$ of eq.\ (\ref{ss20}) such that $p(T)=p(0)$, is the reconstructed path $(g_t,p(t))$ closed? If not, is there a way to measure the difference between the initial configuration, $g_0$, and the final one, $g_T$? As we now show, the path $g_t$ is typically {\it not} closed even when $p(t)$ is (see fig.\ \ref{s33}), and the degree to which it fails to close is the combination of a dynamical phase, proportional to energy and period, and a Berry phase. Following \cite{Montgomery}, we shall prove this by integrating the Liouville one-form (\ref{ss30}) along a closed path in $G\times\mg^*$ given by Lie-Poisson reconstruction. We will then argue that this integral can be interpreted in two ways: (i) as a Berry phase, (ii) as the sum of a dynamical phase and an observable rotation angle after one period. The centrally extended version of that argument is postponed to section \ref{sec24}.

\paragraph{Integrating the Liouville one-form.} Let $p(t)=g_t\cdot k$ be a closed path, with period $T$, in the orbit $\cO_k$ (notation as in (\ref{s23})). At this stage we do not yet assume that $p(t)$ solves the Lie-Poisson equation (\ref{ss20}), nor that $g_t$ satisfies the reconstruction condition (\ref{ss31b}). Instead, we introduce a loop in $G\times\mg^*$ (see fig.\ \ref{s33}) given by
\be
\gamma_t
=
\big(\bar g_t,\bar g_t\cdot k\big),
\label{ss34}
\ee
where $\bar g_t$ is the concatenation of $g_t$ with a path $h_t$ in the stabilizer of $k$, such that $\bar g_t$ closes:
\be
\bar g_t
=
\left\{
\begin{array}{ll}
g_t & \text{for }0\leq t\leq T,\\
g_T\,h_t & \text{for }T\leq t\leq T'.
\end{array}
\right.
\label{s32t}
\ee
Here $h_t\cdot k=k$ for any $t\in[T,T']$; the starting point of $h_t$ is the identity $h_T=\II$, and its endpoint is $h_{T'}=g_T^{-1}g_0$, which indeed ensures that $\bar g_{T'}=\bar g_0$. The fact that $h_t$ fixes $k$ also ensures that the momentum in (\ref{ss34}) is constant on $[T,T']$, where it equals $p(0)=p(T)=g_T\cdot k$. Aside from these constraints, the choice of path $h_t$ and time $T'$ is arbitrary, and will not affect the final results in any way.\\[-.3cm]%%PARBREAK%%

Having said this, one key assumption is implicit in our definition (\ref{s32t}) of the closed curve $\bar g_t$: we need the group element $h_{T'}=g_T^{-1}g_0$ to belong to the component of the identity in the stabilizer of $k$ (otherwise there would be no way to connect $h_T=\II$ to $h_{T'}=g_T^{-1}g_0$ by a path $h_t$ fixing $k$). Unless the stabilizer is connected, there is no {\it a priori} reason for this to be true: even though $g_T^{-1}g_0$ is certainly connected to the identity in $G$ (namely by the curve $g_t^{-1}g_0$), the connection may not remain if the path is confined to the stabilizer. It might be interesting indeed to find reconstructed Lie-Poisson equations where $g_T^{-1}g_0$ is not stabilizer-connected to the identity, but we will not attempt to do that. Instead, let us introduce a crucial simplifying assumption: {\it from now on, the stabilizer of $k$ is assumed to be a U(1) group}. This may seem restrictive, but it will turn out to be enough for any Lie-Poisson equation based on the Virasoro group --- including KdV.\\[-.3cm]%%PARBREAK%%

Since the stabilizer of $k$ is now connected by assumption, the path $h_t$ needed to close (\ref{s32t}) exists. Accordingly, let us integrate the Liouville one-form (\ref{ss30}) along the loop (\ref{ss34}):
\be
\oint_{\gamma}\cA
=
\int_0^T\!\dd t\,\bigl<k,\Ad_{g^{-1}}\dot g\,g^{-1}\bigr>
+
\int_T^{T'}\!\dd t\,\Bigl<
k,
\Ad_{h^{-1}g_T^{-1}}
\dd(R_{h^{-1}g_T^{-1}})_{g_Th}
\dd(L_{g_T})_h\dot h
\Bigr>.
\ee
Here we split the integral in two pieces coming from the two parts of the path (\ref{ss34}), and $L_g$, $R_g$ respectively denote left and right multiplications by $g$. In the first term, we relate the right Maurer-Cartan form to the left one: $\Ad_{g^{-1}}\dot g\,g^{-1}=g^{-1}\dot g$. In the second term, the integrand simplifies into $\langle k,h^{-1}\dot h\rangle$, which yields
\be
\oint_{\gamma}\cA
=
\int_0^T\!\dd t\,\langle k,g^{-1}\dot g\rangle
+
\int_T^{T'}\!\dd t\,\langle k,h^{-1}\dot h\rangle.
\label{s35}
\ee
This expression is a (known \cite{Marsden90,Marsden,Montgomery}) key result for the rest of this paper. We stress once more that its derivation did not require the path $p(t)=g_t\cdot k$ to solve the Lie-Poisson equation (\ref{ss20}); all we needed was that $p(t)$ be closed (with period $T$) and entirely contained in a single coadjoint orbit.\\[-.3cm]%%PARBREAK%%

Still following \cite{Montgomery}, we shall now provide two interpretations of the integral (\ref{s35}). On the one hand, it will turn out to be the flux of a symplectic form through a surface enclosed by the path $p(t)$, allowing us to think of it as a Berry phase associated with adiabatic changes of reference frames $g_t$. On the other hand, {\it when $p(t)$ solves the Lie-Poisson equation (\ref{ss20}) and provided $g_t$ satisfies the reconstruction condition (\ref{ss31b})}, eq.\ (\ref{s35}) will be the sum of a dynamical phase and a rotation angle $\Delta\phi$ in the $\text{U}(1)$ stabilizer of $k$. Here is the detailed argument:

\paragraph{Eq.\ (\ref{s35}) \texorpdfstring{$=$}~~Berry phase.} To prove this statement, we relate (\ref{s35}) to the symplectic structure of coadjoint orbits of $G$. Indeed, the line integral of $\cA$ along the path (\ref{ss34}) can be rewritten in the group manifold alone, without reference to $\mg^*$:
\be
\oint_{\gamma}\cA
=
\oint_{(\bar g,\bar g\cdot k)}\big(\langle\bar g\cdot k,\dd\bar g\,\bar g^{-1}\rangle,0\big)
=
\oint_{\bar g}\langle k,{\bar g}^{-1}\dd\bar g\rangle
=
\int_{\Sigma_{\bar g}}\dd\langle k,g^{-1}\dd g\rangle.
\label{s41}
\ee
In the last equality we used Stokes' theorem, with $\Sigma_{\bar g}$ an oriented two-dimensional surface in $G$ whose boundary is the closed path $\bar g$. The integrand on the far right-hand side of this expression is a two-form on $G$, and it can be shown (see \eg \cite[sec.\ 5.3.2]{Oblak:2016eij}) that it coincides with the pullback of the Kirillov-Kostant symplectic form on $\cO_k$ by the projection $\Pi:G\rightarrow\cO_k:g\mapsto g\cdot k$. In formulas, this means that
\be
\dd\langle k,g^{-1}\dd g\rangle
=
-(\Pi^*\Omega)_g
\label{random}
\ee
where the symplectic form $\Omega$, defined on the coadjoint orbit of $k$, is such that the Poisson bracket (\ref{ss18}) reads $\{\cF,\cG\}(p)=\Omega_p(\dd\cF_p,\dd\cG_p)$ for any $p\in\cO_k$. Plugging (\ref{random}) back in (\ref{s41}), one finds
\be
\oint_{\gamma}\cA
=
-\int_{\Sigma_{\bar g}}\Pi^*\Omega
=
-\int_{\Pi(\Sigma_{\bar g})}\Omega
=
-\int_{\Sigma_{g\cdot k}}\Omega
\label{s42}
\ee
where $\Sigma_{g\cdot k}$ is any surface in $\cO_k$ whose boundary is the curve $g_t\cdot k=p(t)$. Thus, the integral (\ref{s35}) is the flux (\ref{s42}) of the Kirillov-Kostant symplectic form, and this flux, in turn, can be interpreted as a Berry phase. In fact, it is often true that quantizing a coadjoint orbit $\cO_k$ produces unitary representations of $G$ in which a coherent state, acted upon by transformations tracing a closed path $\bar g_t$, has a Berry connection whose curvature is the symplectic form $\Omega$ \cite{boya2001berry,Oblak:2017ect}. The phase (\ref{s42}) provides a classical analogue of that statement.\\[-.3cm]%%PARBREAK%%

Note that eq.\ (\ref{s42}) is expressed solely in terms of the path $p(t)=g_t\cdot k$ in momentum space --- there is no longer any reference to the path $g_t$ by itself. This fact will play an essential role for the evaluation of the geometric phase (\ref{s35}) in section \ref{secEKO}: it implies that the phase's value is independent of the choice of $g_t$, as long as $p(t)=g_t\cdot k$. In particular, it will allow us to evaluate (\ref{s35}) with relatively simple choices of paths, as opposed to the generally complicated paths produced by the reconstruction condition (\ref{ss31b}).

\paragraph{Eq.\ (\ref{s35}) \texorpdfstring{$=$}~~Dynamical phase \texorpdfstring{$+$}~~Rotation.} So far, since introducing the path (\ref{ss34}), we did not need to assume that $p(t)$ solves the Lie-Poisson equation (\ref{ss20}) or that $g_t$ satisfies the reconstruction condition (\ref{ss31b}). We now enforce both of these assumptions and work out their consequences for the integral (\ref{s35}). To begin, using eqs.\ (\ref{gkk})-(\ref{ss31b}), the integrand of the first term in (\ref{s35}) can be recast as
$\langle k,g^{-1}\dot g\rangle
=
\langle g\cdot k,\dot g g^{-1}\rangle
=
\langle g\cdot k,\cI^{-1}(g\cdot k)\rangle
=
\langle p,\cI^{-1}(p)\rangle
=
2 H(p)$, where $H(p)$ is the Hamiltonian (\ref{tt20}). Since energy $E$ is conserved, $H(p)$ is constant and the first term of (\ref{s35}) becomes a dynamical phase:
\be
\int_0^T\!\dd t\,\langle k,g^{-1}\dot g\rangle
=
2ET.
\label{s36}
\ee
As to the second part of (\ref{s35}), it is a boundary term because the one-form $\langle k,h^{-1}\dd h\rangle$ is exact. In fact, it is essentially the difference between $g_0$ and $g_T$: since the stabilizer is U(1) by assumption, we may label its elements by an angle $\phi$ and write $h^{-1}\dot h=-\dot\phi\,\xi_0$, where $\xi_0\in\mg$ generates the stabilizer. The normalization of $\xi_0$ is fixed so that $e^{2\pi\xi_0}=\II$ be the identity in $G$, but $e^{t\xi_0}\neq\II$ for $t\in(0,2\pi)$. Then the second integral in (\ref{s35}) reads
\be
\int_T^{T'}\!\dd t\,\langle k,h^{-1}\dot h\rangle
=
-\langle k,\xi_0\rangle\big(\phi(T')-\phi(T)\big)
\equiv
-\langle k,\xi_0\rangle\Delta\phi,
\label{Data}
\ee
where $\Delta\phi$ is the angle of the rotation $g_0^{-1}g_T$. As a result, one can write the integral (\ref{s35}) as the sum of the dynamical phase (\ref{s36}) and the angle (\ref{Data}). Equivalently \cite{Marsden90,Marsden,Montgomery},
\be
\Delta\phi
=
\frac{2ET}{\langle k,\xi_0\rangle}
-\frac{1}{\langle k,\xi_0\rangle}\oint_{\gamma}\cA.
\label{s37}
\ee
Note that the value of $\Delta\phi$ depends on the normalization of $\xi_0$, but the product $\xi_0\Delta\phi$ does not, so this ambiguity is merely a matter of `units'. In particular, if $\phi$ is normalized so that one turn corresponds to an angle of $2\pi$ radians (as stated above eq.\ (\ref{Data})), then the normalization of $\xi_0$ becomes fixed uniquely.\\[-.3cm]%%PARBREAK%%

Formula (\ref{s37}) makes it manifest that the complete rotation $\Delta\phi$ is the sum of two very different contributions. The first, proportional energy and period, is a dynamical phase. The second is a geometric phase (\ref{s35}) that coincides with the symplectic flux (\ref{s42}). In section \ref{sec24}, we will apply this statement to centrally extended groups. As we shall see, the extension will affect the Berry phase (\ref{s35}) and add an extra term to the right-hand side of (\ref{s37}). Both modifications have observable consequences in the KdV equation (and more generally in any Lie-Poisson equation for the Virasoro group).

\paragraph{Remark on generic Hamiltonians.} Much of the derivation shown so far carries over to Lie-Poisson systems with {\it arbitrary}, non-quadratic, Hamiltonians (see appendix \ref{appb} for details). Indeed, owing to eq.\ (\ref{s20}), it is still true that $p(t)$ lies entirely in a single coadjoint orbit, so one may ask how the reconstructed path $g_t$ behaves when $p(t)$ is periodic. Thinking of $\mg^*$ as `one half' of the symplectic manifold $T^*G\cong G\times\mg^*$, one can extend the Hamiltonian $H$ on $\mg^*$ into a right-invariant function $\cH(g,p)=H(p)$ on the full phase space $G\times\mg^*$. The ensuing reconstruction condition generalizes eq.\ (\ref{ss31b}):
\be
\dot g_t\,g_t^{-1}
=
\dd H_{p(t)}.
\label{recong}
\ee
The resulting curve $g_t$ is generally not a geodesic; aside from this difference, the arguments of section \ref{sec22} remain valid. The same is true of geometric phases --- up to a point: if $p(t)$ is periodic, one can still define the loop (\ref{s32t}), evaluate the holonomy (\ref{s35}) of the symplectic potential, and interpret the term (\ref{Data}) as a one-period rotation. But a new phenomenon arises in the dynamical phase: in contrast to eq.\ (\ref{s36}), one now has
\be
\int_0^T\!\dd t\,\langle k,g^{-1}\dot g\rangle
=
\int_0^T\!\dd t\,\langle g\cdot k,\dot g\,g^{-1}\rangle
\stackrel{\text{(\ref{recong})}}{=}
\int_0^T\!\dd t\,\langle p(t),\dd H_{p(t)}\rangle,
\label{dypaw}
\ee
where the integrand is generally {\it not} proportional to $H(p(t))$. In fact, proportionality only holds if the Hamiltonian is a homogeneous function, \ie if $H(\lambda p)=\lambda^n\,H(p)$ for any $\lambda\in\RR$. One can then rewrite the dynamical phase as $nET$, and eq.\ (\ref{s37}) holds with $2ET$ replaced by $nET$. However, if $H(p)$ is {\it not} homogeneous, the best one can say is that the dynamical phase is an integral (\ref{dypaw}). It might be interesting to find physical systems of Lie-Poisson type described by non-quadratic Hamiltonians, where dynamical phases would take the somewhat exotic form outlined here. We stress, however, that none of this applies to the KdV equation, whose Hamiltonian is quadratic.

\subsection{Reconstruction for centrally extended groups}
\label{sec24}

We are eventually interested in the reconstructed dynamics of Lie-Poisson equations for the Virasoro group. The latter is a central extension of the group of diffeomorphisms of the circle, so we now describe the extended analogue of sections \ref{sec22} and \ref{secDYBE}. We start with some general preliminaries on centrally extended groups and their Lie-Poisson equations, then briefly analyse their reconstruction, and finally write general formulas for the geometric phases of reconstructed dynamics. The key differences with respect to unextended groups will consist of (i) an extra contribution to the Berry phase and (ii) an `anomalous phase' to be added to dynamical and Berry phases when evaluating $\Delta\phi$. This anomalous phase has, to our knowledge, never appeared in the literature (although reconstruction on the Virasoro group has been studied in \cite{HolmTyr}).

\paragraph{Central extensions and Lie-Poisson equations.} Let $\hG=G\times\RR$ be a central extension of a Lie group $G$. Its elements are pairs $(f,\alpha)$, $(g,\beta)$, {\it etc.}\ with a group law
\be
(f,\alpha)*(g,\beta)
=
\big(fg,\alpha+\beta+\sfC(f,g)\big)
\label{GOREP}
\ee
where $\sfC(f,g)\in\RR$ is a cocycle.\footnote{See \eg \cite[chap.\ 2]{Oblak:2016eij} for this terminology. In short, $\sfC$ is such that the product (\ref{GOREP}) is associative.} The corresponding Lie algebra is $\hg=\mathfrak{g}\oplus\RR$, and its dual space $\hg^*=\mathfrak{g}^*\oplus\RR$ consists of pairs $(p,c)$, where $p\in\mathfrak{g}^*$ and $c\in\RR$, the latter being a central charge. The pairing between $\hg$ and its dual reads $\langle(p,c),(\xi,\alpha)\rangle=\langle p,\xi\rangle+c\alpha$. As before, the coadjoint representation (\ref{coadef}) will play a key role; it turns out to read
\be
(f,\alpha)\cdot(p,c)
=
\Big(\Ad^*_f(p)-\frac{c}{12}\sfS[f^{-1}]\,,\,c\Big),
\label{KIP}
\ee
where the $\Ad^*$ on the right is the coadjoint representation of $G$ (without extension) and $\sfS[f]$ is the {\it Souriau cocycle} associated with $\sfC$, defined by
\be
\langle\sfS[f],\xi\rangle
\equiv
-12
\left.\frac{\dd}{\dd t}\right|_{t=0}
\Big[
\sfC\big(f,e^{t\xi}\big)
+
\sfC\big(f\,e^{t\xi},f^{-1}\big)
\Big].
\label{SORO}
\ee
In the Virasoro group, $\sfS[f]$ will be the Schwarzian derivative that plays an important role in CFT (see section \ref{sec3}). We also need the coadjoint representation of $\hg$, obtained by differentiating (\ref{KIP}), or equivalently given by eq.\ (\ref{coadefa}). One thus finds
\be
\widehat{\text{ad}}{}_{(\xi,\alpha)}^*(p,c)
=
\Big(
\ad^*_{\xi}p+\frac{c}{12}\mathsf{s}[\xi],0
\Big),
\label{cecorep}
\ee
where $\mathsf{s}[\xi]\equiv\der_t|_0\sfS[e^{t\xi}]$ is the infinitesimal Souriau cocycle.\\[-.3cm]%%PARBREAK%%

Having settled these preliminaries, let us turn to Lie-Poisson dynamics. In order to build a quadratic Hamiltonian (\ref{tt20}), consider an extended inertia operator
\be
\widehat{\cI}:
\hg\to\hg^*:
(\xi,\alpha)\mapsto\big(\cI(\xi),J\alpha\big)
\label{INOPP}
\ee
where $\cI$ is an inertia operator (\ref{t20}) for $\mathfrak{g}$, while $J>0$ is a number (whose value will eventually turn out to be irrelevant).\footnote{More generally, one could define `non-diagonal' inertia operators that mix the central extension and the centreless algebra, but we do not need to worry about such possibilities here.} Using the coadjoint representation (\ref{cecorep}), the corresponding Lie-Poisson equation (\ref{ss20}) reads
\be
\big(\dot p,\dot c\big)
=
\Big(
\ad^*_{\cI^{-1}(p)}(p)+\frac{c}{12}\mathsf{s}\big[\cI^{-1}(p)\big],0
\Big).
\label{easport}
\ee
Note, in particular, that the central charge $c$ is a fixed parameter --- this really just follows from its being left invariant by the coadjoint action (\ref{KIP}).

\paragraph{Reconstruction conditions.} Suppose one is given a solution $(p(t),c)$ of eq.\ (\ref{easport}). Lie-Poisson reconstruction consists in finding a path $\big(g_t,\alpha_t\big)$ in $\hG$ such that
\be
\big(p(t),c\big)
=
\big(g_t,\alpha_t\big)\cdot(k,c)
\label{CODODO}
\ee
for some fixed coadjoint vector $k$, where the dot denotes the coadjoint representation (\ref{KIP}) of $\hG$. In addition, the path must be such that the reconstruction condition (\ref{ss31b}) holds. For a centrally extended group, this means that
\be
\big(p(t),c\big)
=
\widehat{\cI}\Big[\der_{\tau}\Big|_{t}\Big(\big(g_{\tau},\alpha_{\tau}\big)*\big(g_t,\alpha_t\big)^{-1}\Big)\Big]
=
\Big(\cI\big(\dot g\,g^{-1}\big),J\dot\alpha+J\,\der_{\tau}\Big|_{t}\sfC\big(g_{\tau},g_t^{-1}\big)\Big).
\label{cereco}
\ee
On the one hand, this yields the expected reconstruction equation $p=\cI(\dot gg^{-1})$, exactly as in the unextended case (\ref{ss31b}). On the other hand, it gives an ordinary differential equation for $\alpha_t$, which is readily solved thanks to the constancy of the central charge $c$ \cite{HolmTyr}:
\be
\alpha_t
=
\alpha_0+\frac{ct}{J}-\int_0^t\!\dd\tau\,\der_s\Big|_{s=\tau}\sfC\big(g_s,g_{\tau}^{-1}\big).
\label{APATE}
\ee
This result will turn out to be crucial for evaluating $\Delta\phi$ in the reconstructed KdV equation. To lighten notation, we define a one-form $\dd\sfC$ on $G$ by $(\dd\sfC)_g(\dot g)=\der_{\tau}\big|_t\sfC(g_{\tau},g_t^{-1})$, whereby the last term of (\ref{APATE}) becomes an integral of $\dd\sfC$ along the path $g_t$.

\paragraph{Geometric phases.} Now let $p(t)$ be a periodic solution of eq.\ (\ref{easport}), with period $T$ and central charge $c$, lying in a coadjoint orbit $\cO_{(k,c)}$. As before, we assume that the stabilizer of $(k,c)$ is a U(1) group consisting of unextended elements of the form $(h,0)$.\footnote{Since the group $\widehat{G}$ is extended, the coadjoint action (\ref{KIP}) ensures that $(k,c)$ is trivially invariant under all `central translations' $(\II,\alpha)$. We mod these out when referring to our `U(1) stabilizer'.} We further assume, without loss of generality, that the cocycle $\sfC$ vanishes on that U(1) subgroup. Our goal is to rewrite eqs.\ (\ref{s35}) and (\ref{s37}) in the centrally extended case.\\[-.3cm]%%PARBREAK%%

To do this, first note that the Berry connection (\ref{ss30}) now gets a central contribution: 
\be
\widehat\cA{}_{((g,\alpha),(p,c))}
=
\big<(p,c),\dd(g,\alpha)\;(g,\alpha)^{-1}\big>
=
\langle p,\dd g\,g^{-1}\rangle
+c\big(\dd\alpha+(\dd\sfC)_g\big).
\label{celi}
\ee
Similarly to (\ref{s32t}), we introduce a loop $\big(\bar g_t,\bar\alpha_t\big)$ in $\hG$ by concatenating the path $(g,\alpha)$, which satisfies (\ref{CODODO}), with a curve $(h,\beta)$ that fixes $k$ and ensures that $(\bar g,\bar\alpha)$ closes. In particular, $\beta_T=0$ and $\beta_{T'}=\alpha_0-\alpha_T+\sfC(g_T^{-1},g_0)$. The integral of the Liouville one-form (\ref{celi}) along the path $\gamma=((\bar g,\bar\alpha),(\bar g,\bar\alpha)\cdot(k,c))$ is then found to be \cite{Oblak:2017ect}
\be
\boxed{
\oint_{\gamma}\widehat\cA
=
\int_0^T\!\dd t\,\Big[\langle k,g^{-1}\dot g\rangle+c\,\der_{\tau}\Big|_{\tau=t}\sfC\big(g_t^{-1},g_{\tau}\big)\Big]
+\int_T^{T'}\!\dd t\,\langle k,h^{-1}\dot h\rangle.
}
\label{cethe}
\ee
This Berry phase is a straightforward generalization of (\ref{s35}), involving just one extra contribution due to the central extension. As before, we stress that its value depends neither on the specific path $g_t$, nor on the parametrization of time. It only depends on the image of the loop $p(t)\in\mg^*$ contained in a certain coadjoint orbit, as in (\ref{s42}). However, in order to interpret (\ref{cethe}) as the sum of a dynamical phase and an observable rotation angle, we need to enforce the reconstruction conditions (\ref{cereco}) and write
\be
\oint_{\gamma}\widehat\cA
=
\int
\big<
(k,c),(g,\alpha)^{-1}\dd(g,\alpha)
\big>
+
\int
\big<
(k,c),(h^{-1}\dd h,\dd\beta)
\big>
\label{KATA}
\ee
where we have split the path $(\bar g,\bar\alpha)$ into a piece $(g,\alpha)$ on the interval $[0,T]$ and a piece $(h,\beta)$ in the stabilizer. The former yields a dynamical phase analogous to (\ref{s36}):
\be
\int\!\!
\big<
(k,c),(g,\alpha)^{-1}\dd(g,\alpha)
\big>
=
\int_0^T\!\!\dd t\,\big<(p,c),\widehat\cI^{-1}(p,c)\big>
=
T\Big(\langle p,\cI^{-1}(p)\rangle+\frac{c^2}{J}\Big)
=
2ET.
\label{ENGIE}
\ee
On the other hand, the second part of (\ref{KATA}) contains several extra terms with respect to the unextended expression (\ref{Data}). Indeed, it reads
\be
\int
\big<
(k,c),(h^{-1}\dd h,\dd\beta)
\big>
=
\int\langle k,h^{-1}\dd h\rangle
-c\,\alpha_T+c\,\alpha_0+c\,\sfC\big(g_T^{-1},g_0\big)
\ee
owing to the conditions $\beta_T=0$ and $\beta_{T'}=\alpha_0-\alpha_T+\sfC(g_T^{-1},g_0)$. Using now the solution (\ref{APATE}) of the reconstruction conditions to evaluate $\alpha_T$, one can rewrite (\ref{KATA}) as
\be
\oint_{\gamma}\widehat\cA
=
2T\Big(E-\frac{c^2}{2J}\Big)
+c\int_g\dd\sfC
+c\,\sfC\big(g_T^{-1},g_0\big)
+\int\langle k,h^{-1}\dd h\rangle.
\ee
Finally, as in (\ref{Data}), we interpret the last term as a rotation angle: recall that we assumed the stabilizer of $k$ to be a U(1) group, generated by $\xi_0\in\mg$. The result is
\be
\boxed{%
\Bigg.
\langle k,\xi_0\rangle
\Delta\phi
=
\underbrace{\bigg.2T\Big(E-\frac{c^2}{2J}\Big)}_{\text{Dynamical}}
\,
\underbrace{-\,\oint_{\gamma}\widehat\cA}_{\text{Berry}}
\,
+
\underbrace{c\int_g\dd\sfC+c\,\sfC\big(g_T^{-1},g_0\big).}_{\text{Anomalous}}
}
\label{KAX}
\ee
This is the centrally extended version of eq.\ (\ref{s37}), and it takes the anticipated form (\ref{DBIB}):
\begin{enumerate}[i]
\setlength\itemsep{0em}
\item[(i)] The first term is the expected dynamical phase, with a subtraction of $c^2/(2J)$ ensuring that $\Delta\phi$ does not depend on the irrelevant parameter $J$ (since $E$ is given by (\ref{ENGIE})). This is as it should be, since $J$ does not affect the Lie-Poisson equation (\ref{easport}).
\item[(ii)] The second term is the Berry phase (\ref{cethe}), given by a loop integral of the Liouville one-form. It is {\it universal} in the sense that it does not depend on the inertia operator.
\item[(iii)] The third and fourth terms are proportional to the central charge, are directly due to the extension $\sfC$, and are also universal. In particular, the third term is a line integral of $c(\dd\sfC)_g=-\langle(0,c),(g,0)\dd(g,0)^{-1}\rangle$. This is the Berry connection on the coadjoint orbit of $(0,c)$, evaluated at the point $g^{-1}$. Thus, the anomalous phase in (\ref{KAX}) is akin to an `inverse Berry phase', except that the path $(g_t^{-1},0)\cdot(0,c)$ is {\it not} closed in general.
\end{enumerate}
We now apply the result (\ref{KAX}) to Lie-Poisson equations based on the Virasoro group, taking KdV as our main example. In general, all three terms of (\ref{KAX}) will be non-zero.

\section{Geometric phases and drift in reconstructed KdV}
\label{secEKO}

This section is devoted to the first key statement of our work. Namely, we focus on wave profiles that are amenable (\ie that can be mapped on a constant thanks to suitable diffeomorphisms) and satisfy a Lie-Poisson equation for the Virasoro group, such as KdV. We then show that, for periodic waves $p(x,t)$, the master equations (\ref{cethe})-(\ref{KAX}) apply and reflect a non-trivial one-period rotation of the reconstructed dynamics. The corresponding angle $\Delta\phi$ measures the drift velocity of `fluid particles' governed by eq.\ (\ref{sb1}). Furthermore, $\Delta\phi$ is the sum of a dynamical phase, a Berry phase and an anomalous phase, all of which can be written explicitly as functionals of either the reconstructed path, or its projection on the coadjoint orbit of $p$. The Berry and anomalous phases are universal (they follow solely from the Virasoro group structure), and the Berry phase in particular takes the form described in \cite{Oblak:2017ect}.\\[-.3cm]%%PARBREAK%%

Before displaying these results, we briefly review some elementary properties of the Virasoro group and its relation to the KdV equation. For more background material, we refer \eg to \cite{Oblak:2019llc} and its appendix A; our notation and conventions will follow those of that paper. Much more detailed, pedagogical accounts of the Virasoro group and its coadjoint orbits \cite{Lazutkin} can be found \eg in \cite{Witten:1987ty,Balog:1997zz,Guieu,Khesin,Oblak:2016eij}. Finally, note that applications of the formulas of this section to travelling waves are postponed to section \ref{secNUM}.

\subsection{Virasoro group and KdV equation}
\label{sec3}

Here we briefly recall how the KdV equation can be derived as a Lie-Poisson equation (\ref{ss20})-(\ref{easport}) associated with the Virasoro group. For many more details on this relation and its generalizations, see \cite{Khesin}. Up to a different choice of inertia operator, the same construction leads to the Hunter-Saxton and Camassa-Holm equations \cite{Khesin2003}.\\[-.3cm]%%PARBREAK%%

The {\it Virasoro group}, which we denote $\hDiff$, is the central extension of the group $\Diff$ of diffeomorphisms of the circle. Accordingly, let $x\in\RR$ be a $2\pi$-periodic coordinate. An element of the Virasoro group is a pair $(f,\alpha)$, where $\alpha$ is a real number while the function $f\in\Diff$ is an (orientation-preserving) diffeomorphism, such that $f'(x)>0$ and $f(x+2\pi)=f(x)+2\pi$.\footnote{To be precise, what we are describing here are the {\it universal covers} of $\Diff$ and $\hDiff$.} For example, a rotation by $\theta$ reads $f(x)=x+\theta$, which we denote as $\cR_{\theta}(x)$ from now on (rotations will soon play a prominent role). The group law is, by definition, of the form (\ref{GOREP}):
\be
(f,\alpha)*(g,\beta)
=
\big(%
f\circ g,
\alpha+\beta
+\sfC(f,g)\big),
\qquad
\sfC(f,g)
\equiv
-\frac{1}{48\pi}\int_0^{2\pi}\!\!\dd x\,\log(f'\circ g)\frac{g''}{g'}
\label{vigolaw}
\ee
where $\circ$ denotes composition and $\sfC$ is the {\it Bott cocycle} \cite{Bott}. For future reference, note that $\sfC$ vanishes on rotations: if either $f$, $g$ or $f\circ g$ is a rotation, then $\sfC(f,g)=0$.\\[-.3cm]%%PARBREAK%%

We let $\hVect$ denote the Lie algebra of $\hDiff$ --- the {\it Virasoro algebra}. Its elements are pairs $(\xi,\alpha)$, where $\xi=\xi(x)\der_x\in\Vect$ is a vector field on the circle and $\alpha\in\RR$ as before. Its dual space, $(\hVect)^*$, consists of pairs $(p,c)$, where $p=p(x)\dd x^2$ is a quadratic density and $c\in\RR$ is a central charge.\footnote{Both vector fields and quadratic densities are $2\pi$-periodic: $\xi(x+2\pi)=\xi(x)$, $p(x+2\pi)=p(x)$.} The pairing between $\hVect$ and its dual is
\be
\big<(p,c),(\xi,\alpha)\big>
\equiv
\frac{1}{2\pi}\int_0^{2\pi}\dd x\,p(x)\xi(x)+c\,\alpha.
\label{PARA}
\ee
In two-dimensional conformal field theory (CFT), $p$ is interpreted as a (chiral component of the) stress tensor and (\ref{PARA}) is the Noether charge of the conformal generator $\xi$. In the KdV context and its cousins, $p(x)$ is a wave profile, governed by a Lie-Poisson equation (\ref{easport}). From that perspective, $p(x)$ is a `momentum vector', which justifies our notation.\\[-.3cm]%%PARBREAK%%

We now derive KdV from Virasoro group theory. To begin, we need the coadjoint representation (\ref{KIP}), which we write as $(f,\alpha)\cdot(p,c)=(f\cdot p,c)$ thanks to the fact that the central charge is invariant. Using the Bott cocycle (\ref{vigolaw}) and the Souriau construction (\ref{SORO}), one can show that \cite{Guieu,Oblak:2016eij}
\be
\big(f\cdot p\big)(x)
=
\big[\big(f^{-1}\big)'(x)\big]^2\,p\big(f^{-1}(x)\big)
-\frac{c}{12}\bigg[\frac{(f^{-1})'''}{(f^{-1})'}-\frac{3}{2}\bigg(\frac{(f^{-1})''}{(f^{-1})'}\bigg)^2\bigg]\bigg|_x
\label{sa66}
\ee
where $f^{-1}$ is the inverse of $f$, such that $f^{-1}(f(x))=f(f^{-1}(x))=x$. This is the standard transformation law of a CFT stress tensor under conformal transformations \cite[sec.\ 5.4]{DiFran}. In particular, the combination of derivatives of $f^{-1}$ multiplying $c/12$ is the {\it Schwarzian derivative} of $f^{-1}$: the Virasoro version of the Souriau cocycle (\ref{SORO}). As a result, the coadjoint representation (\ref{cecorep}) of the Virasoro algebra reads
\be
\widehat{\text{ad}}{}^*_{(\xi,\alpha)}(p,c)
=
\Big(-\xi p'-2\xi'p+\frac{c}{12}\xi''',0\Big).
\label{s65b}
\ee
The vanishing second entry confirms that the central charge is constant in time, for any choice of inertia operator. By contrast, $p(x)$ transforms non-trivially under the Virasoro group, so it will generally have a non-trivial time evolution. Specifically, we choose the inertia operator to be the simplest possible map of the form (\ref{INOPP}):
\be
\widehat\cI:
\hVect\to\hVect{}^*:
\big(\xi(x)\der_x,\alpha\big)\mapsto\big(\xi(x)\dd x^2,J\alpha\big),
\label{s66}
\ee
where $J$ is an arbitrary (and ultimately irrelevant) positive constant. This choice ensures that $\widehat\cI$ is invertible and self-adjoint (recall the definition around (\ref{t20})), so it is indeed an inertia operator. It is also anisotropic in the sense of eq.\ (\ref{b20}), since the adjoint and coadjoint representations of Virasoro are inequivalent. More complicated inertia operators yield different wave equations, such as Hunter-Saxton and Camassa-Holm \cite{Khesin2003,Khesin}. We will not consider these more general cases in detail, but our approach also applies to them up to straightforward modifications of all expressions involving $\widehat\cI$. (See in particular the remark at the very end of section \ref{secEPP}.)\\[-.3cm]%%PARBREAK%%

From now on, all integrals over $x$ are implicitly evaluated on the interval $[0,2\pi]$. The quadratic Hamiltonian (\ref{tt20}) induced by the inertia operator (\ref{s66}) then reads
\be
H[p]
=
\frac{1}{4\pi}\int\!\dd x\,p(x)^2
+\frac{c^2}{2J}.
\label{ss66}
\ee
As $p(x)\dd x^2$ transforms according to eq.\ (\ref{sa66}) (or (\ref{s65b}) in infinitesimal terms), this expression is manifestly not Virasoro-invariant. This implies that the resulting Lie-Poisson equation (\ref{ss20})-(\ref{easport}) is non-trivial; using (\ref{s65b}), one finds indeed
\be
\dot p+3pp'-\frac{c}{12}p'''
=
0,
\label{s67}
\ee
where, as in (\ref{easport}), the central charge $c$ is a constant parameter. This is the {\it Korteweg-de Vries equation} (\ref{KADA}) for the field $p(x,t)$, derived here as a Lie-Poisson equation of Virasoro.

\subsection{Reconstruction and phases for periodic waves}
\label{secEPP}

We now describe reconstructed dynamics in the (cotangent bundle of the) Virasoro group when $p(x,t)$ is a periodic solution of (\ref{s67}), such that $p(x,t+T)=p(x,t)$. This is the setup of section \ref{sec2}, so eqs.\ (\ref{cethe}) and (\ref{KAX}) apply. At this stage, we adopt an abstract viewpoint without reference to particle motion, and without assuming that $p(x,t)$ is a travelling wave --- these problems will be addressed in sections \ref{secSTOKES} and \ref{secNUM}, respectively. We refer again to \cite[chap.\ 4-6]{Guieu} and \cite[chap.\ 6-7]{Oblak:2016eij} for some background on the Virasoro group, especially its coadjoint orbits \cite{Lazutkin,Witten:1987ty}, which will now start playing an important role.

\paragraph{Amenable profiles.} As in section \ref{sec2}, the motion of $p(x,t)$ determines a path $(g_t,\alpha_t)$ --- actually a geodesic --- in the Virasoro group, and our task is to find the difference between $g_T$ and $g_0$ when $p$ has period $T$. Before doing that, however, we need to state one key simplifying assumption: from now on, we require the wave profile $p(x,t)$ to be {\it amenable}, that is, conformally equivalent to a {\it uniform} ($x$-independent) configuration $k$. In other words, we assume there exists a constant $k$ and a diffeomorphism $g_0\in\Diff$ such that, at time $t=0$,
\be
p(x,0)
=
(g_0\cdot k)(x),
\label{sa6}
\ee
where the dot denotes the coadjoint action (\ref{sa66}). One can show that the value of $k$ is uniquely fixed by $p$.\footnote{\label{tobemode}That value follows from the trace of the monodromy matrix of Hill's equation given by $p$ (see \cite{Balog:1997zz}). In section \ref{secCNO}, we display $k$ explicitly for cnoidal waves \cite{Oblak:2019llc}. As mentioned in footnote \ref{tobemade} and shown in appendix \ref{appb}, $k$ is the value of a momentum map evaluated along a solution $(g_t,p(t))$ of reconstructed KdV. This map is the `Noether charge' associated with the right action of the Virasoro group on its cotangent bundle, \ie with `particle relabelling symmetry' in the language of fluid mechanics.} The map $g_0$ can then be interpreted as a `boost' (analogously to Lorentz boosts) that sends the uniform profile $k$ on $p(x,0)$. The ensuing path $g_t$ consists of diffeomorphisms that may be seen, from the fluid dynamics perspective, as changes of coordinates mapping the `Lagrangian' reference frame, where fluid particles are uniformly distributed on the circle, to the `Eulerian' one, where the wave profile is non-uniform.\\[-.3cm]%%PARBREAK%%

Since $p(x,t)$ solves the KdV equation, it is confined to a coadjoint orbit (\ref{s23}) throughout time evolution, so eq.\ (\ref{sa6}) guarantees that $p(x,t)$ is conformally equivalent to $k$ at any time. Note that the requirement of amenability is restrictive: a great many wave profiles are {\it not} conformally equivalent to uniform configurations; a prominent example is provided by cnoidal waves with sufficient pointedness \cite{Oblak:2019llc}, to which we shall return in section \ref{secCNO}. Despite this, we do wish to stick to the assumption that $p\in\cO_k$ for some constant $k$, since it implies that the stabilizer of the orbit is isomorphic to $\text{U}(1)$, as in section \ref{sec2}. Indeed, for almost all values of the uniform profile $k$, the set of $\Diff$ elements leaving it fixed (in the sense that $h\cdot k=k$) is exactly the group of rigid rotations $h(x)=x+\theta$, generated by $\xi_0=\der_x$. As a result, the time periodicity of $p(x,t)$ guarantees that the reconstructed diffeomorphism $g_0^{-1}\circ g_T$ is a rotation. Our task is to express the angle of that rotation in terms of observable wave data.\\[-.3cm]%%PARBREAK%%

This reasoning suffers from a minor caveat. Indeed, there exists a discrete family of amenable profiles whose stabilizer is {\it not} isomorphic to U(1): given an integer $n\neq0$, the {\it exceptional constant} $k=-n^2c/24$ is fixed by a three-parameter group locally isomorphic to SL$(2,\RR)$ \cite{Lazutkin,Witten:1987ty}. In that case, it is no longer true that the one-period reconstruction $g_0^{-1}\circ g_T$ is a mere rotation. Furthermore, the Bott cocycle (\ref{vigolaw}) does {\it not} vanish on such SL$(2,\RR)$ transformations, so several of the key formulas derived in section \ref{sec24} need to be modified. To avoid this problem in what follows, it will always be understood that any amenable profile is also {\it generic} in the sense that the corresponding constant $k$ does {\it not} equal $-n^2c/24$ for some non-zero integer $n$; genericity thus ensures that $k$ has a U(1) stabilizer, as described above. The extension of our arguments to exceptional profiles appears to be a rich and subtle issue, which we leave for future work.

\paragraph{Geometric phases in KdV.} In order to find the angle of the rotation $g_0^{-1}\circ g_T$, let us apply Lie-Poisson reconstruction to a (generic) amenable periodic solution $(p(x,t),c)$ of KdV. Given this wave, the reconstruction condition (\ref{cereco}) reads
\be
\Big(p(t),\frac{c}{J}\Big)
=
\der_{\tau}\Big|_t\Big[\big(g_{\tau},\alpha_{\tau}\big)*\big(g_t,\alpha_t\big)^{-1}\Big]
=
\bigg(
\dot g_t\circ g_t^{-1},\dot\alpha_t
-\frac{1}{48\pi}\int\!\dd x\,\dot g_t'\circ g_t^{-1}\,(g_t^{-1})''
\bigg)
\label{KAKOR}
\ee
where we used the multiplication (\ref{vigolaw}) of the Virasoro group. On the far right-hand side, the first entry yields the reconstruction condition that one would find, without central extension, in the group $\Diff$:
\be
p(x,t)
=
\der_{\tau}\big|_{\tau=t}\,g_{\tau}\big(g_t^{-1}(x)\big),
\qquad\text{\ie}\qquad
\der_tg_t^{-1}(x)+p(x,t)\der_xg_t^{-1}(x)=0.
\label{sa4}
\ee
The initial condition $g_0(x)$ is free in principle, but we choose it to satisfy eq.\ (\ref{sa6}). As for the path $\alpha_t\in\RR$, one can solve eq.\ (\ref{KAKOR}) similarly to (\ref{APATE}) and find \cite{HolmTyr}
\be
\alpha_t
=
\alpha_0+\frac{ct}{J}
+\frac{1}{48\pi}\int_0^t\dd\tau\int\!\dd x\,\dot g_{\tau}'\circ g_{\tau}^{-1}\,(g_{\tau}^{-1})''.
\ee
This will eventually contribute to $\Delta\phi$ through the `anomalous phase' of eq.\ (\ref{KAX}).\\[-.3cm]%%PARBREAK%%

As stressed earlier, choosing $g_0$ to satisfy (\ref{sa6}) for some generic constant $k$, along with the periodicity of $p$, ensures that $g_0^{-1}\circ g_T$ is a rotation (since $g_0\cdot k=g_T\cdot k$, and $k$ is only stabilized by rotations). We now use eq.\ (\ref{KAX}) to compute the angle of that rotation as the sum of a dynamical phase, a Berry phase and an anomalous term:
\begin{enumerate}[i]
\item[(i)] The dynamical phase (\ref{ENGIE}) involves the energy, given by the Hamiltonian (\ref{ss66}):
\be
E-\frac{c^2}{2J}
=
\frac{1}{4\pi}\int\!\dd x\,p(x,0)^2.
\label{sa9}
\ee
Energy being conserved, one may evaluate it at any time $t$; here we chose $t=0$. As stressed below (\ref{KAX}), the right-hand side of this expression is independent of $J$, as it should be, since $J$ affects neither $p(x,t)$, nor the reconstruction condition (\ref{sa4}).
\item[(ii)] The Berry phase (\ref{cethe}) is, in fact, standard: as was shown in \cite{Alekseev:1988ce}, line integrals of the symplectic potential can be computed in closed form; they are `geometric actions' for the Virasoro group, and were later interpreted as Berry phases associated with adiabatic conformal transformations \cite{Oblak:2017ect}. Importantly, these phases only depend on the image of the path $p(t)$ in momentum space --- not on the reconstructed curve $g_t$ in the group manifold. As a result, one is free to express the Berry phase in terms of any path $f_t$ in $\Diff$ such that the image of $f_t\cdot k$ coincides with $p(t)$. We choose such a path $f$. Then, adapting the notation of \cite{Oblak:2017ect} to the case at hand,\footnote{What we call $k$ here is denoted as $h-c/24$ in \cite{Oblak:2017ect}, where $h$ is the weight of a primary state.} eq.\ (\ref{cethe}) becomes
\be
\oint_{\gamma}\widehat\cA
=
\int\frac{\dd t\,\dd x}{2\pi}\frac{\dot f}{f'}
\bigg[k+\frac{c}{24}\bigg(\frac{f''}{f'}\bigg)'\bigg]
-kf_0^{-1}(f_T(0)),
\label{sa8}
\ee
being understood that the integrals over $t$ and $x$ run from $0$ to $T$ and $2\pi$, respectively. 
\item[(iii)] Finally, since $g_T^{-1}\circ g_0$ is a rotation and since the cocycle (\ref{vigolaw}) vanishes on rotations, the very last term of eq.\ (\ref{KAX}) does not contribute. The only non-zero contribution to the anomalous phase comes from the integral of the derivative of $\sfC$, namely
\be
\label{KLO}
\int\!\dd t\,\der_{\tau}\Big|_t\sfC\big(g_{\tau},g_t^{-1}\big)
\stackrel{(\text{\ref{vigolaw}})}{=}
-\int\frac{\dd t\,\dd x}{48\pi}\,\dot g'\circ g^{-1}(g^{-1})''
=
-\int\frac{\dd t\,\dd x}{48\pi}\,\frac{\der_tg^{-1}}{(g^{-1})'}\bigg(\frac{(g^{-1})''}{(g^{-1})'}\bigg)',
\ee
where we used the properties $\dot g\circ g^{-1}+g'\circ g^{-1}\,\der_tg^{-1}=0$ and $g'\circ g^{-1}\,(g^{-1})'=1$, along with an integration by parts.
\end{enumerate}
Combining eqs.\ (\ref{sa9}), (\ref{sa8}) and (\ref{KLO}), we can finally write the angle $\Delta\phi$, given by (\ref{KAX}), as
\be
\!\!\!\!\!\!\!\boxed{%
\Bigg.
k\Delta\phi
=
\underbrace{\frac{T}{2\pi}\int\!\!\dd x\,p^2}_{\text{Dynamical}}
+\underbrace{kf_0^{-1}(f_T(0))-\int\!\frac{\dd t\,\dd x}{2\pi}\frac{\dot f}{f'}\bigg[k+\frac{c}{24}\bigg(\frac{f''}{f'}\bigg)'\bigg]}_{\text{Berry}}
-\underbrace{\frac{c}{24}\int\!\frac{\dd t\,\dd x}{2\pi}\frac{\der_tg^{-1}}{(g^{-1})'}\bigg(\frac{(g^{-1})''}{(g^{-1})'}\bigg)'}_{\text{Anomalous}}}
\label{DPHI}
\ee
where we used $\xi_0=\der_x$ to simplify $\langle k,\xi_0\rangle=k$. This takes the anticipated form (\ref{DBIB}) and contains two geometric phases: one is a Virasoro Berry phase \cite{Oblak:2017ect}, the other is an `anomalous phase' which we deliberately wrote in a way that exhibits its similarity with the Berry term. We stress, however, that the path $g$ in the anomalous phase is the reconstructed curve that satisfies (\ref{sa4}), whereas the Berry phase involves any path $f$ such that the curve $f_t\cdot k$ coincides with $p(t)$. This distinction will allow us to evaluate the Berry phase easily for travelling waves, while the anomalous one will require a bit more work.\\[-.3cm]%%PARBREAK%%

Two concluding remarks are in order before relating $\Delta\phi$ to a drift velocity:
\begin{enumerate}[i]
\setlength\itemsep{0em}
\item[(i)] The overall factor $k$ on the left-hand side of (\ref{DPHI}) implies that, at $k=0$, the right-hand side of (\ref{DPHI}) vanishes. However, since $c\neq0$ in general, one may well have $\Delta\phi\neq0$ even for $k=0$; cnoidal waves (section \ref{secCNO}) will provide an example of this.
\item[(ii)] Eq.\ (\ref{DPHI}) is quite universal: aside from the model-dependent dynamical phase, it applies to any Lie-Poisson system based on the Virasoro group, such as the Hunter-Saxton and Camassa-Holm equations \cite{Hunter}. To illustrate this point, note that both cases have an inertia operator $\cI(\xi)=\big(a\xi(x)-b\xi''(x)\big)\dd x^2$ with positive constants $a,b$ \cite{Khesin2003}, generalizing the choice $a=1$, $b=0$ in (\ref{s66}).\footnote{The Hunter-Saxton equation has $a=0$, leading to a degenerate metric on $\Diff$ \cite{Khesin}. This issue may complicate the interpretation of $\Delta\phi$, but we will not address it here.} Then the inverse $\cI^{-1}$ maps $p(x)$ on its convolution with the $2\pi$-periodic Green's function $\sfG(x)$ such that
\be
\sfG(x)
\equiv
\frac{1}{2\sqrt{ab}}
\frac{\cosh\big(\sqrt{\frac{a}{b}}(x-\pi)\big)}{\sinh\big(\sqrt{\frac{a}{b}}\pi\big)}
\qquad
\text{for }x\in[0,2\pi].
\ee
The corresponding dynamical phase involves an energy whose expression in terms of `velocity' $\xi$ is a local functional $\propto\int\dd x(a\xi^2+b\xi'^2)$, \ie a non-local functional
\be
E-\frac{c^2}{2J}
=
\int
\frac{\dd x}{4\pi}\,
p(x)\big(\cI^{-1}(p)\big)(x)
=
\int
\frac{\dd x\,\dd y}{4\pi}\,p(x)\sfG(x-y)p(y)
\ee
in terms of `momentum' $p$. Excepting this difference, eq.\ (\ref{DPHI}) remains unchanged.
\end{enumerate}

\subsection{Drift velocity as a Poincar\'e rotation number}
\label{secSTOKES}

We now return to the reconstruction equation (\ref{sa4}) to explain how the angle (\ref{DPHI}) can be observed by monitoring the motion of `fluid particles' as defined by eq.\ (\ref{sb1}). Again, we impose no restrictions on $p(x,t)$ other than amenability and periodicity in space and time (so $p(x,t)$ could, for instance, consist of colliding periodic solitons with rational phase shift). At the end of this section, we will comment further on the (in)applicability of our approach to actual fluid dynamics, owing to a subtlety in reference frames that we already alluded to in the \hyperref[secIN]{introduction}. The application of our arguments to travelling waves is postponed to section \ref{secNUM}.

\paragraph{Particle drift as reconstruction.} Consider a `fluid particle' on the real line whose position $x(t)$ satisfies the equation of motion (\ref{sb1}) in terms of the (given) wave profile $p$. We claim that this equation is equivalent to the reconstruction condition (\ref{sa4}).\footnote{For the record, this is a well known fact; see \eg \cite[sec.\ 2]{HolmTyr}.} Indeed, let $X(t,x_0)$ be the unique solution of (\ref{sb1}) with initial condition $X(0,x_0)=x_0$. One can think of this solution as a time-dependent diffeomorphism $g_t$, with an arbitrary initial configuration $g_0$, acting on a suitable starting point: $X(t,x_0)\equiv g_t\big(g_0^{-1}(x_0)\big)$. Then, in terms of $g_t$, eq.\ (\ref{sb1}) becomes
$\dot g_t\big(g_0^{-1}(x_0)\big)
=
\big(p(t)\circ g_t\big)(g_0^{-1}(x_0))$.
Since this holds for all $x_0$, we may remove the argument $g_0^{-1}(x_0)$ and deduce that $g_t$ satisfies the reconstruction condition (\ref{sa4}), as announced. Conversely, the condition (\ref{sa4}) may be seen as an equation of motion for (comoving) fluid particles. This is true for any $g_0$, but from now on we always let $g_0$ be a uniformizing map that satisfies eq.\ (\ref{sa6}).\\[-.3cm]%%PARBREAK%%

Relating reconstruction to particle motion suggests a way to observe the angle $\Delta\phi$ computed in (\ref{DPHI}). Indeed, consider the following question: given a particle with initial position $x_0$ and equation of motion (\ref{sb1}), what is the particle's position after one period? We can certainly write $x(t)=g_t(g_0^{-1}(x_0))$ in terms of the reconstructed curve $g_t$, since this is the unique solution of (\ref{sb1}) with initial condition $x_0$. After one period, one has
\be
x(T)
=
g_0\Big(g_0^{-1}\circ g_T\big(g_0^{-1}(x_0)\big)\Big).
\label{OneQuestion}
\ee
Now recall the crucial fact, due to the periodicity and amenability of $p$, that $g_0^{-1}\circ g_T$ is a rotation by $\Delta\phi$ (given by eq.\ (\ref{DPHI}) when $p$ solves KdV). As a result, after $N$ periods,
\be
x(NT)
=
g_0\big(g_0^{-1}(x_0)+N\Delta\phi\big),
\label{GAZO}
\ee
and we may identify the map $F$ in eq.\ (\ref{FEQ}) with the composition $g_0\circ\cR_{\Delta\phi}\circ g_0^{-1}$, where $\cR_{\Delta\phi}(x)\equiv x+\Delta\phi$. Stroboscopic particle motion is thus a discrete-time dynamical system governed by iterations of the diffeomorphism $F=g_0\circ\cR_{\Delta\phi}\circ g_0^{-1}$. The latter is conjugate to a rotation by $\Delta\phi$, which allows us to exploit a key result on circle dynamics: the {\it Poincar\'e rotation number} of $F$, as defined in (\ref{DAFA}), is conjugation-invariant\footnote{One also says that the rotation number is a {\it class function} on $\Diff$.} and therefore coincides with $\Delta\phi$ \cite[sec.\ 4.4.3]{Guieu}. In terms of the particle's position, this statement reads
\be
\Delta\phi
=
\lim_{N\to+\infty}\frac{x(NT)-x(0)}{N}
\equiv v_{\text{Drift}}\,T,
\label{VATA}
\ee
where the drift velocity is defined as in (\ref{DAFA}). From that perspective, $\Delta\phi$ yields the average rotation of a particle during one period --- which answers the question raised above (\ref{OneQuestion}). Note that $\Delta\phi$ is independent of the particle's initial position $x(0)$, as is the drift velocity.\\[-.3cm]%%PARBREAK%%

We have thus shown that `particle motion' in the sense of eq.\ (\ref{sb1}) provides a system whose late-time behaviour is directly sensitive to the angle (\ref{DPHI}) through the drift velocity (\ref{DAFA})-(\ref{VATA}). In particular, this velocity contains a contribution due to a Virasoro Berry phase \cite{Oblak:2017ect}, somewhat analogously to the crest slowdown found in \cite{Banner} for breaking waves whose envelope is described by the nonlinear Schr\"odinger equation. Note that this prediction is independent of the uniformizing map $g_0$ that satisfies (\ref{sa6}); in fact, that map is generally unknown (even when the value of $k$ is known for a given $p(x,t)$). In the next section we will study the drift velocity in travelling waves satisfying KdV, and in that case we will actually manage to find $g_0$ analytically.

\paragraph{Comparison with fluid dynamics.} At this point, it is worth comparing our approach, and in particular the drift velocity defined in (\ref{DAFA})-(\ref{VATA}), to the Stokes drift of fluid particles in shallow water \cite{Longuet53}. Indeed, within the KdV approximation of fluid mechanics in a shallow layer (see \eg \cite{Ockendon}), eq.\ (\ref{KADA}) describes the slow time evolution of a right-moving wave $p(x,t)$. Here, $x$ is emphatically {\it not} a fix laboratory coordinate, but rather a (dimensionless) `light-cone', or comoving, coordinate
\be
x=X-Ct
\label{XXT}
\ee
where $X$ is a static laboratory coordinate, $t$ is the (dimensionless) slow time variable, and $C\gg 1$ is a dimensionless remnant of the standard velocity $\sqrt{gh}$ of gravity waves of average depth $h$ in a gravitational field $g$. In fact, $C\propto L^2/h^2$, where $L$ is the (dimensionful) wavelength. The KdV approximation then holds in the `non-relativistic' limit $h/L\to0$, \ie $C\to\infty$. In that limit, the leading velocity of fluid particles is purely horizontal and given by an equation of motion that closely resembles, yet is crucially different from, eq.\ (\ref{sb1}) above. Indeed, in terms of the static laboratory coordinate $X$, particle motion reads
\be
\dot X(t)
=
2\,p\big(X(t)-Ct,t\big),
\label{XDOT}
\ee
which differs from eq.\ (\ref{sb1}) in two respects: (i) the spatial argument of $p$, and (ii) the factor $2$ on the right-hand side. Equivalently, in terms of the comoving coordinate $x$, one has $\dot x=2p(x,t)-C$, which obviously differs from eq.\ (\ref{sb1}).\\[-.3cm]%%PARBREAK%%

One can then repeat the question raised above: given a periodic wave train, what is the drift velocity of $X(t)$? This is the velocity that would presumably be seen in a laboratory, and it is tempting to hope that it is related to the one we introduced (\ref{VATA}). However, the latter was defined from the equation of motion (\ref{sb1}) in the comoving (`lightcone') frame, and it is quite clear that the drift velocity in the laboratory frame, due to eq.\ (\ref{XDOT}), will take a very different form because of the extra dominant term $C\propto L^2/h^2$. For instance, at leading order in $h/L$, the particle satisfying (\ref{XDOT}) sees a fast average of the wave profile, and its position at time $t$ (assuming $t$ is of order one) is simply
\be
X(t)\sim X(0)+\frac{t}{2\pi}\int_0^{2\pi}\dd x\,p(x,0)+\cO(h^2/L^2).
\ee
The drift velocity then coincides with the average of the wave profile (this average is constant along KdV time evolution), which is very different indeed from the prediction (\ref{DPHI}).\footnote{Incidentally, in fluid dynamics, one typically lets the average of $p$ vanish, since it must not contribute to the fixed average fluid depth $h$; but this subtlety is irrelevant to the point we wish to make here, namely that the drift velocity of eq.\ (\ref{XDOT}) is completely different from the one predicted by (\ref{sb1}).} In addition, independently of the dominant velocity $C$, the factor $2$ in eq.\ (\ref{XDOT}) may forbid any interpretation of Stokes drift in shallow water as Lie-Poisson reconstruction. (Intuitively, the solution of the reconstruction equation $\dot g\,g^{-1}=p$ can be written as a time-ordered exponential of $p$, which is exceedingly sensitive to the normalization of $p$.) So, while the Berry phases and drift studied here do have some similarities with fluid dynamics, they do not, ultimately, describe the same phenomenology.

\section{Particle drift and phases of travelling waves}
\label{secNUM}

Travelling waves are a prominent class of solutions of the KdV equation (and of wave equations in general): they take the form $p(x,t)=p(x-vt)$ for some velocity $v$, so their shape is constant throughout time evolution. When the profile $p(x)$ is $2\pi$-periodic in space, such travelling waves are automatically time-periodic with period $T=2\pi/|v|$.\\[-.3cm]%%PARBREAK%%

In this section, we investigate the reconstruction equations (\ref{sb1})-(\ref{sa4}) for travelling waves that solve KdV. For amenable profiles, we show that these equations are {\it integrable} --- they can be solved exactly in terms of known wave data ---, and we build an explicit uniformizing map from which an exact expression for the drift velocity (\ref{DAFA}) follows. We then use this to derive a simplified formula for the rotation angle (\ref{DPHI}). Finally, we apply this to cnoidal waves and obtain a detailed picture of drift velocity throughout the cnoidal parameter space. In particular, we exhibit `orbital bifurcations' that occur at the boundaries of a resonance wedge anticipated in \cite{Oblak:2019llc}: in the wedge, particle motion is locked to the wave and $v_{\text{Drift}}=v$, while no such locking occurs outside of the wedge and $v_{\text{Drift}}\sim v/3$ at large $|v|$. Thus, drift velocity emerges in this picture as a diagnostic of wave amenability, \ie of the nature of Virasoro coadjoint orbits.

\subsection{Exact reconstruction for travelling waves}
\label{secuni}

Here we study the reconstruction equations (\ref{sb1})-(\ref{sa4}), without reference to geometric phases for now. Our goal is to show that, for amenable travelling waves, these equations can be solved exactly in terms of readily accessible wave data. An explicit expression for the drift velocity will follow. The comparison to geometric phases and formula (\ref{DPHI}) is postponed to section \ref{secGET}.

\paragraph{Amenable travelling waves.} Consider a travelling wave $p(x,t)=p(x-vt)$ that solves the KdV equation (\ref{KADA}) with some non-zero velocity $v$. Since $\dot p=-vp'$, the KdV equation can be written entirely in terms of the time-independent profile $p(x)$. Upon integrating KdV once, one finds that there must exist a constant $A$ such that
\be
-vp+\frac{3}{2}p^2-\frac{c}{12}p''=A.
\label{s1}
\ee
Multiplying this equation by $p'$ and integrating again, one concludes that there also exists a constant $B$ such that
\be
-\frac{v}{2}p^2+\frac{1}{2}p^3-\frac{c}{24}p'^2=Ap+B.
\label{b3}
\ee
The parameters $A,B$ will eventually turn out to be closely related to the drift velocity (\ref{DAFA}).\\[-.3cm]%%PARBREAK%%

As before, we assume $p(x)$ to be amenable in the sense that there exists a generic\footnote{Recall (section \ref{secEPP}) that {\it genericity} means $k\neq-n^2c/24$; the stabilizer thus consists of rotations.} constant $k$ and a map $g_0\in\Diff$ satisfying eq.\ (\ref{sa6}). We now show that this assumption, along with the fact that $p(x,t)$ is a travelling wave, yields a tremendous simplification of the reconstruction equation (\ref{sa4}). Indeed, since $p(x)=(g_0\cdot k)(x)$, we can write
\be
p(x,t)
=
\big((\cR_{vt}\circ g_0)\cdot k\big)(x)
\label{s5}
\ee
where $\cR_{\theta}(x)\equiv x+\theta$ is a rotation by $\theta$. On the other hand, we know, by definition of reconstruction, that $p(x,t)=(g_t\cdot k)(x)$. Combining this with eq.\ (\ref{s5}), it follows that the diffeomorphism $g_t^{-1}\circ\cR_{vt}\circ g_0$ stabilizes $k$. Since $k$ is uniform and generic, its stabilizer consists of rotations only, so there must exist a function $\theta(t)\in\RR$ such that
\be
g_t
=
\cR_{vt}\circ g_0\circ\cR_{\theta(t)}.
\label{t5}
\ee
We have thus `factorized' the dependence of $g_t(x)$ on $t$ and $x$. Indeed, rewriting (\ref{t5}) as
\be
g_t^{-1}(x)
=
g_0^{-1}(x-vt)-\theta(t),
\label{s6}
\ee
the `advection' form of eq.\ (\ref{sa4}) yields $\dot\theta(t)=[p(x-vt)-v]\times(g_0^{-1})'(x-vt)$, which holds for all $x,t$. Since $t$ and $x-vt$ are independent coordinates on the plane, we may just as well rename $x-vt$ into $x$ and rewrite the equation for $\theta(t)$ as
\be
\dot\theta(t)=[p(x)-v]\,(g_0^{-1})'(x).
\label{t6}
\ee
This is a crucial result, as we now explain.

\paragraph{Uniformization and drift velocity.} Several striking implications follow from the separate dependence of the left- and right-hand sides of (\ref{t6}) on $t$ and $x$. First, differentiating (\ref{t6}) with respect to $x$, we conclude that there exists a constant $\cV\neq0$ such that
\be
\big(g_0^{-1}\big)'(x)
=
\frac{\cV}{p(x)-v}.
\label{s7}
\ee
We will soon see that this constant contributes to the drift velocity of fluid particles, justifying the notation $\cV$. Since $g_0\in\Diff$, eq.\ (\ref{s7}) readily implies that, if the profile $p$ is amenable, then $p(x)-v$ has no roots (so its sign is constant); in fact, we will show (right before section \ref{secGET}) that the implication also goes the other way around. Furthermore, the condition $g_0^{-1}(x+2\pi)=g_0^{-1}(x)+2\pi$ fixes the value of $\cV$:
\be
\cV
=
\frac{2\pi}{\ds\int_0^{2\pi}\frac{\dd x}{p(x)-v}}.
\label{t7}
\ee
Together with eq.\ (\ref{s7}), this determines the uniformizing map $g_0^{-1}$exactly and uniquely, up to an arbitrary rotation by $\phi$:
\be
g_0^{-1}(x)
=
\phi
+\int_0^x\frac{\cV\,\dd y}{p(y)-v}.
\label{tt7}
\ee
We have thus found $g_0^{-1}$, hence $g_0$, in terms of the wave profile. We will use this below (section \ref{secCNO}) to obtain an explicit expression for cnoidal uniformizing maps.\\[-.3cm]%%PARBREAK%%

Returning now to eq.\ (\ref{t6}) and differentiating with respect to $t$, one finds $\ddot\theta=0$. In fact, owing to eq.\ (\ref{s7}), one has $\dot\theta=\cV$, so $\theta(t)=\cV t$ (the integration constant vanishes by eq.\ (\ref{t5}) and $g_{t=0}=g_0$). From this, one deduces the exact reconstruction $g_t$, hence the solution $x(t)$ of (\ref{sb1}), hence the drift velocity (\ref{VIDI}). Here we go: from (\ref{s6}) we read off
\be
g_t^{-1}(x)
=
g_0^{-1}(x-vt)-\cV t
\qquad\Leftrightarrow\qquad
g_t(x)
=
g_0(x+\cV t)+vt,
\label{t8}
\ee
which is an {\it exact} geodesic in the Virasoro group (with respect to the right-invariant metric induced by the inertia operator (\ref{s66})). The ensuing particle motion reads
\be
x(t)
=
g_t\big(g_0^{-1}(x_0)\big)
=
g_0\big(g_0^{-1}(x_0)+\cV t\big)+vt.
\label{tt8}
\ee
Again, this is an {\it exact} solution of the equation of motion (\ref{sb1}).\\[-.3cm]%%PARBREAK%%

The drift velocity (\ref{VATA}) is obtained by computing the Poincar\'e rotation number of $g_0\circ g_T^{-1}$: using (\ref{t8}), we find that $g_T^{-1}\circ g_0$ is a rotation by $-\cV T\mp2\pi$, where $\pm=\text{sign}(v)$. Since the rotation number is conjugation-invariant \cite[sec.\ 4.4.3]{Guieu}, one has
\be
\text{Poincar\'e rotation number of $g_T\circ g_0^{-1}$}
\equiv\Delta\phi
=
\pm2\pi+\cV T,
\ee
where $g_T\circ g_0^{-1}$ was called $F$ in (\ref{FEQ}) and below (\ref{GAZO}). Hence the drift velocity (\ref{DAFA})-(\ref{VATA}) is
\be
\boxed{
\bigg.
v_{\text{Drift}}
=
\frac{\Delta\phi}{T}
=
v+\cV.}
\label{t9}
\ee
This confirms that the normalization constant $\cV$ of eq.\ (\ref{s7}) is a velocity, as anticipated. All these expressions are exact. We shall apply them to cnoidal waves in section \ref{secCNO}.

\paragraph{An accidental identity.} We have just seen that amenable travelling waves solving KdV can be mapped on a constant profile $k$ through a diffeomorphism $g_0$ given by the inverse of (\ref{tt7}), with $\cV$ as written in (\ref{t7}). We now show that this implies one constraint on the constants $A,B$ defined by (\ref{b3}). Indeed, recall that the coadjoint action of $g_0$ on $k$ is given by (\ref{sa66}); using eq.\ (\ref{s7}) for $(g_0^{-1})'$, we can write $g_0\cdot k$ as
\be
\big(g_0\cdot k\big)(x)
=
\frac{1}{[p(x)-v]^2}\Big[k\cV^2-\frac{c}{24}p'(x)^2+\frac{c}{12}\big(p(x)-v\big)p''(x)\Big].
\label{inter}
\ee
Here the right-hand side is supposed to equal $p(x)$, which, at this stage, is not obvious at all. To make progress, we use the fact that $p$ solves KdV, so that eqs.\ (\ref{s1})-(\ref{b3}) hold. This allows us to eliminate all derivatives of $p$ in eq.\ (\ref{inter}) and yields
\be
g_0\cdot k
=
p+\frac{1}{(p-v)^2}\big[k\cV^2+B+Av\big].
\label{codamir}
\ee
Since the right-hand side must equal $p$, we conclude that the constants $A,B,k,\cV$ are related through
\be
\boxed{\big.k\cV^2
=
-B-Av}
\label{s11}
\ee
where all coefficients are determined by the wave profile $p(x-vt)$. Thus, we have now proved that any amenable travelling wave solution of the KdV equation satisfies identity (\ref{s11}). This can be used \eg to find $k$ once $A,B,\cV$ are known, or to find $\cV$ when $A,B,k$ are known. In practice, eq.\ (\ref{s11}) provides a consistency check to be used once $A,B,k,\cV$ have been found by independent means. That will be our point of view below for cnoidal waves.\\[-.3cm]%%PARBREAK%%

Incidentally, eq.\ (\ref{s11}) allows us to prove the following point: {\it if $p(x-vt)$ solves KdV, then $p(x)$ is amenable (in the generic sense) if and only if $p(x)-v$ has no roots.} Indeed, we have already shown below eq.\ (\ref{s7}) that (generic) amenability implies the absence of roots. Conversely, if $p(x)-v$ has no roots, then one can define a map $g_0\in\Diff$ by eq.\ (\ref{tt7}), and the resulting coadjoint action on any constant $k$ is given by eq.\ (\ref{codamir}). Upon choosing $k$ to satisfy eq.\ (\ref{s11}), one finds $g_0\cdot k=p$, proving that $p(x)$ is amenable.

\subsection{Geometric phases of travelling waves}
\label{secGET}

Having shown that particle motion is integrable for amenable travelling waves, we now return to eq.\ (\ref{DPHI}) for the rotation angle $\Delta\phi$ and use the properties of travelling waves to simplify it. We treat separately the dynamical and Berry phases on the one hand, and the anomalous phase on the other hand, then verify that the resulting prediction of $v_{\text{Drift}}$ coincides with eq.\ (\ref{t9}).

\paragraph{Dynamical and Berry phases.} For a travelling wave, the dynamical phase in (\ref{DPHI}) is readily evaluated as the integral of $p(x)^2$. As for the Berry phase, it is greatly simplified by the fact that the path $f_t(x)$ need not be the reconstructed one, $g_t(x)$. Owing to the fact that $p(x,t)=p(x-vt)$ is a travelling wave, one may choose $f_t(x)=g_0(x)+vt$, where $g_0$ satisfies (\ref{sa6}). Plugging this into the Berry phase (\ref{sa8}) yields
\be
\oint_{\gamma}\widehat\cA
=
\pm
\int\frac{\dd x}{g_0'}\bigg[k+\frac{c}{12}\Big(\frac{g_0'''}{g_0'}-\frac{3}{2}\Big(\frac{g_0''}{g_0'}\Big)^2\Big)\bigg]
\mp2\pi k
=
\pm\int\!\dd x\,p(x)
\mp2\pi k,
\label{SIMBA}
\ee
where $\pm=\text{sign}(v)$ and the coadjoint representation (\ref{sa66}) has been used to recognize the integrand as $(g_0\cdot k)(x)=p(x)$. Thus, up to a sign and a term $2\pi k$, the Berry phase is the zero-mode (the average) of the profile $p$.

\paragraph{Anomalous phase.} The anomalous phase (\ref{KLO}) explicitly depends on the reconstructed path $g_t$, so, in contrast to the dynamical and Berry phases, one really needs to solve eq.\ (\ref{sa4}) in order to simplify it. Fortunately, we have already done that: we showed in section \ref{secuni} that the equation of motion (\ref{sb1}) can be integrated exactly for amenable travelling waves. Accordingly, we use the solution (\ref{t8}) to rewrite the anomalous phase (\ref{KLO}) as
\be
c\int_g\dd\sfC
=
\frac{c\,T}{48\pi}\int\!\dd x\,\frac{\cV}{(g_0^{-1})'}\bigg(\frac{(g_0^{-1})''}{(g_0^{-1})'}\bigg)'
=
\frac{c\,T}{48\pi}\int\!\dd x\,\frac{p'(x)^2}{p(x)-v},
\label{CITA}
\ee
where we also relied on eq.\ (\ref{s7}) to express $(g_0^{-1})'$ in terms of $p$. Combining this with the Berry phase (\ref{SIMBA}) and the dynamical phase, we obtain an expression of $\Delta\phi$ that only involves the (time-independent) profile $p(x)$ and its derivative, without any other wave data. In practice, it is simpler to express the formula as a drift velocity (\ref{DAFA}) instead of $\Delta\phi$, so as to absorb the awkward signs of eq.\ (\ref{SIMBA}). The result reads
\be
\boxed{%
\Bigg.
v_{\text{Drift}}
=
\frac{\Delta\phi}{T}
=
\underbrace{\frac{1}{2\pi k}\int\!\dd x\,p(x)^2}_{\ds v_{\text{Dynamical}}}
+
\underbrace{v-\frac{v}{2\pi k}\int\!\dd x\,p(x)}_{\ds v_{\text{Berry}}}
+
\underbrace{\frac{c}{48\pi k}\int\!\dd x\,\frac{p'(x)^2}{p(x)-v}}_{\ds v_{\text{Anomalous}}}}
\label{SANTO}
\ee
where we have grouped the contributions of various phases as in (\ref{DBIB})-(\ref{DPHI}). Note that, from this perspective, the anomalous phase looks like a correction to the dynamical phase (both contribute terms that are not proportional to the velocity $v$, as opposed to the Berry phase). However, this is really specific to travelling waves: for other kinds of profiles $p(x,t)$, the simplification (\ref{CITA}) would not hold.

\paragraph{Consistency check.} At this point, one should compare the geometric phase prediction (\ref{SANTO}) to the previously derived exact result (\ref{t9}). Indeed, it is not manifest that these expressions coincide. The fact that they do follows from a series of formulas derived earlier: first, upon writing (\ref{SANTO}) as $v_{\text{Drift}}=v+\cV$, one finds the condition
\be
2\pi k\cV
=
\int\dd x\bigg[p(x)\big(p(x)-v\big)+\frac{c}{24}\frac{p'(x)^2}{p(x)-v}\bigg].
\ee
Establishing this equality then proves that (\ref{SANTO}) and (\ref{t9}) coincide. To this end, one can use eq.\ (\ref{b3}) to express $p'(x)^2$ in terms of $p$, and eq.\ (\ref{s1}) to write the remaining $p^2$ term as a linear combination of constants, $p$ and $p''$. The latter does not contribute to the integral (by periodicity), and, after various cancellations, one ends up having to prove the identity
\be
2\pi k\cV
=
(-Av-B)\int\frac{\dd x}{p(x)-v}.
\ee
Owing to eq.\ (\ref{t7}), this is equivalent to the formula $-Av-B=k\cV^2$, which we encountered in eq.\ (\ref{s11}). We conclude that eq.\ (\ref{SANTO}) does coincide, as expected, with eq.\ (\ref{t9}).\\[-.3cm]%%PARBREAK%%

We have now completed a full conceptual circle: we first argued on symplectic grounds that particle motion, in the sense of eq.\ (\ref{sb1}), has a drift velocity determined by the sum of phases (\ref{DPHI}). We then showed, independently, that $v_{\text{Drift}}$ is given by (\ref{t9}) for amenable travelling waves, and we just proved that this formula is consistent with the geometric phase prediction. In practice, it is much easier to compute the drift velocity using eq.\ (\ref{t9}), with $\cV$ given by (\ref{t7}), than in terms of the phases (\ref{SANTO}). For travelling waves, the main virtue of (\ref{SANTO}) is that it neatly isolates the various geometric contributions to $v_{\text{Drift}}$. For more complicated wave profiles, however, a simple formula such as (\ref{t9}) is not available, so the general expression (\ref{DPHI}) is required to compute the drift velocity. We will consider such more general cases elsewhere.

\subsection{Drift in cnoidal waves and orbital bifurcations}
\label{secCNO}

Here we apply the tools of the previous pages to cnoidal waves --- the periodic solitons of KdV. The main results in that context are explicit formulas for the cnoidal uniformizing map and drift velocity, both given in terms of an elliptic integral of the third kind. Importantly, not all cnoidal waves are amenable \cite{Oblak:2019llc}: the profiles that have no uniform orbit representative span a `resonance wedge' in the cnoidal parameter space, with $v_{\text{Drift}}=v$ in the wedge. By contrast, outside of the wedge, $v_{\text{Drift}}\neq v$ is such that $v_{\text{Drift}}\sim v/3$ at large $|v|$, corresponding to an average one-period rotation by $\Delta\phi\sim\pm2\pi/3$ (see eq.\ (\ref{asyvev})). As we explain, all these results are consequences of the (symplectic) geometry of coadjoint orbits of the Virasoro group.\\[-.3cm]%%PARBREAK%%

We will use the same notation and conventions as ref.\ \cite{Oblak:2019llc}, to which we refer for a detailed description of coadjoint orbits of cnoidal waves. In practice, this subject relies on the theory of elliptic functions, which will not be reviewed here at all; for an introduction, see \eg \cite{Whittaker,Lawden} or appendix B of \cite{Oblak:2019llc}.

\begin{SCfigure}[2][t]
\centering
\includegraphics[width=0.55\textwidth]{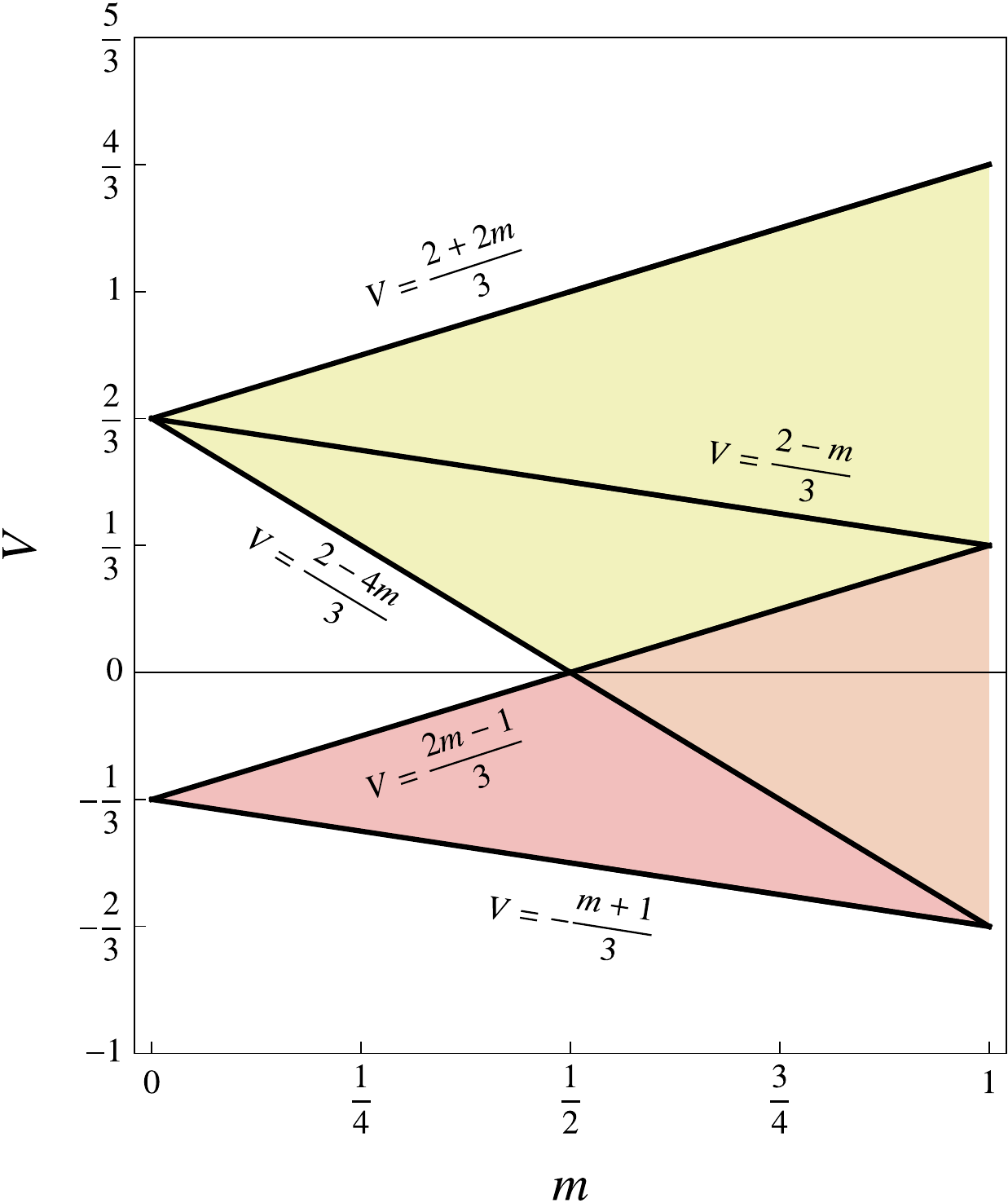}
\caption{The $(m,V)$ plane of cnoidal parameters, including the (yellow) root wedge (\ref{WAA}) and the (red) resonance wedge (\ref{WOO}). They partly overlap (orange) when $m>0.5$. For completeness, we also display the line $V=(2-m)/3$, along which $k=0$. As we shall see below, the contributions of dynamical, Berry and anomalous phases all diverge on that line, but these divergences cancel out so that the sum (\ref{SANTO}) is finite even when $k=0$. A more detailed picture can be found in fig.\ 7 of \cite{Oblak:2019llc}.\label{VAM}}
\end{SCfigure}%

\begin{figure}[t]
\centering
\includegraphics[width=0.50\textwidth]{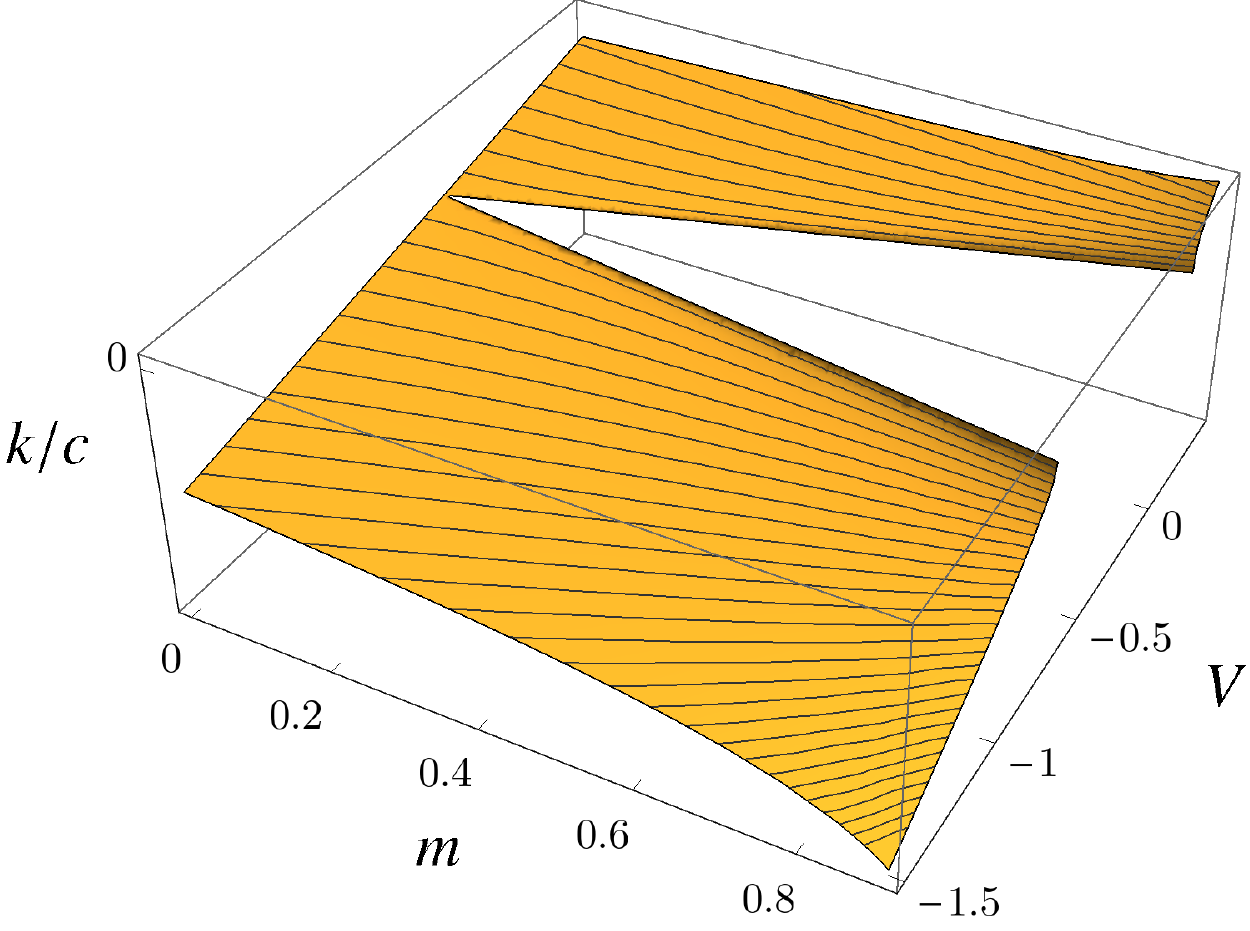}
\caption{The uniform orbit representative (\ref{KEQ}) of a cnoidal wave with parameters $(m,V)$; the black lines are level curves of $k$. At fixed $m$, $k/c$ is a monotonously increasing function of $V$ (and the dependence is roughly linear at large $|V|$). Note the wedge, with boundaries $V=[(1\pm3)m-2]/6$, in which (\ref{KEQ}) becomes complex: all waves in the wedge are non-amenable. Adapted from fig.\ 4 in \cite{Oblak:2019llc}.\label{KACH}}
\end{figure}

\paragraph{Cnoidal waves.} A {\it cnoidal wave} is a $2\pi$-periodic travelling wave solution of the KdV equation (\ref{KADA}). It is specified by two parameters: a `pointedness' $m\in[0,1)$ and a (rescaled) velocity $V\in\RR$ (see fig.\ \ref{VAM}). In these terms, the wave reads
\be
p(x,t)
=
\frac{cK(m)^2}{3\pi^2}
\biggl[
\frac{V}{2}
-\frac{m+1}{3}
+m\sn^2\Bigl(\frac{K(m)}{\pi}(x-vt)\Big|m\Bigr)
\biggr]
\label{cnop}
\ee
where $K(m)$ is the complete elliptic integral of the first kind, $\sn$ is the Jacobi elliptic sine, and the wave's velocity is a function of $(m,V)$ given by
\be
v
=
\frac{cK(m)^2}{2\pi^2}V.
\label{ss12}
\ee
Several qualitative aspects of the equation of motion (\ref{sb1}) for `fluid particles' can be read off from simple properties of the profile (\ref{cnop}). For example, $\dot x(t)$ has a constant sign if and only if $p(x)$ has no roots; such roots occur in the wedge
\be
\frac{2-4m}{3}
<V<
\frac{2+2m}{3}
\qquad
\text{(root wedge).}
\label{WAA}
\ee
Much more importantly, the results of section \ref{secuni} imply that the key object is not quite $p(x)$, but rather $p(x)-v$\,: as shown below eq.\ (\ref{s7}), $p(x)-v$ has roots if and only if $p(x)$ is non-amenable. Using the cnoidal profile (\ref{cnop}) and the velocity (\ref{ss12}), one readily sees that such roots only occur in the following {\it resonance wedge}:
\be
-\frac{m+1}{3}
<V<
\frac{2m-1}{3}
\qquad
\text{(resonance wedge).}
\label{WOO}
\ee
This is consistent with the classification of Virasoro orbits of cnoidal waves described in \cite{Oblak:2019llc} (and closely related to the band structure of the Lam\'e equation \cite{Whittaker}). Indeed, any profile with labels $(m,V)$ outside of the wedge (\ref{WOO}) is amenable, with a uniform orbit representative\footnote{Eq.\ (\ref{KEQ}) is roughly the square of the crystal momentum for a state with energy $\cE\propto\text{cst}-V$ in a Lam\'e lattice. From that viewpoint, the resonance wedge is the Lam\'e band gap. See \cite{Oblak:2019llc} for details.}
\be
k
=
\frac{c}{6\pi^2}
\Bigl[
K(m)\zeta\bigl(\wp^{-1}(V)\bigr)
-
\zeta\bigl(K(m)\bigr)\wp^{-1}(V)
\Bigr]^2
\label{KEQ}
\ee
that becomes complex (hence nonsensical) once $(m,V)$ enter into the resonance wedge. Here $\wp^{-1}$ is the inverse Weierstrass function and $\zeta$ is the Weierstrass zeta function (both specified by half-periods $K(m)$ and $iK(1{-}m)$). On the boundaries of the wedge, where $V=[(1\pm3)m-2]/6$, the constant (\ref{KEQ}) takes the exceptional value $k=-c/24$. See fig.\ \ref{KACH} for a plot of $k$ in the $(m,V)$ plane. For many more details about this, see \cite{Oblak:2019llc}.%\\[-.3cm]%%PARBREAK%%

\begin{figure}[t]
\centering
\begin{subfigure}{0.35\textwidth}
	\centering
	\includegraphics[width=\linewidth]{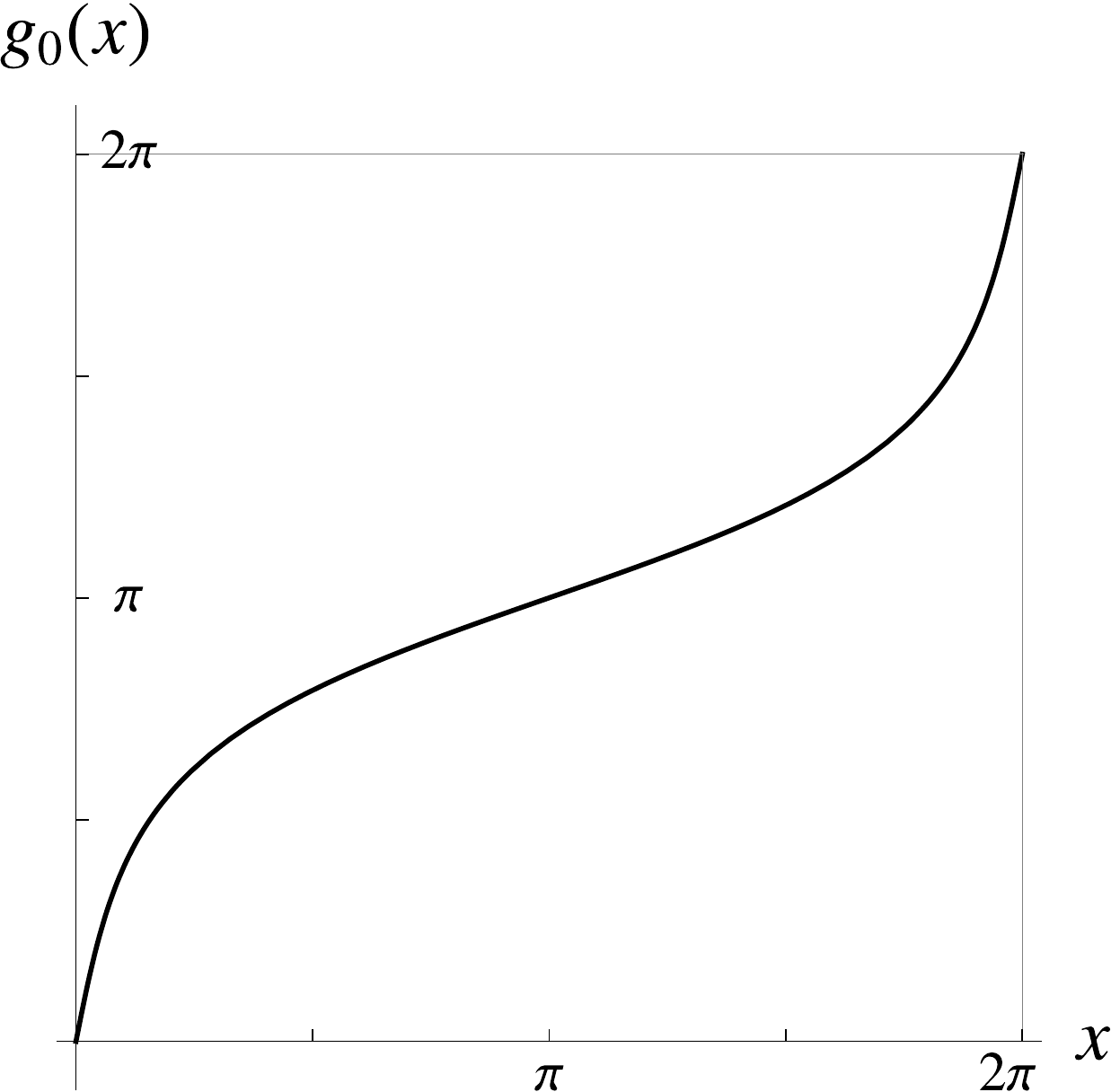}
\end{subfigure}
\hfill
\begin{subfigure}{0.20\textwidth}
	\centering
	\includegraphics[width=\linewidth]{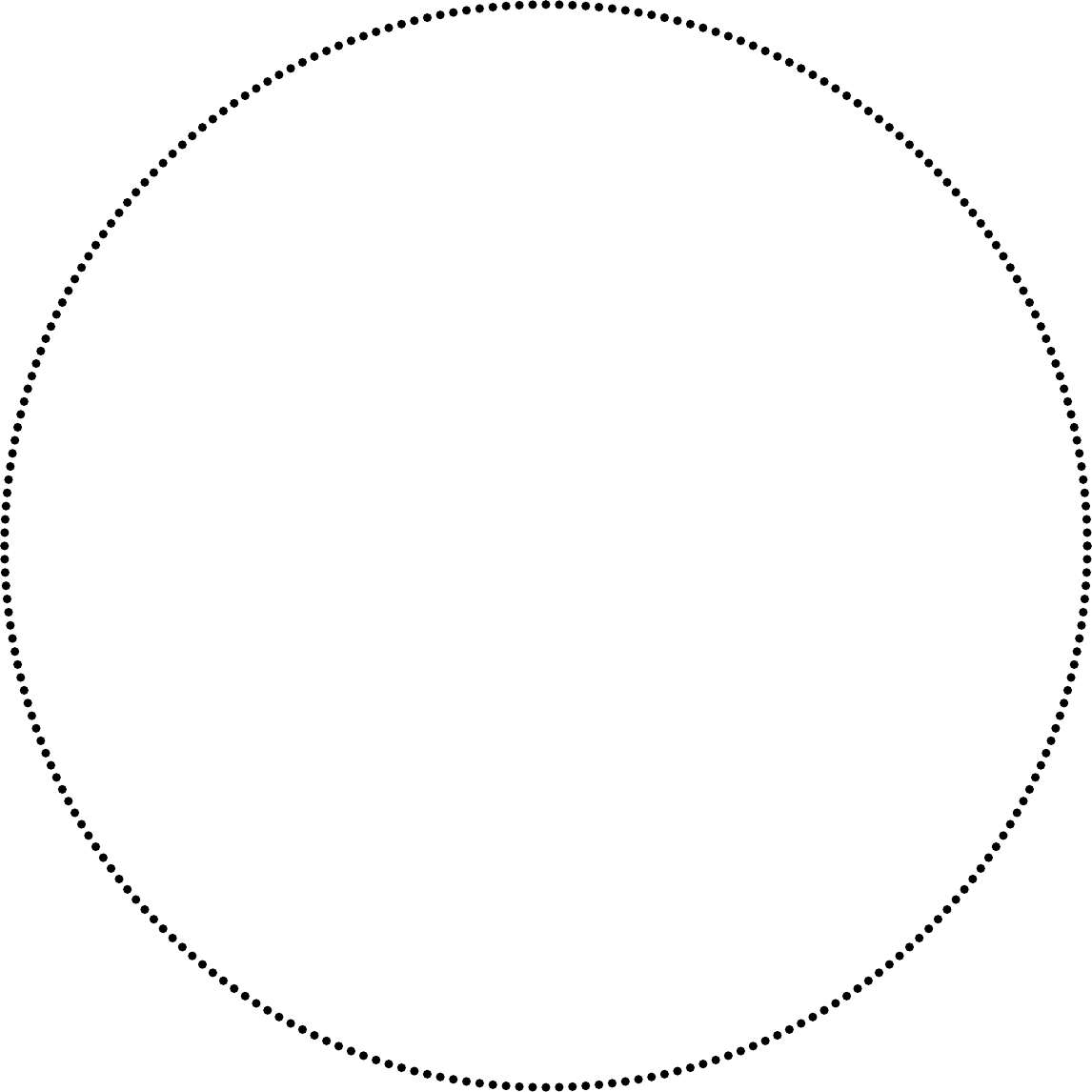}
\end{subfigure}
$\quad\xrightarrow{~~~~~~}\quad$%
\begin{subfigure}{0.20\textwidth}
	\centering
	\includegraphics[width=\linewidth]{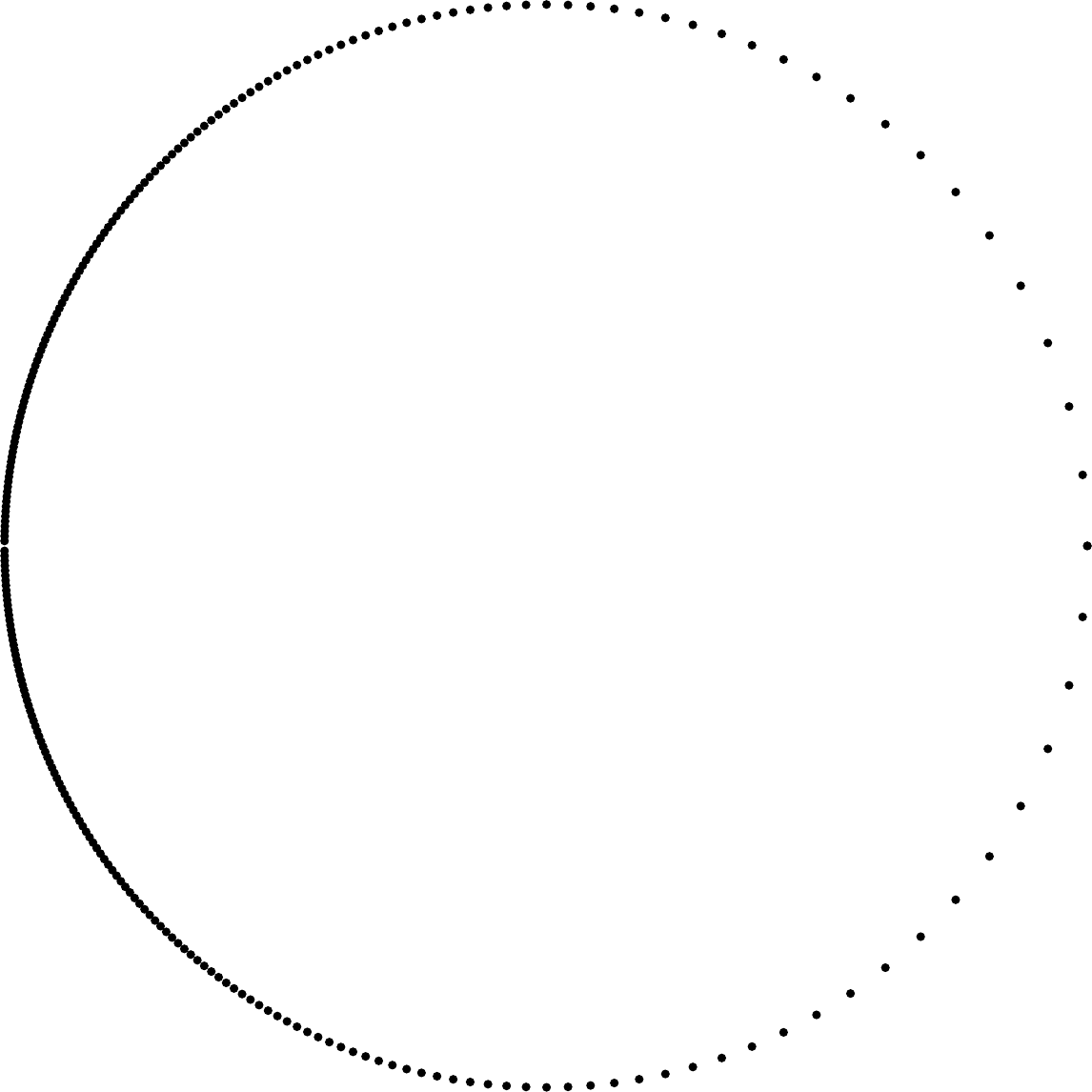}
\end{subfigure}
\caption{A plot of the cnoidal boost $g_0(x)$, the inverse of (\ref{cnoboo}), for $\phi=0$, $m=0.9$, $V=1/3$. On the right, we represent the boost's effect on uniformly distributed points on a circle with angular coordinate $x\sim x+2\pi$. The region with the highest density of points is $x=\pi$, while the lowest density occurs at $x=0$. These points respectively correspond to the maximum and minimum of $p(x)/c$, when $p(x)$ is a cnoidal profile (\ref{cnop}).\label{BOOST}}
\end{figure}

\begin{figure}[t]
\centering
~~~
\includegraphics[width=0.40\textwidth]{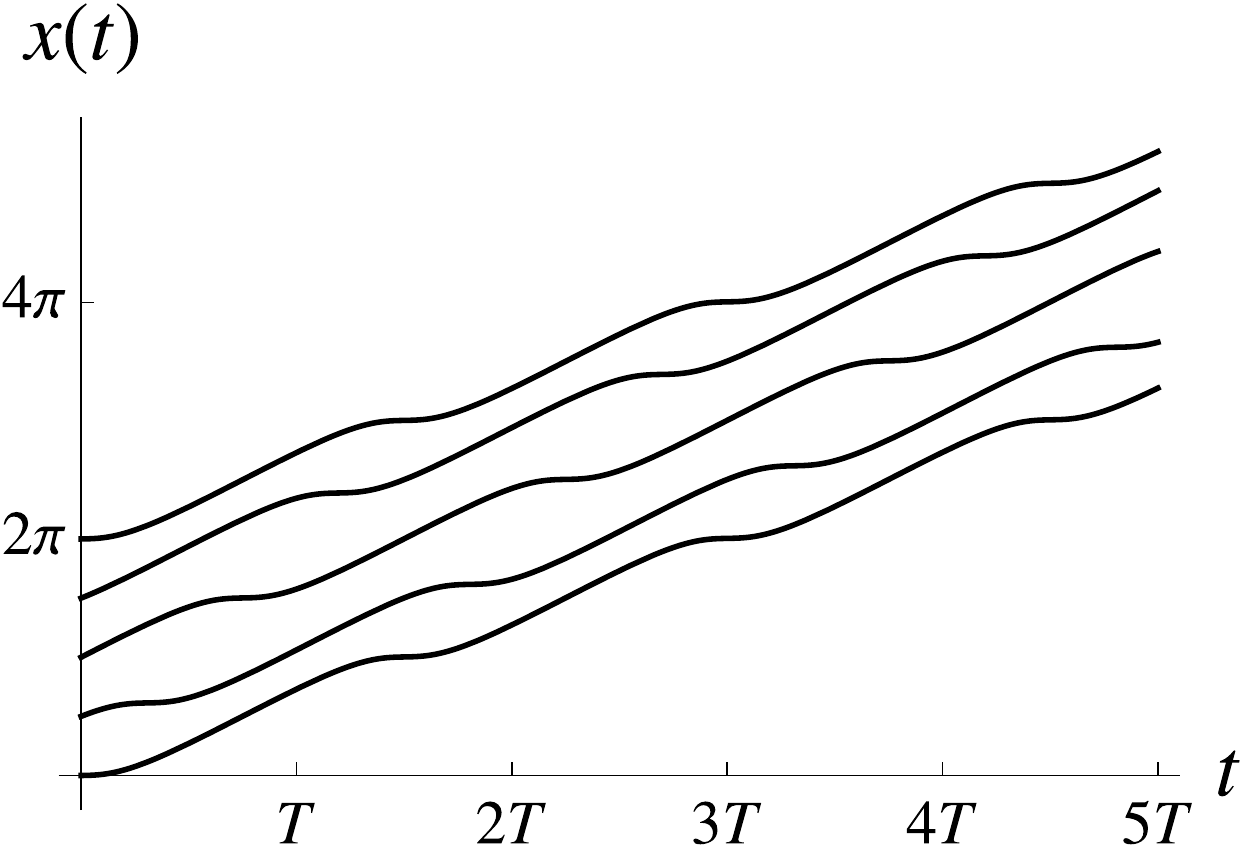}
\hfill
\includegraphics[width=0.40\textwidth]{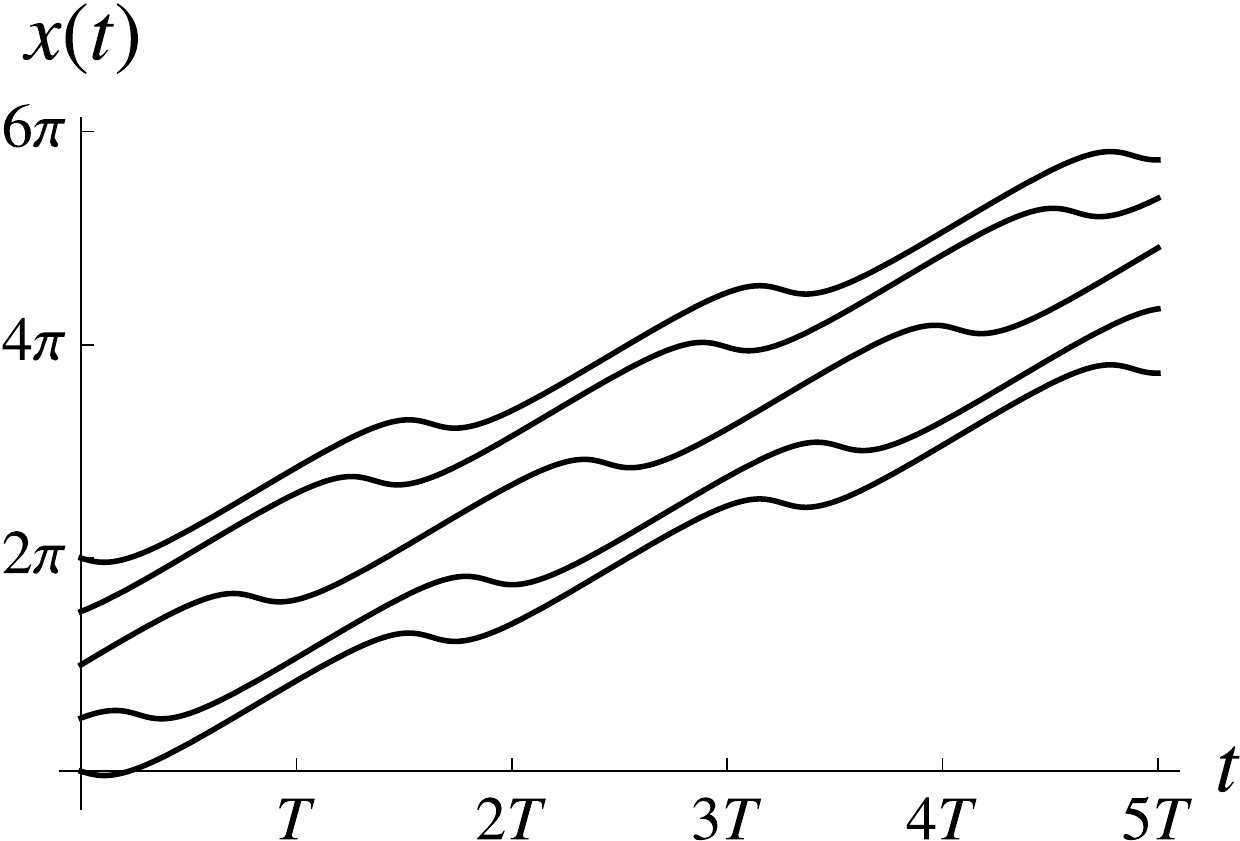}
~~~
\caption{A few solutions of the equation of motion (\ref{sb1}) for a cnoidal wave (\ref{cnop}) at $c=1$, $m=0.9$, $V=4/3$ (left panel) and $V=2/3$ (right panel). Both sets of parameters are well outside of the resonance wedge (\ref{WOO}), so cnoidal waves are amenable and the motion is given by eq.\ (\ref{tt8}), with $g_0$ written in (\ref{cnoboo}) and $\cV$ in eq.\ (\ref{s13}). Looking at the plots `from far away', particle motion is approximately linear in time, with a drift velocity (\ref{s14}). In the right panel, note the regions where $\dot x<0$: this is because $p(x)$ has roots for $(m,V)=(0.9,2/3)$, which lies in the root wedge (\ref{WAA}). Such a behaviour does not occur in the left panel, as $(m,V)=(0.9,4/3)$ lies outside of the root wedge.\label{FIMO}}
\end{figure}

\paragraph{Particle motion and drift.} The tools of section \ref{secuni} readily apply to amenable cnoidal waves. For instance, the velocity $\cV$ defined by (\ref{t7}) reads
\be
\cV
=
-\frac{cK(m)^3}{3\pi^2}\;\frac{\ds V+(m+1)/3}{\ds\Pi\Big(\frac{m}{V+(m+1)/3}\Big|m\Big)}
\label{s13}
\ee
where $\Pi(x|m)$ is the complete elliptic integral of the third kind. As a result, the uniformizing map (\ref{tt7}) can be written as
\be
g_0^{-1}(x)
=
\phi+
\pi\,\frac{\ds\Pi\Big(\frac{m}{V+(m+1)/3},\text{am}\big(K(m)x/\pi\big|m\big)\Big|m\Big) }{\ds\Pi\Big(\frac{m}{V+(m+1)/3}\Big|m\Big)}
\label{cnoboo}
\ee
where $\Pi(x,y|m)$ is the incomplete elliptic integral of the third kind and $\text{am}(x|m)$ is the Jacobi amplitude. This is an explicit `cnoidal boost': by construction, $g_0$ maps a uniform profile $k$ (`at rest') on a cnoidal (`boosted') one. The effect of such a boost on a circle is plotted in fig.\ \ref{BOOST}. The corresponding particle motion is given by eq.\ (\ref{tt8}); see fig.\ \ref{FIMO} for a few examples. As is manifest there, the large-scale behaviour of $x(t)$ is approximately linear in $t$, with a drift velocity given by eq.\ (\ref{t9}):
\be
\boxed{v_{\text{Drift}}
=
\frac{cK(m)^2}{\pi^2}\left[\frac{V}{2}-\frac{K(m)}{3}\,\frac{\ds V+(m+1)/3}{\ds\Pi\Big(\frac{m}{V+(m+1)/3}\Big|m\Big)}\right].}
\label{s14}
\ee
The asymptotics of this expression at large $|V|$ follow from $\Pi(x|m)\sim K(m)+\cO(x)$ for $x\to0$, which yields
\be
v_{\text{Drift}}\sim\frac{v}{3}\qquad\text{as }v\to\pm\infty.
\label{asyvev}
\ee
This is the asymptotic formula announced above. Writing $v_{\text{Drift}}=\Delta\phi/T$, it reflects an average rotation angle $\Delta\phi\sim\pm2\pi/3$ during each period.\\[-.3cm]%%PARBREAK%%

\begin{figure}[t]
\centering
\begin{tabular}{@{}lr@{}}
\includegraphics[width=0.47\textwidth]{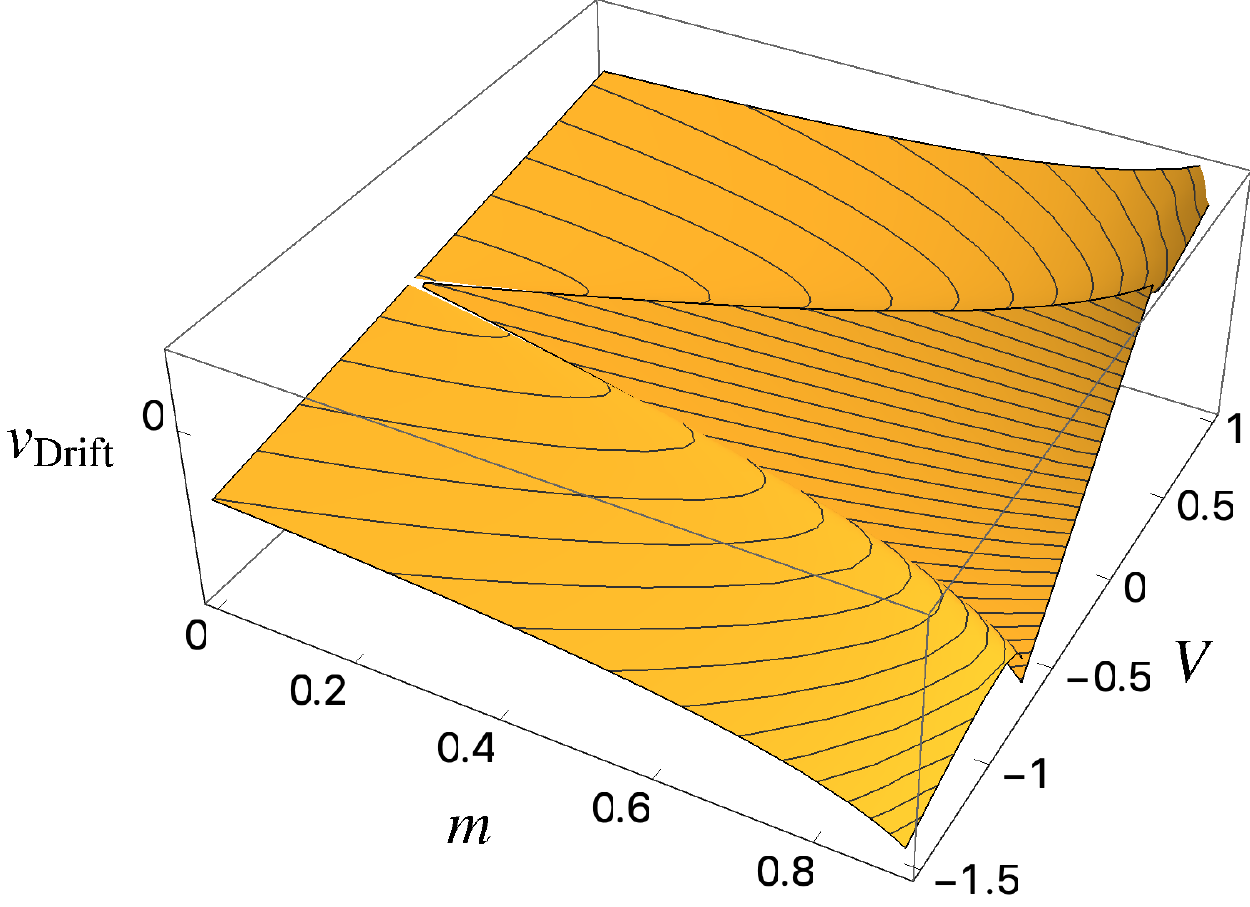}~~~~ & \includegraphics[width=0.47\textwidth]{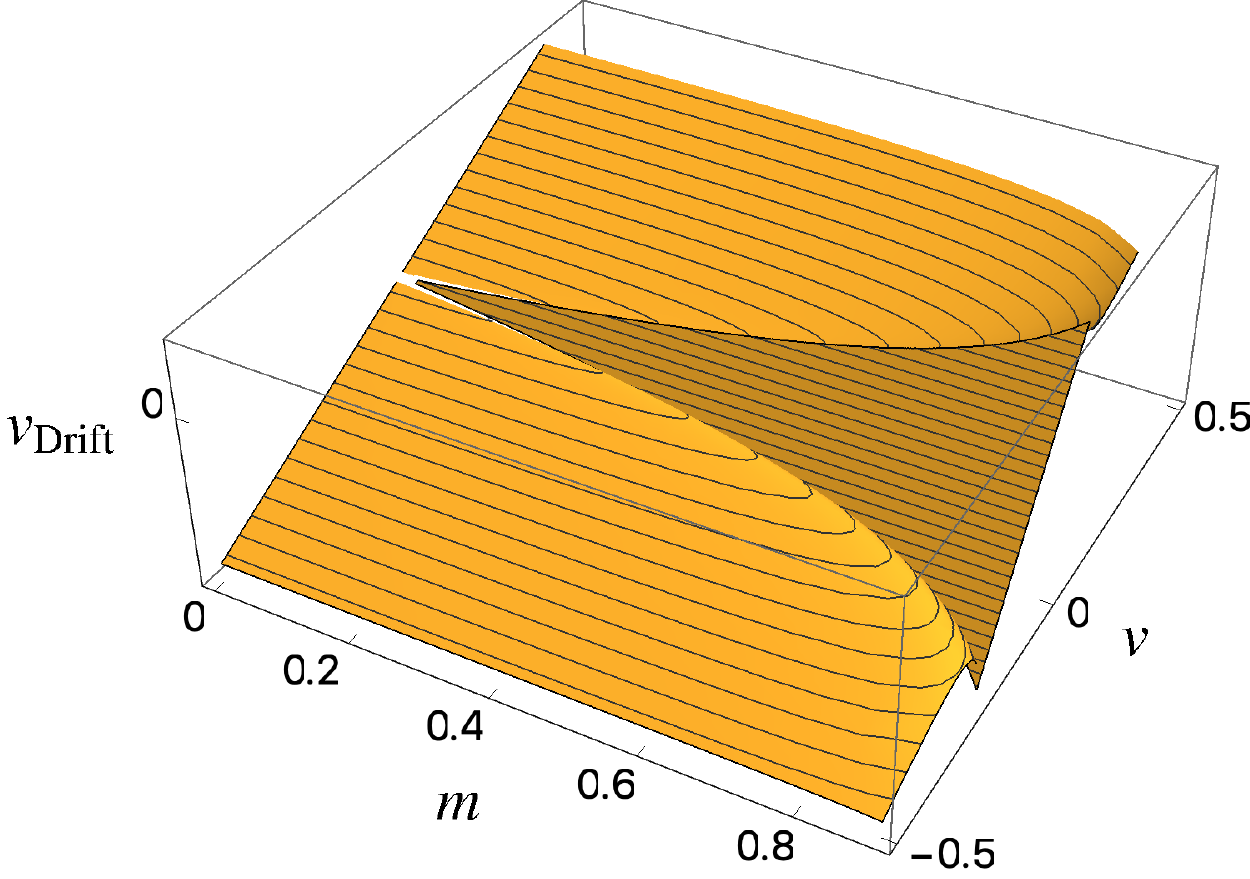} \\[.5cm]
\includegraphics[width=0.47\textwidth]{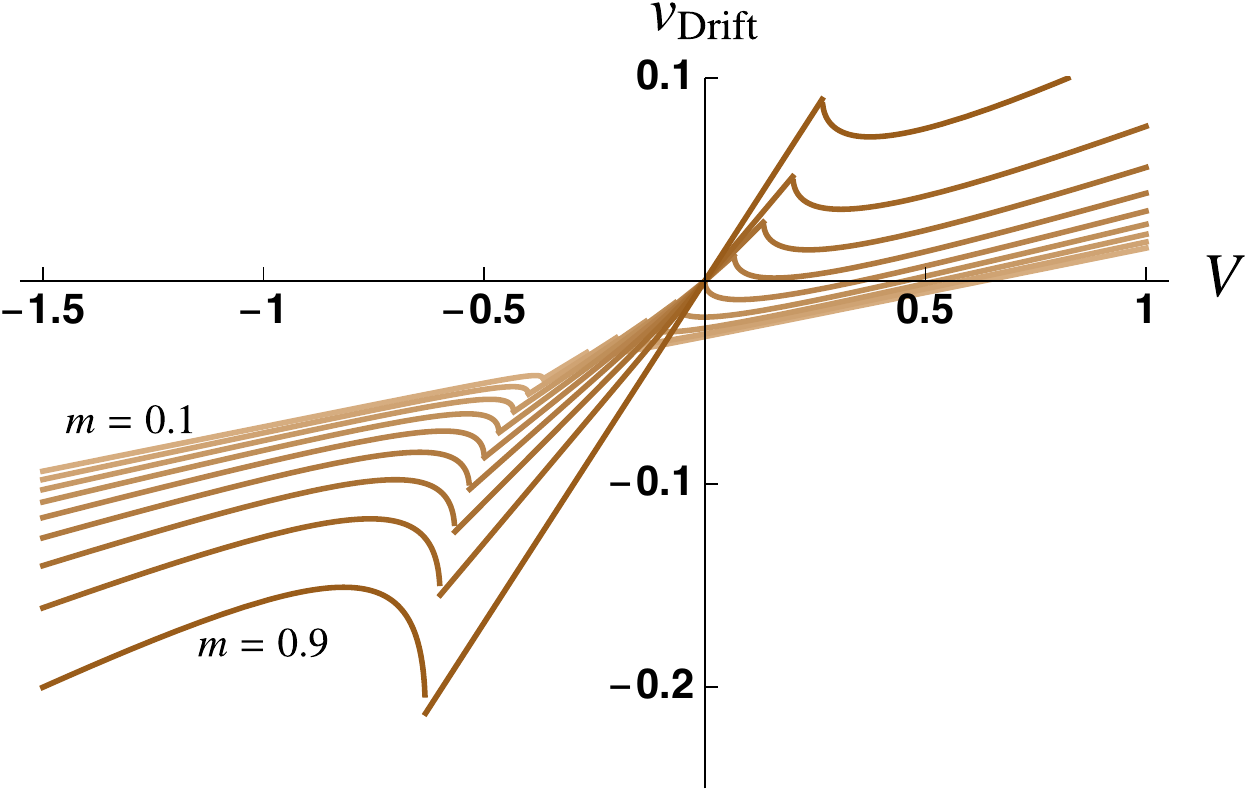}~~~~ & \includegraphics[width=0.47\textwidth]{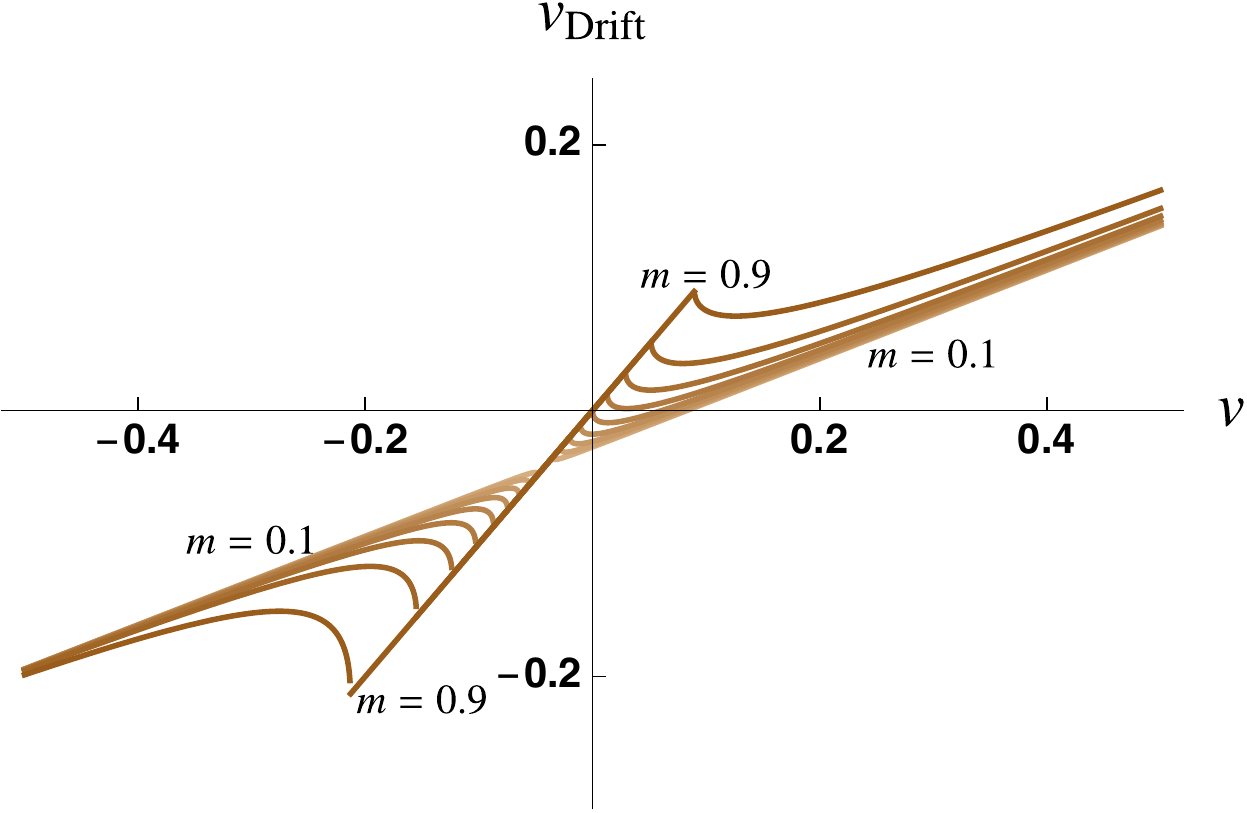}
\end{tabular}
\caption{The cnoidal drift velocity given by (\ref{s14}) and (\ref{VIVA}), along with level curves of $v_{\text{Drift}}$, at $c=1$. In the left panel, $v_{\text{Drift}}$ is plotted as a function of $(m,V)$, with the wave velocity given by (\ref{ss12}), and the resonance wedge has straight boundaries given by (\ref{WOO}). In the right panel, $V$ is traded for $v$, so that the boundaries of the resonance wedge are no longer straight lines, but the values $v_{\text{Drift}}=v$ in the wedge, and $v_{\text{Drift}}\sim v/3$ outside of the wedge, are apparent. Note that, despite the orbital bifurcations along wedge boundaries, $v_{\text{Drift}}$ is a continuous function on the cnoidal parameter space.\label{VIPLOT}}
\end{figure}

We stress that eq.\ (\ref{s14}) only holds in the region of amenable cnoidal waves, \ie outside of the resonance wedge (\ref{WOO}). In order to obtain a complete picture of $v_{\text{Drift}}$ as a function of $(m,V)$, throughout the entire cnoidal parameter space, we need to study the equation of motion (\ref{sb1}) in the resonance wedge. This will be done below, and the resulting final shape of $v_{\text{Drift}}(m,V)$ is displayed in fig.\ \ref{VIPLOT}. For now, we study eq.\ (\ref{s14}) from the point of view of geometric phases.

\paragraph{Geometric phases of cnoidal waves.} As explained in section \ref{secGET}, eq.\ (\ref{t9}) for the drift velocity can be written in a `decomposed' form (\ref{SANTO}) that exhibits the separate contributions of dynamical, Berry and anomalous phases. We now list the values of these three quantities for cnoidal waves in terms of parameters $(m,V)$, being understood that all formulas only hold outside of the resonance wedge (\ref{WOO}):
\begin{align}
\label{VIDY}
v_{\text{Dynamical}}
&=
\frac{c^2K(m)^4}{9\pi^3 k}
\left[
\frac{V^2}{2}+V\left(\frac{4-2m}{3}-\frac{2E(m)}{K(m)}\right)+\frac{2-2m+2m^2}{9}\right],\\[.4cm]
\label{VIBER}
v_{\text{Berry}}
&=
\frac{cVK(m)^2}{2\pi^2}
\left[%
1-\frac{2cK(m)^2}{6\pi^2k}
\left(\frac{V}{2}+\frac{2-m}{3}-\frac{E(m)}{K(m)}\right)
\right],\\[.4cm]
\label{VIAN}
v_{\text{Anomalous}}
&=
-\frac{c^2m^2K(m)^4}{144\pi^4 k}
\left[
1
-
\frac{V+\frac{m-2}{3}}{V+\frac{m+1}{3}}\,
\frac{\Pi\left(\left.\frac{m}{V+(m+1)/3}\right|m\right)}{K(m)}
\right].
\end{align}
Here $K(m)$, $E(m)$ and $\Pi(x|m)$ are respectively complete elliptic integrals of the first, second and third kind, and $k$ is the function of $(m,V)$ written in eq.\ (\ref{KEQ}). It is not particularly illuminating to plot these velocities as functions of $(m,V)$. Their overwhelmingly dominant feature is a divergence on the line $V=(2-m)/3$, where $k=0$. This divergence cancels out when the three velocities are added together, since the total drift velocity (\ref{s14}) is finite even when $k=0$ (as it should be); see fig.\ \ref{VIPLOT} below. Note also that the Berry velocity (\ref{VIBER}) is the only one that vanishes on the line $V=0$.\\[-.3cm]%%PARBREAK%%

As shown in section \ref{secGET}, the coincidence between formulas (\ref{t9}) and (\ref{SANTO}) for the drift velocity hinges on the identity (\ref{s11}) satisfied by amenable travelling waves. We now verify this identity for cnoidal waves: using the profile (\ref{cnop}) and the definition (\ref{b3}) of the constants $A,B$, one finds
\begin{align}
A &= \frac{1}{6}\Big(\frac{cK(m)^2}{\pi^2}\Big)^2\bigg[-\frac{V^2}{4}+\frac{m^2-m+1}{9}\bigg],\\
B &= \frac{1}{2}\Big(\frac{cK(m)^2}{\pi^2}\Big)^3\bigg[\frac{V^3}{8}-\frac{m^2-m+1}{6}V+\frac{2m^3-3m^2-3m+2}{27}\bigg].
\end{align}
The value of $-B-Av$ follows. Upon plugging it in eq.\ (\ref{s11}) and using the value (\ref{KEQ}) of $k$, one encounters the formula
\be
\begin{split}
&V^3-\frac{m^2-m+1}{3}V-\frac{2m^3-3m^2-3m+2}{27}
=\\
&=\left(
\Big[K(m)\zeta\big(\wp^{-1}(V)\big)-\zeta\big(K(m)\big)\wp^{-1}(V)\Big]
\frac{\ds V+(m+1)/3}{\ds\Pi\Big(\frac{m}{V+(m+1)/3}\Big|m\Big)}
\right)^{\ds 2}.
\end{split}
\label{awter}
\ee
This turns out to be a known, albeit somewhat obscure, identity that arises in the Lam\'e band structure: it coincides with eq.\ (7.20) of \cite{Doman} upon identifying the crystal momentum as $q(\cE)\equiv\tfrac{\pi}{K(m)}\sqrt{-6k/c}$ in terms of the uniform representative (\ref{KEQ}), along with $e=V$, $e_1=(2-m)/3$, $e_2=(2m-1)/3$ and $e_3=-(m+1)/3$.\footnote{Up to permutation, those are the standard values of Weierstrass roots $e_i$ appropriate for Jacobi elliptic functions (see \eg \cite[app.\ B]{Oblak:2019llc}). The left-hand side of (\ref{awter}) then coincides with $-(e_1-e)(e_2-e)(e_3-e)$.} Cnoidal waves thus satisfy eq.\ (\ref{s11}), as they should.

\begin{figure}[t]
\centering
~~~
\includegraphics[width=0.40\textwidth]{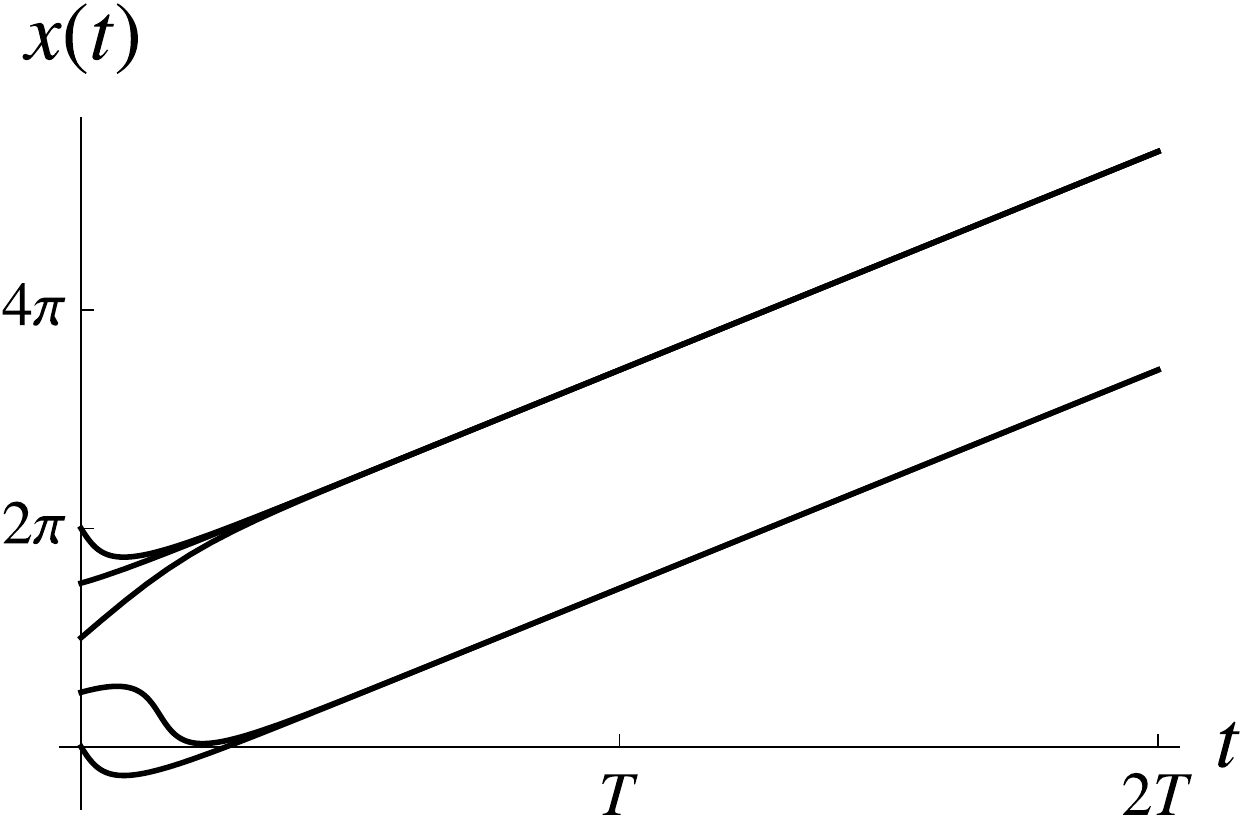}
\hfill
\includegraphics[width=0.40\textwidth]{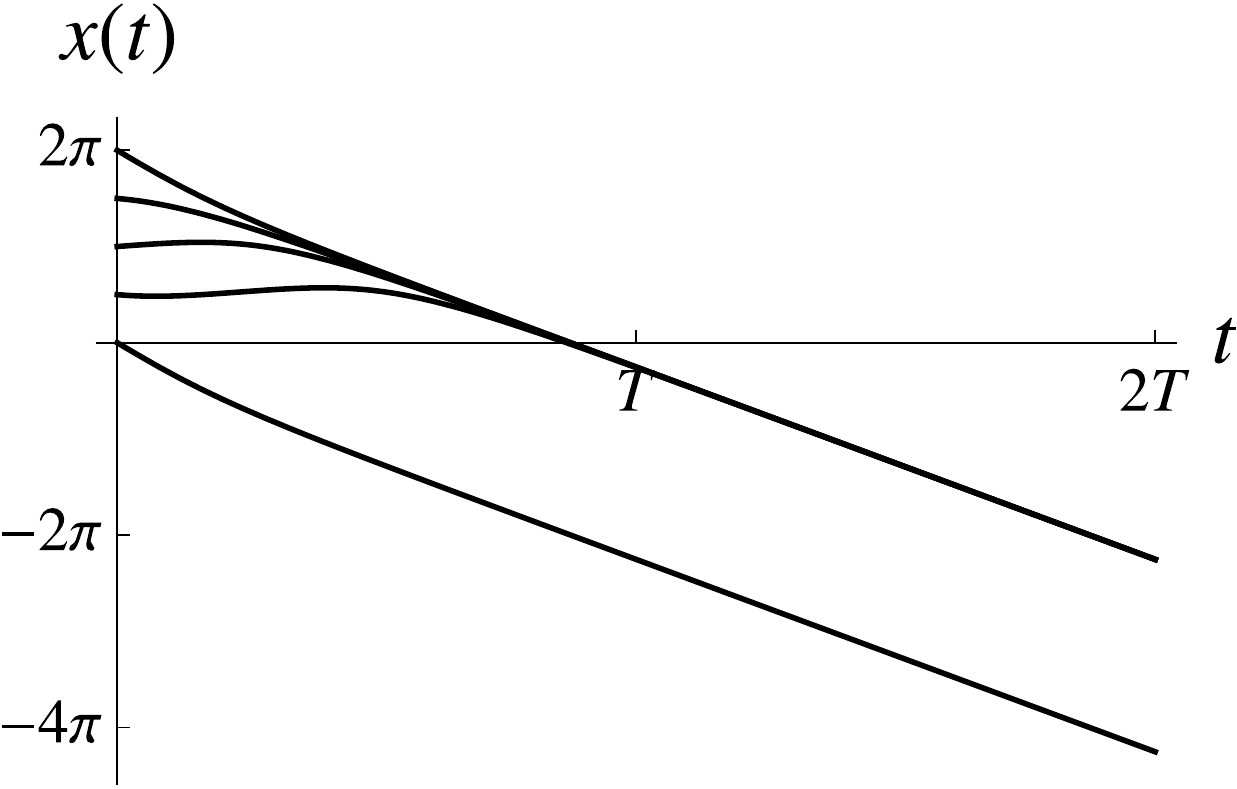}
~~~
\caption{A few solutions of (\ref{sb1}) when $p(x,t)$ is a cnoidal wave (\ref{cnop}) at $c=1$, $m=0.9$, $V=0.1$ (left panel) and $V=-1/3$ (right panel). These parameters lie in the resonance wedge (\ref{WOO}), so eq.\ (\ref{tt8}) does not apply and the plotted values $x(t)$ were obtained by numerical integration of (\ref{sb1}). Note the monotonous behaviour of $x(t)/t$, radically different from the oscillating one in fig.\ \ref{FIMO}. The rotation number $\Delta\phi=\pm2\pi$ is manifest.\label{TOXIC}}
\end{figure}%

\paragraph{Particle motion in the resonance wedge.} The cnoidal waves whose parameters $(m,V)$ belong to the resonance wedge (\ref{WOO}) are not amenable \cite{Oblak:2019llc}, so the solution of the equation of motion (\ref{sb1}) is no longer given by eq.\ (\ref{tt8}). It is, nonetheless, easy to compute the particle drift velocity as defined in (\ref{VIDI}). Indeed, for a travelling wave $p(x,t)=p(x-vt)$, the equation of motion (\ref{sb1}) can be recast as
\be
\dot X(t)
=
p(X(t))-v
\label{XOP}
\ee
in terms of $X\equiv x-vt$. Thus, if $p(X)-v$ has roots (which precisely occurs in the resonance wedge), then at least one of them, say $X^*$, is a stable fixed point of the system (\ref{XOP}). It follows that $X(t)\to X^*$ at late times, which is to say that $x(t)\sim X^*+vt$ at large $t>0$. This is manifest in fig.\ \ref{TOXIC}, where we plot a few solutions of (\ref{sb1}) for a wave in the resonance wedge. As a result, the drift velocity (\ref{VIDI}) is
\be
v_{\text{Drift}}
=
v
\qquad\text{in the resonance wedge (\ref{WOO}).}
\label{VIVA}
\ee
This finally justifies the name `resonance wedge': for parameters $(m,V)$ that satisfy (\ref{WOO}), particle motion is `locked' to the wave --- it `resonates'. This is akin to the {\sc sniper} bifurcation of the Adler equation \cite{Adler}. The resulting, complete function $v_{\text{Drift}}(m,V)$, on the entire cnoidal parameter space, is shown in fig.\ \ref{VIPLOT}.\\[-.3cm]%%PARBREAK%%

We stress again that eq.\ (\ref{VIVA}) could not have been deduced from a rotation angle such as (\ref{DPHI}). If anything, upon declaring that (\ref{VIVA}) is to be written in the form $v_{\text{Drift}}=\Delta\phi/T$, one would find a (piecewise) constant value
\be
\Delta\phi
=
2\pi\,\text{sign($v$)}
\qquad\text{in the resonance wedge.}
\label{FATOP}
\ee
This could not have emerged from the phase formula (\ref{DPHI}) since $k$, given by eq.\ (\ref{KEQ}), is complex in the resonance wedge. Despite this, it is in fact possible to predict that $\Delta\phi/(2\pi)$ must be an integer in the resonance wedge by relying on the symplectic arguments of section \ref{sec2} --- the exact same line of thought that led to eq.\ (\ref{DPHI}). Indeed, we argued in section \ref{secSTOKES} that the equation of motion (\ref{sb1}) implements Lie-Poisson reconstruction for the Virasoro group, and this holds whether or not $p(x,t)$ is amenable. Furthermore, when $p(x,t)$ is periodic in time, it still remains true that the reconstructed path $g_t\in\Diff$ satisfies $\big((g_0^{-1}\circ g_T)\cdot k\big)(x)=k(x)$ for a suitable orbit representative $k(x)$. In non-amenable orbits, $k(x)$ cannot be uniform (it must depend on $x$), but the fact remains that $g_0^{-1}\circ g_T$ leaves it fixed. Now, crucially, the stabilizer of non-amenable profiles is {\it never} a group of rotations \cite{Witten:1987ty,Balog:1997zz,Guieu,Khesin,Oblak:2016eij}. Instead, the stabilizer is a non-compact group, isomorphic to $\RR$ (up to further finite factors), and consists of circle diffeomorphisms whose rotation number vanishes modulo $2\pi$. It follows that, in that case,
\be
\Delta\phi
=
\text{Rotation number of }g_T\circ g_0^{-1}
\in
2\pi\ZZ,
\ee
which is consistent with the observation (\ref{FATOP}). In this sense, the `frozen' drift velocity (\ref{VIVA}) confirms that cnoidal profiles in the resonance wedge are not amenable \cite{Oblak:2019llc}.

\Needspace{10\baselineskip}
\section{Conclusion and outlook}
\label{sec5}

The purpose of this paper has been to initiate a geometric study of nonlinear wave equations, such as KdV, Hunter-Saxton or Camassa-Holm, from a point of view that strongly relies on group theory and symplectic geometry. This follows a long tradition in mathematical physics \cite{Marsden90,Marsden,Montgomery,HolmTyr,Arnold} (see also the very recent \cite{Modin}), but it seems to have been somewhat overlooked in the mainstream physics literature despite the relevance of geometric objects, such as Berry phases, in a plethora of systems ranging from condensed matter to nonlinear dynamics (see \eg the classic \cite{Shapere:1989kp} or the more contemporary review \cite{Cohen}, and references therein). Our goal has been to start filling that gap.\\[-.3cm]%%PARBREAK%%

Specifically, we considered a spatially periodic KdV equation (\ref{KADA}) and the equation of motion (\ref{sb1}), seen as a model for `fluid particles' dragged along by the wave profile $p(x,t)$. Using the fact that (\ref{sb1}) effectively coincides with the equation for Lie-Poisson reconstruction used in symplectic geometry, we predicted the value of the particle's drift velocity in terms of a Poincar\'e rotation number $\Delta\phi$. The latter turned out to be a sum of phases (\ref{DBIB}) that we wrote explicitly in eq.\ (\ref{DPHI}) and that crucially involves a Virasoro Berry phase in the sense of \cite{Oblak:2017ect}. We then turned to travelling waves, for which the assumption of amenability yielded a striking simplification of the equation of motion (\ref{sb1}). In fact, we showed that particle motion becomes integrable in that case, leading to the simple formula (\ref{t9}) for the drift velocity. An equivalent way to write that velocity in terms of geometric phases was displayed in eq.\ (\ref{SANTO}), and we showed that the two expressions, while not manifestly identical, do in fact coincide thanks to the KdV equation. Finally, we applied our tools to cnoidal waves and exhibited `orbital bifurcations' occurring at the boundaries of the resonance wedge (\ref{WOO}), inside of which particle motion is locked to the wave. These bifurcations are a direct observation of the sharp change in nature of Virasoro orbits of cnoidal waves, as investigated in \cite{Oblak:2019llc}.\\[-.3cm]%%PARBREAK%%

As stressed in section \ref{secSTOKES}, the equation of motion (\ref{sb1}) that we studied does not quite coincide with the actual equation of motion for a fluid particle in shallow water. In particular, while conceptually similar, it is not clear whether the drift velocity studied here has anything to do with the standard notion of Stokes drift \cite{Longuet53}. A natural question that follows from our work, therefore, is whether any actual experiment could exhibit the rotation $\Delta\phi$, and in particular Virasoro Berry phases. For instance, it is almost trivially true that a quantum particle on a circle, subjected to the periodic Hamiltonian $\widehat H=p(\hat x,t)\hat P-\frac{i\hbar}{2}p'(\hat x,t)$, would have a wavefunction that rotates by an average angle $\Delta\phi$ during each period of the profile $p(x,t)$. This example, however, is somewhat artificial, as one is merely restating our construction in the language of one-body quantum mechanics. It would be much more satisfactory to find a classical mechanical system that naturally reproduces the equation of motion (\ref{sb1}) --- for example, a plasma in one dimension subjected to a suitable external magnetic field. We hope to address this issue elsewhere.\\[-.3cm]%%PARBREAK%%

Setting aside the question of experimental signatures, one can think of numerous follow-ups of our work. For example, remaining within the confines of the KdV equation, it is natural to apply formula (\ref{DPHI}) to non-travelling waves, such as profiles containing multiple colliding cnoidal waves with a rational phase shift \cite{Its}. The phase shift being rational ensures that the profile as a whole is periodic in time. Indeed, in terms of the phase shift $\Delta\theta$ (which has {\it a priori} nothing to do with $\Delta\phi$!), the profile satisfies $p(x,t+\tau)=p(x-\Delta\theta,t)$ for some quasi-period $\tau$. If the phase shift is rational in the sense that $\Delta\theta=2\pi p/q$, with $p,q$ coprime integers, then the time period of the profile is $T=q\tau$, and our results of section \ref{secEKO} apply. It would be illuminating to display the resulting Berry phase spectrum (as a function of the wave profile parameters), especially as one might hope to even extend the picture, formally, to non-periodic profiles (for which $\Delta\theta$ is irrational). Other potential extensions of our paper include the quantum version of $\Delta\phi$, in the sense of the quantum KdV equation \cite{Sasaki}, and the effects of stochasticity. Regarding the latter, see \eg \cite{Debou} and references therein, or \cite{HolmStoch} for a language very close to ours.\\[-.3cm]%%PARBREAK%%

The symplectic approach of this paper applies to a host of other nonlinear wave equations, and it may be interesting to investigate the analogue of the angle $\Delta\phi$ (and its observable effects) in such setups. As examples, we already mentioned the Hunter-Saxton and Camassa-Holm equations \cite{Hunter}, for which the geometric and anomalous phases of eq.\ (\ref{DPHI}) remain unchanged. Perhaps more interestingly, one could also extend the strategy to Lie-Poisson wave equations not based on the Virasoro group, such as Hirota-Satsuma dynamics \cite{Hirota} or the nonlinear Schr\"odinger equation, which may be seen as a Lie-Poisson equation based on the loop group of $\text{SO}(3)$ \cite{Lakshmanan}. We hope to contribute to some of these research avenues in the future.

\section*{Acknowledgements}

The authors are indebted to T.\ Erneux and N.\ Goldman for illuminating discussions, and to anonymous referees for insightful comments on the initial manuscript. B.O.\ also thanks J.\ Bosma, J.\ Cotler, B.\ Estienne, M.\ Gaberdiel, S.\ Huber and K.\ Jensen for discussions and support. Finally, many thanks to A.\ Alekseev and J.\ Sonner for organizing the conference `Chaos, Holography and Coadjoint Orbits', where a preliminary version of our results was presented. The work of B.O.\ was mostly carried out at ETH Z\"urich, where it was supported by the Swiss National Science Foundation and the NCCR SwissMAP; his current funding is the ANR grant {\it TopO}, number ANR-17-CE30-0013-01. G.K.\ is a Research Associate with {\it Fonds de la Recherche Scientifique}-FNRS (Belgium).

\appendix

\section{Groups and geometry}
\label{appa}
\setcounter{equation}{0}
\renewcommand{\theequation}{A.\arabic{equation}}

In this appendix, we review some of the elementary material on Lie groups and symplectic geometry needed in section \ref{sec2}. For a broader and more pedagogical introduction, we refer \eg to \cite{Abraham,Marsden90,Marsden}. Lie-Poisson dynamics and reconstruction are covered in appendix \ref{appb}.

\paragraph{Lie groups.} Let $G$ be a Lie group with Lie algebra $\mg=T_{\II}G$; we denote elements of the former as $f$, $g$, {\it etc.}\ and those of the latter as $\xi$, $\zeta$, {\it etc.}, while $\II$ is the identity in $G$. The group acts on its algebra through the {\it adjoint representation}\footnote{Here $e^{t\xi}$ is the exponential of $t\xi$, such that $e^0=\II$.}
\be
\Ad_g(\xi)
\equiv
\left.\frac{\dd}{\dd t}\right|_0
\Big(
g\cdot e^{t\xi}\cdot g^{-1}
\Big),
\label{s16}
\ee
which, for matrix groups, boils down to $g\,\xi\,g^{-1}$. The corresponding infinitesimal action, \ie the {\it adjoint representation of the Lie algebra}, coincides with the Lie bracket:
\be
\ad_{\xi}\zeta
\equiv
\dt\Ad_{e^{t\xi}}(\zeta)
=
[\xi,\zeta],
\label{ss16}
\ee
which, for matrix groups, is just a commutator. In the context of Lie-Poisson equations (\ref{ss20}), one thinks of the group manifold $G$ as the space of classical configurations of a dynamical system, while the Lie algebra $\mg$ consists of all infinitesimal motions (deformations) of $G$, such as angular velocities. Equivalently, $G$ is the group of all possible changes of reference frames that a dynamical system is allowed to go through, while the adjoint representation (\ref{s16}) says how angular velocity varies under changes of frames. For instance, the rotation group $G=\text{SO}(3)$ describes all possible orientations of a rigid body (with respect to a frame whose origin lies at the body's centre of mass).\\[-.3cm]%%PARBREAK%%

The Lie algebra $\mg$ is a vector space whose dual $\mg^*$ consists of elements that we write as $p$, $q$, {\it etc.}, each of which is a linear form on $\mg$,
\be
p:\mg\rightarrow\RR:
\xi\mapsto\langle p,\xi\rangle.
\label{s18}
\ee
We refer to such maps as {\it coadjoint vectors}, and their transformation law under the action of the group $G$ is the {\it coadjoint representation} introduced in eq.\ (\ref{coadef}):
\be
g\cdot p
\equiv
\Ad^*_g(p)
\equiv
p\circ\Ad_g^{-1}
\qquad\forall\, 
g\in G,\,p\in\mg^*.
\label{s17b}
\ee
This definition is dual to (\ref{s16}) and ensures that the pairing (\ref{s18}) is invariant under `changes of reference frames' in the sense that $\langle\Ad^*_g(p),\Ad_g(\xi)\rangle=\langle p,\xi\rangle$. Analogously to (\ref{ss16}), the {\it coadjoint representation of the Lie algebra} is
\be
\ad^*_{\xi}(p)
\equiv
\dt\Ad^*_{e^{t\xi}}(p)
=
-p\circ\ad_{\xi}.
\label{ss17b}
\ee
As mentioned in eq.\ (\ref{coadefa}), this is to say that $\langle\ad^*_{\xi}p,\zeta\rangle=-\langle p,[\xi,\zeta]\rangle$ for all $\xi,\zeta\in\mg$, $p\in\mg^*$. Any algebra admitting a non-degenerate invariant bilinear form has equivalent adjoint and coadjoint representations, whose distinction is then inconsequential. However, for the algebra of vector fields that are needed for KdV, adjoint and coadjoint representations differ, which is why the intrinsic definitions (\ref{s17b})-(\ref{ss17b}) are important for our purposes.

\paragraph{The phase space \texorpdfstring{$T^*G$}~.} What now follows is a technical preliminary to section \ref{sec22}. Given a Lie group $G$, we show that its cotangent bundle $T^*G=\bigsqcup_{g\in G} T_g^*G$ is a trivial bundle. This will allow us to think of the product $G\times\mg^*$ as a symplectic manifold.\\[-.3cm]%%PARBREAK%%

\noindent{\it Lemma:} $T^*G$ is diffeomorphic to a direct product $G\times\mg^*$.\\[-.3cm]%%PARBREAK%%

\noindent{\it Proof:} The key point is that group translations can be used to map any tangent space $T_gG$ on the Lie algebra $\mg=T_{\II}G$. Accordingly, any element of $T_g^*G$, \ie a map $u:T_gG\rightarrow\RR$, can be turned into a map from $\mg$ to $\RR$, \ie an element of $\mg^*$. Concretely, let us write right translations in $G$ as $R_g:G\rightarrow G:f\mapsto fg$. Then define a map
\be
\Phi:
T^*G\rightarrow G\times\mg^*:
(g,u)\mapsto\Big(g,\big((R_g)^*u\big)_{\II}\Big)
\label{s28}
\ee
where $\big((R_g)^*u\big)_{\II}$ is a coadjoint vector such that $\left<\big((R_g)^*u\big)_{\II},\xi\right>=\left< u,\dd(R_g)_{\II}\xi\right>$ for any $\xi\in\mg$. The map (\ref{s28}) is a smooth bijection whose inverse
\be
\Phi^{-1}:
G\times\mg^*\rightarrow T^*G:
(g,p)\mapsto\big(g,(R_{g^{-1}}^*p)_g\big)
\label{t29}
\ee
is also smooth, so it is a diffeomorphism. This concludes the proof.\hfill$\blacksquare$\\[-.3cm]%%PARBREAK%%

Having established that $T^*G\cong G\times\mg^*$ is a trivial bundle, we now look at its symplectic form. First recall how the Liouville symplectic form is built on $T^*G$: one has a projection $\pi:T^*G\rightarrow G:(g,u)\mapsto g$ whose differential at $(g,u)$ projects any tangent vector of $T^*G$ on its part tangent to $G$ alone: $\dd\pi_{(g,u)}:T_{(g,u)}T^*G\rightarrow T_gG:(V,\cV)\mapsto V$. The {\it Liouville} (or {\it tautological}) {\it one-form} on $T^*G$ then reads
\be
\tilde\cA_{(g,u)}
\equiv
u\circ\dd\pi_{(g,u)},
\qquad\text{\ie}\qquad
\tilde\cA_{(g,u)}(V,\cV)
=
\langle u,V\rangle,
\label{s29}
\ee
and the symplectic form of $T^*G$ is its exterior derivative: $\tilde\omega=-\dd\tilde\cA$. We put tildes on these objects because we are eventually interested in the counterpart of (\ref{s29}) in $G\times\mg^*$, which will be tilde-free. To find this counterpart, one pulls back the Liouville one-form thanks to the inverse diffeomorphism (\ref{t29}):
\begin{align}
\big[(\Phi^{-1})^*\tilde\cA\big]_{(g,p)}(V,X)
&=
\tilde\cA_{(g,((R_g^{-1})^*p)_g)}\Big(\dd(\Phi^{-1})_{(g,p)}(V,X)\Big)\\
&\stackrel{\text{(\ref{s29})}}{=}
\left<(R_{g^{-1}}^*p)_g,V\right>
=
\left<p,\dd(R_{g^{-1}})_gV\right>.
\label{s30}
\end{align}
Thus, in terms of the right Maurer-Cartan form $\dd(R_{g^{-1}})_g\equiv\dd g\,g^{-1}$, and denoting the (pulled-back) Liouville one-form on $G\times\mg^*$ by $\cA\equiv(\Phi^{-1})^*\tilde\cA$, eq.\ (\ref{s30}) yields
\be
\cA_{(g,p)}
=
\big(\langle p,\dd g\,g^{-1}\rangle,0\big).
\label{ss3030}
\ee
This is the result announced in eq.\ (\ref{ss30}). One can write it more intrinsically as $\cA_{(g,p)}=\langle p,\dd g\,g^{-1}\rangle\circ\dd\Pi_{(g,p)}$, where $\Pi:G\times\mg^*\to G$ is the projection on the first component.

\section{Reconstructed dynamics and momentum map}
\label{appb}
\setcounter{equation}{0}
\renewcommand{\theequation}{B.\arabic{equation}}

This second appendix completes section \ref{sec22} by presenting reconstructed dynamics from a perspective where phase space is the cotangent bundle $T^*G$ throughout, without `reduction' to $\mg^*$. Along the way, we derive the momentum map for the (right) action of $G$ on $T^*G$ and show that its constant value along a solution $(g_t,p(t))$ of eqs.\ (\ref{s20})-(\ref{recong}) (resp.\ (\ref{ss20})-(\ref{ss31b}) for quadratic Hamiltonians) coincides with the coadjoint vector $k$ introduced in eq.\ (\ref{gkk}). This was mentioned in footnotes \ref{tobemade} and \ref{tobemode}, and provides an intrinsic geometric interpretation of the `orbit representative' $k$. In the KdV equation, right translations by $\Diff$ implement the `particle relabelling symmetry' of fluid mechanics and the ensuing momentum map defines `Noether charges' associated with the symmetry. As explained in section \ref{secEPP}, our considerations on geometric phases apply to those wave profiles for which it is possible to choose a {\it uniform}, \ie $x$-independent, value $k$; in that case, all these Noether charges vanish except the one corresponding to rigid rotations.\\[-.3cm]%%PARBREAK%%

The plan of this part of the appendix is as follows. We first define left and right actions on the cotangent bundle $T^*G$ and find the ensuing momentum maps. Then we use the diffeomorphism (\ref{s28}) to express everything in terms of the product manifold $G\times\mg^*$. Finally, we compute the equations of motion associated with a right-invariant Hamiltonian on $G\times\mg^*$, prove that they coincide with the Lie-Poisson equation (\ref{s20}) and its reconstruction (\ref{recong}), and show that the momentum map associated with right translations takes a constant value $k\in\mg^*$ on any solution of these equations.

\paragraph{Actions of \texorpdfstring{$G$}~~on \texorpdfstring{$T^*G$}~.} Recall that the cotangent bundle $T^*G$ has a symplectic form that can be written as $\tilde\omega=-\dd\tilde\cA$ in terms of the Liouville one-form (\ref{s29}). The group $G$ acts on itself through left and right multiplications, respectively denoted as $L$ and $R$. The corresponding left and right actions of $G$ on $T^*G$ are\footnote{As in appendix \ref{appa}, the tildes in $\tilde\sfL,\tilde\sfR$ stress that we will soon define tilde-free actions on $G\times\mg^*$.}
\be
\tilde\sfL_f(g,u)
\equiv
\big(fg,u\circ\dd(L_{f^{-1}})_{fg}\big),
\qquad
\tilde\sfR_f(g,u)
\equiv
\big(gf^{-1},u\circ\dd(R_f)_{gf^{-1}}\big)
\label{lerac}
\ee
for all $f,g\in G$, and any $u\in T_g^*G$. One readily verifies that these are indeed group actions in the sense that $\tilde\sfL_f\circ\tilde\sfL_g=\tilde\sfL_{fg}$ and $\tilde\sfR_f\circ\tilde\sfR_g=\tilde\sfR_{fg}$. Furthermore, they preserve the Liouville one-form since $(\tilde\sfL_f)^*\tilde\cA=(\tilde\sfR_f)^*\tilde\cA=\tilde\cA$, so they also preserve the symplectic form.\\[-.3cm]%%PARBREAK%%

The infinitesimal generators of the left and right actions (\ref{lerac}) are vector fields $X^{L,R}_{\xi}$ on $T^*G$, defined for any Lie algebra element $\xi\in\mg=T_{\II}G$ by
\be
(X^L_{\xi})_{(g,u)}
\equiv
%\left.\frac{\dd}{\dd t}\right|_0
\der_t\big|_0
\tilde\sfL_{e^{t\xi}}(g,u),
%=
%\big(\dd(R_g)_{\II}(\xi),...\big),
\qquad
(X^R_{\xi})_{(g,u)}
\equiv
%\left.\frac{\dd}{\dd t}\right|_0
\der_t\big|_0
\tilde\sfR_{e^{t\xi}}(g,u).
%=
%\big(-\dd(L_g)_{\II}(\xi),...\big),
\ee
From the definitions (\ref{lerac}) of left and right actions, one readily finds
\be
(X^L_{\xi})_{(g,u)}
=
\big(\dd(R_g)_{\II}(\xi),...\big),
\qquad
(X^R_{\xi})_{(g,u)}
=
\big(-\dd(L_g)_{\II}(\xi),...\big),
\label{ifige}
\ee
where we only write the components of $X^{L,R}_{\xi}$ that are tangent to $G$; the ellipses are components tangent to the cotangent spaces in $T^*G$, and will be unimportant.

\paragraph{Momentum maps.} Given a Hamiltonian action of a group $G$ on a symplectic manifold $(\cM,\omega)$, a {\it momentum map} $J:\cM\to\mg^*:p\mapsto J(p)$ is such that, for any $\xi\in\mg$, the function $J_{\xi}\equiv\langle J(\cdot),\xi\rangle$ satisfies
\be
\iota_{X_{\xi}}\omega
=
\dd J_{\xi}.
\label{momap}
\ee
Thus, $J_{\xi}$ is the Hamiltonian function for the infinitesimal generator $X_{\xi}$; equivalently, $J_{\xi}(p)$ is the `Noether charge' of the point $p$, associated with the `symmetry generator' $\xi$. When the dynamics on $\cM$ (specified by some Hamiltonian) is invariant under the $G$ action, Noether charges are constant on solutions of the equations of motion: $J_{\xi}(p(t))=\text{const}$.\\[-.3cm]%%PARBREAK%%

When $\omega=-\dd\cA$ and the group action preserves $\cA$ (as is the case for left and right actions of $G$ on $T^*G$), the Lie derivative $\cL_{X_{\xi}}\cA=0$ vanishes and the solution of (\ref{momap}) is $J_{\xi}=\iota_{X_{\xi}}\cA$ up to an irrelevant integration vector in $\mg^*$, here set to zero. This readily applies to the actions (\ref{lerac}) with infinitesimal generators (\ref{ifige}), so the momentum maps associated with left and right actions of $G$ on $T^*G$ are
\be
\tilde J^{L}(g,u)
=
u\circ\dd(R_g)_{\II},
\qquad
\tilde J^R(g,u)
=
-u\circ\dd(L_g)_{\II}.
\label{momar}
\ee
At this point, this does not look remotely similar to any of the expressions encountered in the main text. Accordingly, we now use the diffeomorphism (\ref{s28}) to rewrite left and right actions, as well as their momentum maps, in terms of the product manifold $G\times\mg^*$.

\paragraph{\texorpdfstring{$G$}~-actions on \texorpdfstring{$G\times\mg^*$}~ and momentum maps.} We saw above that the cotangent bundle $T^*G$, with its canonical symplectic form $\tilde\omega$, is symplectomorphic to the product $G\times\mg^*$ with symplectic form $\omega=-\dd\cA$ in terms of the Berry connection (\ref{ss3030}). Let us therefore define left and right Hamiltonian actions of $G$ on $G\times\mg^*$ by $\sfL_f\equiv\Phi\circ\tilde\sfL_f\circ\Phi^{-1}$ and $\sfR_f\equiv\Phi\circ\tilde\sfR_f\circ\Phi^{-1}$, where $\tilde\sfL$ and $\tilde\sfR$ are given by (\ref{lerac}) while $\Phi$ is written in (\ref{s28}). Using these definitions, one finds
\be
\sfL_f(g,p)
=
(fg,f\cdot p),
\qquad
\sfR_f(g,p)
=
(gf^{-1},p)
\label{leracto}
\ee
for all $f,g\in G$ and any $p\in\mg^*$, with $f\cdot p\equiv\Ad^*_f(p)$ the coadjoint representation (\ref{s17b}). Notice that the right action only affects the factor $G$ in $G\times\mg^*$, leaving `momenta' in $\mg^*$ untouched. This is not true of the left action, which acts non-trivially on $\mg^*$. The distinction will eventually be crucial for Lie-Poisson dynamics.\\[-.3cm]%%PARBREAK%%

Since the actions $\tilde\sfL,\tilde\sfR$ preserve the Liouville form $\tilde\cA$ of $T^*G$, it is also true that the actions $\sfL,\sfR$ preserve the Berry connection $\cA$ on $G\times\mg^*$. Accordingly, there are momentum maps associated with these actions, namely $J^L=\tilde J^L\circ\Phi^{-1}$ and $J^R=\tilde J^R\circ\Phi^{-1}$. Since the maps $\tilde J^{L,R}$ are given by (\ref{momar}), one obtains
\be
J^L(g,p)=p,
\qquad
J^R(g,p)
=
-g^{-1}\cdot p.
\label{jiji}
\ee
These expressions are somewhat more illuminating than eqs.\ (\ref{momar}). In particular, both momentum maps now send a phase space point $(g,p)$ on an actual momentum --- respectively $p$ and $g^{-1}\cdot p$ for the left and right actions.\\[-.3cm]%%PARBREAK%%

At this point, one can anticipate what happens for Lie-Poisson equations: if a time-dependent momentum vector is written as $p(t)=g_t\cdot k$ for some time-dependent configuration $g_t\in G$ and some time-independent coadjoint vector $k$, then the value of the right momentum map is constant along the path $(g_t,p(t))$:
\be
J^R(g_t,p(t))
=
J^R(g_t,g_t\cdot k)
\stackrel{\text{(\ref{jiji})}}{=}
-g_t^{-1}\cdot g_t\cdot k
=
-k.
\label{jik}
\ee
This is the essence of footnote \ref{tobemade}, where we claimed that $k$ may be seen as the constant image of the momentum map associated with right $G$ translations. To complete the picture, it now remains to derive full reconstructed dynamics in $G\times\mg^*$ from a right-invariant Hamiltonian function, confirming that the equality between the `$k$' of the main text and the momentum map (\ref{jik}) is more than a mere coincidence.

\paragraph{Lie-Poisson dynamics on \texorpdfstring{$G\times\mg^*$}~.} Consider once more the product manifold $G\times\mg^*$ with symplectic form $\omega=-\dd\cA$. Suppose, as in section \ref{sec21}, that one is given a (not necessarily quadratic) Hamiltonian function $H$ on $\mg^*$. Since the right action (\ref{leracto}) of $G$ on $G\times\mg^*$ leaves momenta untouched, it is straightforward to extend $H$ into a right-invariant function $\cH$ on the full phase space $G\times\mg^*$: just declare that $\cH(g,p)\equiv H(p)$. (Incidentally, this confirms the interpretation of $H$ as `kinetic energy', mentioned below eq.\ (\ref{tt20}): since the Hamiltonian on $G\times\mg^*$ is configuration-independent, there is no `potential energy'.)\\[-.3cm]%%PARBREAK%%

Now, the equations of motion associated with the Hamiltonian $\cH$ are given, by definition, by the flow of the vector field $X_{\cH}$ such that $\iota_{X_{\cH}}\omega=\dd\cH$. In the case at hand, we pick a point $(g,p)\in G\times\mg^*$ and an arbitrary tangent vector $(V,\cV)$, and write the components of $(X_{\cH})_{(g,p)}$ as $(\dot g,\dot p)$ since we are interested in the flow of $X_{\cH}$. Using the fact that $\omega=-\dd\cA$ with $\cA$ given by eq.\ (\ref{ss3030}), along with $\dd\cH=(0,\dd H)$, we find
\be
\iota_{(\dot g,\dot p)}\omega_{(g,p)}(V,\cV)
=
-\langle\dot p,(\dd R_{g^{-1}})_gV\rangle
-\langle p,[\dot g\,g^{-1},\dd g\,g^{-1}(V)]\rangle
+\langle\cV,\dot g\,g^{-1}\rangle
\stackrel{!}{=}
%(\dd\cH)_{(g,p)}(V,\cV)
%=
\dd H_p(\cV).
\label{dedepo}
\ee
Thus, removing the linearly independent arguments $V$ and $\cV$, the defining relation $\iota_{X_{\cH}}\omega=\dd\cH$ holds if the following two equations are satisfied:
\be
\dot p
=
\ad^*_{\dot g\,g^{-1}}(p),
\qquad
\dot g\,g^{-1}
=
\dd H_p.
\label{eoma}
\ee
Replacing $\dot g\,g^{-1}$ by $\dd H$ in the first equation, one readily recognizes the Lie-Poisson equation (\ref{s20}) and its reconstruction (\ref{recong}). For a quadratic Hamiltonian $H(p)=\tfrac{1}{2}\langle p,\cI^{-1}(p)\rangle$, this reduces to eqs.\ (\ref{ss20})-(\ref{ss31b}). Furthermore, since the Hamiltonian on $G\times\mg^*$ is right-invariant, it readily follows that the momentum map $J^R$ in (\ref{jiji}) is constant along any solution of the equations of motion. As argued around (\ref{jik}), this constant is precisely the orbit representative $k$ that has played a key role for geometric phases.

\addcontentsline{toc}{section}{References}

\providecommand{\href}[2]{#2}


\begin{thebibliography}{99}

{\small
\bibitem{Oblak:2017ect}
B.~Oblak, ``{Berry Phases on Virasoro Orbits},'' {\em JHEP} \textbf{10} (2017)
  114,
\href{http://www.arXiv.org/abs/1703.06142}{\texttt{1703.06142}}.
%%CITATION = ARXIV:1703.06142;%%.

\bibitem{Berry:1984jv}
M.~V. Berry, ``{Quantal phase factors accompanying adiabatic changes},'' {\em
  Proc.\ Roy.\ Soc.\ Lond.} \textbf{A392} (1984)
45--57.
%%CITATION = PRSLA,A392,45;%%.

\bibitem{berry1985classical}
M.~V.~Berry, ``{Classical adiabatic angles and quantal adiabatic phase},'' {\em
  J.\ Phys.\ A} \textbf{18} (1985), no.~1, 15.
\textbullet\
T.~F. Jordan, ``{Berry phases and unitary transformations},'' {\em J.\ Math.\
  Phys.} \textbf{29} (1988), no.~9, 2042--2052.

\bibitem{Thomas:1926dy}
L.~H. Thomas, ``{The motion of a spinning electron},'' {\em Nature}
  \textbf{117} (1926)
514.
%%CITATION = NATUA,117,514;%%.
\textbullet\
P.~Aravind, ``{The Wigner angle as an anholonomy in rapidity space},'' {\em
  Am.\ J.\ Phys.} \textbf{65} (1997), no.~7, 634--636.
\textbullet\
H.~Mathur, ``{Thomas precession, spin-orbit interaction, and Berry's phase},''
  {\em Phys. Rev. Lett.} \textbf{67} (1991) 3325--3327.
\textbullet\
B.~Oblak, ``{Probing Wigner Rotations for Any Group},'' {\em J. Geom. Phys.}
  \textbf{129} (2018) 168--185,
\href{http://www.arXiv.org/abs/1710.06883}{\texttt{1710.06883}}.
%%CITATION = ARXIV:1710.06883;%%.

\bibitem{Nakahara:2003nw}
M.~Nakahara, {\em {Geometry, Topology and Physics}}.
\newblock Taylor \& Francis, 2003.

\bibitem{leek2007observation}
P.~Leek, J.~Fink, A.~Blais, R.~Bianchetti, M.~G{\"o}ppl, J.~Gambetta,
  D.~Schuster, L.~Frunzio, R.~Schoelkopf, and A.~Wallraff, ``{Observation of
  Berry's phase in a solid-state qubit},'' {\em Science} \textbf{318} (2007),
  no.~5858, 1889--1892.

\bibitem{svensmark1994experimental}
H.~Svensmark and P.~Dimon, ``{Experimental Observation of Berry's Phase of the
  Lorentz Group},'' {\em Phys.\ Rev.\ Lett.} \textbf{73} (1994), no.~25, 3387.

\bibitem{AvronSeiler}
J.~E. Avron, R.~Seiler, and P.~G. Zograf, ``{Viscosity of Quantum Hall
  Fluids},'' {\em Phys. Rev. Lett.} \textbf{75} (1995) 697--700.
\textbullet\
P.~L\'evay, ``{Berry phases for Landau Hamiltonians on deformed tori},'' {\em
  {J.\ Math.\ Phys.}} \textbf{36} (1995), no.~6, 2792--2802.

\bibitem{Alekseev:1988ce}
A.~Alekseev and S.~L. Shatashvili, ``{Path Integral Quantization of the
  Coadjoint Orbits of the Virasoro Group and 2D Gravity},'' {\em Nucl.\ Phys.}
  \textbf{B323} (1989)
719--733.
%%CITATION = NUPHA,B323,719;%%.
\textbullet\
A.~Alekseev and S.~L. Shatashvili, ``{From geometric quantization to conformal
  field theory},'' {\em Commun.\ Math.\ Phys.} \textbf{128} (1990) 197--212.
[,22(1990)].
%%CITATION = CMPHA,128,197;%%.
\textbullet\
B.~Rai and V.~G.~J. Rodgers, ``{From Coadjoint Orbits to Scale Invariant {WZNW}
  Type Actions and 2-$D$ Quantum Gravity Action},'' {\em Nucl.\ Phys.}
  \textbf{B341} (1990)
119--133.
%%CITATION = NUPHA,B341,119;%%.

\bibitem{Bradlyn:2015wsa}
B.~Bradlyn and N.~Read, ``{Topological central charge from Berry curvature:
  Gravitational anomalies in trial wave functions for topological phases},''
  {\em Phys.\ Rev.} \textbf{B91} (2015), no.~16, 165306,
\href{http://www.arXiv.org/abs/1502.04126}{\texttt{1502.04126}}.
%%CITATION = ARXIV:1502.04126;%%.

\bibitem{Hannay}
J.~H. Hannay, ``{Angle variable holonomy in adiabatic excursion of an
  integrable Hamiltonian},'' {\em J.\ Phys.\ A} \textbf{18} (1985), no.~2, 221.

\bibitem{Korteweg}
D.~Korteweg and G.~de~Vries, ``{On the change of form of long waves advancing
  in a rectangular canal, and on a new type of long stationary waves},'' {\em
  Lond., Edinb., Dublin Philos.\ Mag.\ J. Sci.} \textbf{39} (1895), no.~240,
  422--443.

\bibitem{Marsden90}
J.~E. Marsden, R.~Montgomery, and T.~S. Rațiu, {\em Reduction, symmetry, and
  phases in mechanics}, vol.~436.
\newblock American Mathematical Soc., 1990.

\bibitem{Marsden}
J.~E. Marsden and T.~S. Ratiu, {\em Introduction to mechanics and symmetry: a
  basic exposition of classical mechanical systems}, vol.~17.
\newblock Springer, 2013.

\bibitem{Holm}
D.~D. Holm, T.~Schmah, and C.~Stoica, {\em Geometric mechanics and symmetry:
  from finite to infinite dimensions}, vol.~12.
\newblock Oxford University Press, 2009.

\bibitem{HolmTyr}
D.~D. Holm and T.~M. Tyranowski, ``{New variational and multisymplectic
  formulations of the Euler-Poincar\'e equation on the Virasoro-Bott group
  using the inverse map},'' {\em Proc.\ Roy.\ Soc.\ A} \textbf{474} (2018),
  no.~2213, 20180052.

\bibitem{Khesin2003}
B.~Khesin and G.~Misiołek, ``{Euler equations on homogeneous spaces and
  Virasoro orbits},'' {\em Adv.\ Math.} \textbf{176} (2003), no.~1, 116--144.

\bibitem{Khesin}
B.~Khesin and R.~Wendt, {\em The geometry of infinite-dimensional groups},
  vol.~51.
\newblock Springer, 2008.

\bibitem{Hunter}
J.~Hunter and R.~Saxton, ``{Dynamics of Director Fields},'' {\em SIAM J.\
  Appl.\ Math.} \textbf{51} (1991), no.~6, 1498--1521.
\textbullet\
R.~Camassa and D.~D. Holm, ``An integrable shallow water equation with peaked
  solitons,'' {\em Phys.\ Rev.\ Lett.} \textbf{71} (1993) 1661--1664.

\bibitem{Montgomery}
R.~Montgomery, ``{How much does the rigid body rotate? A Berry’s phase from
  the 18th century},'' {\em Am.\ J.\ Phys.} \textbf{59} (1991), no.~5,
  394--398.

\bibitem{Natario}
J.~Nat\'ario, ``{An elementary derivation of the Montgomery phase formula for the
  Euler top},''  {\em J.\ Geom.\ Mech.} \textbf{2} (2010),
  no.~1941-4889\_2010\_1\_113, 113,
  \href{http://www.arXiv.org/abs/0909.2109}{\texttt{0909.2109}}.

\bibitem{Alber}
M.~S. Alber and J.~E. Marsden, ``On geometric phases for soliton equations,''
  {\em Commun.\ Math.\ Phys.} \textbf{149} (1992) 217--240.

\bibitem{Abraham}
R.~Abraham and J.~Marsden, {\em {Foundations of Mechanics}}.
\newblock AMS Chelsea Pub./American Mathematical Society, 1978.

\bibitem{Guieu}
L.~Guieu and C.~Roger, {\em {L'alg\`ebre et le groupe de Virasoro}}.
\newblock Publications du CRM, Universit{\'e} de Montr\'eal, 2007.

\bibitem{Oblak:2016eij}
B.~Oblak, {\em {BMS Particles in Three Dimensions}}.
\newblock PhD thesis, Brussels U., 2016.
\newblock \href{http://www.arXiv.org/abs/1610.08526}{\texttt{1610.08526}}.
\newblock {\it{Springer Theses}}, 2017.

\bibitem{Lazutkin}
V.~F. Lazutkin and T.~F. Pankratova, ``{Normal forms and versal deformations
  for Hill's equation},'' {\em Funct.\ Anal.\ Appl.} \textbf{9} (1975)
  306--311.
\textbullet\
A.~A. Kirillov, ``{Orbits of the group of diffeomorphisms of a circle and local
  Lie superalgebras},'' {\em Funct.\ Anal.\ Appl.} \textbf{15} (1981)
  135--137.

\bibitem{Witten:1987ty}
E.~Witten, ``{Coadjoint Orbits of the Virasoro Group},'' {\em Commun.\ Math.\ Phys.} \textbf{114} (1988) 1.
%%CITATION = CMPHA,114,1;%%.

\bibitem{Oblak:2019llc}
B.~Oblak, ``{Orbital Bifurcations and Shoaling of Cnoidal Waves},'' {\em J. Math. Fluid Mech.} \textbf{22}, 29 (2020),
\href{http://www.arXiv.org/abs/1907.01438}{\texttt{1907.01438}}.
%%CITATION = ARXIV:1907.01438;%%.

\bibitem{Adler}
R.~{Adler}, ``{A Study of Locking Phenomena in Oscillators},'' {\em Proc.\ IRE}
  \textbf{34} (1946) 351--357.
\textbullet\
S.~Strogatz, {\em {Nonlinear Dynamics and Chaos: With Applications to Physics,
  Biology, Chemistry, and Engineering}}.
\newblock CRC Press, 2015.

\bibitem{Ockendon}
H.~Ockendon and A.~B. Tayler, {\em Inviscid fluid flows}, vol.~43.
\newblock Springer, 2013.

\bibitem{Longuet53}
M.~S. Longuet-Higgins, ``Mass transport in water waves,'' {\em Philos.\ Trans.\
  Roy.\ Soc.\ Lond.\ A} \textbf{245} (1953), no.~903, 535--581.
\textbullet\
A.~Constantin and J.~Escher, ``{Particle trajectories in solitary water
  waves},'' {\em Bull.\ Am.\ Math.\ Soc.} \textbf{44} (2007) 423--431.
\textbullet\
A.~Constantin, ``On the particle paths in solitary water waves,'' {\em Q.\
  Appl.\ Math.} \textbf{68} (2010), no.~1, 81--90.

\bibitem{Banner}
M.~L. Banner, X.~Barthelemy, F.~Fedele, M.~Allis, A.~Benetazzo, F.~Dias, and
  W.~L. Peirson, ``{Linking Reduced Breaking Crest Speeds to Unsteady Nonlinear
  Water Wave Group Behavior},'' {\em Phys.\ Rev.\ Lett.} \textbf{112} (2014) 114502, \href{http://www.arXiv.org/abs/1305.3980}{\texttt{1305.3980}}.
\textbullet\
F.~Fedele, ``Geometric phases of water waves,'' {\em Europhys.\ Lett.}
  \textbf{107} (2014) 69001,
  \href{http://www.arXiv.org/abs/1406.0051}{\texttt{1406.0051}}.
  
\bibitem{Arnold}
V.~Arnold, ``{Sur la g{\'e}om{\'e}trie diff{\'e}rentielle des groupes de Lie de
  dimension infinie et ses applications {\`a} l'hydrodynamique des fluides
  parfaits},'' in {\em Ann.\ Inst.\ Fourier}, vol.~16, pp.~319--361.
\newblock 1966.
\textbullet\
V.~I. Arnold and B.~A. Khesin, {\em Topological methods in hydrodynamics},
  vol.~125.
\newblock Springer, 1999.
\textbullet\
T.~Ratiu, E.~Sousa~Dias, L.~Sbano, G.~Terra, and R.~Tudoran, ``{A Crash Course
  in Geometric Mechanics},'' in {\em {Geometric mechanics and symmetry: the
  Peyresq lectures}}, N.~Hitchin {\em et al.}, eds., vol.~306.
\newblock Cambridge University Press, 2005.

\bibitem{boya2001berry}
L.~J. Boya, A.~M. Perelomov, and M.~Santander, ``{Berry phase in homogeneous
  K{\"a}hler manifolds with linear Hamiltonians},'' {\em J.\ Math.\ Phys.}
  \textbf{42} (2001), no.~11, 5130--5142.

\bibitem{Balog:1997zz}
J.~Balog, L.~Feher, and L.~Palla, ``{Coadjoint orbits of the Virasoro algebra
  and the global Liouville equation},'' {\em Int.\ J.\ Mod.\ Phys.}
  \textbf{A13} (1998) 315--362,
\href{http://www.arXiv.org/abs/hep-th/9703045}{\texttt{hep-th/9703045}}.
%%CITATION = HEP-TH/9703045;%%.

\bibitem{Bott}
R.~Bott, ``{On the characteristic classes of groups of diffeomorphisms},'' {\em
  Enseign.\ Math.} \textbf{23} (1977), no.~3-4, 209--220.

\bibitem{DiFran}
P.~Di~Francesco, P.~Mathieu, and D.~S{\'e}n{\'e}chal, {\em Conformal field
  theory}.
\newblock Springer, 2012.

\bibitem{Whittaker}
E.~T. Whittaker and G.~N. Watson, {\em A course of modern analysis}.
\newblock Cambridge University Press, 1996.

\bibitem{Lawden}
D.~F. Lawden, {\em Elliptic functions and applications}, vol.~80.
\newblock Springer, 2013.

\bibitem{Doman}
B.~G.~S. Doman, ``{A Mean-field Treatment of Charge Density Waves in a Strong
  Magnetic Field},'' in {\em Essays in Theoretical Physics}, W.~PARRY, ed.,
  pp.~19--42.
\newblock Pergamon, 1984.

\bibitem{Modin}
K. Modin, ``{Geometric Hydrodynamics: from Euler, to Poincar\'e, to Arnold},"
\href{http://www.arXiv.org/abs/1910.03301}{\texttt{1910.03301}}.

\bibitem{Shapere:1989kp}
A.~D. Shapere and F.~Wilczek, ``{Geometric Phases in Physics},'' {\em Adv.\
  Ser.\ Math.\ Phys.} \textbf{5} (1989)
1--509.
%%CITATION = 00304,5,1;%%.

\bibitem{Cohen}
E.~Cohen, H.~Larocque, F.~Bouchard, F.~Nejadsattari, Y.~Gefen, and E.~Karimi,
  ``{Geometric phase from Aharonov-Bohm to Pancharatnam-Berry
  and beyond},'' {\em Nature Rev.\ Phys.} \textbf{1} (2019) 437--449.

\bibitem{Its}
A.~R. Its and V.~B. Matveev, ``{Schr{\"o}dinger operators with finite-gap
  spectrum and N-soliton solutions of the Korteweg-de Vries equation},'' {\em
  Theor.\ Math.\ Phys.} \textbf{23} (1975) 343--355.

\bibitem{Sasaki}
R.~Sasaki and I.~Yamanaka, ``Virasoro algebra, vertex operators, quantum
  sine-Gordon and solvable quantum field theories,'' in {\em Conformal Field
  Theory and Solvable Lattice Models}, pp.~271--296.
\newblock Mathematical Society of Japan, Tokyo, Japan, 1988.

\bibitem{Debou}
A.~De~Bouard, A.~Debussche, and Y.~Tsutsumi, ``{Periodic Solutions of the
  Korteweg--de Vries Equation Driven by White Noise},'' {\em SIAM Journal on
  Mathematical Analysis} \textbf{36} (2005), no.~3, 815--855.

\bibitem{HolmStoch}
D.~D. Holm, ``Variational principles for stochastic fluid dynamics,'' {\em
  Proc.\ Roy.\ Soc.\ A} \textbf{471} (2015), no.~2176, 20140963.

\bibitem{Hirota}
R.~Hirota and J.~Satsuma, ``{Soliton solutions of a coupled Korteweg-de Vries
  equation},'' {\em Phys.\ Lett.\ A} \textbf{85} (1981), no.~8, 407 -- 408.

\bibitem{Lakshmanan}
M.~Lakshmanan, ``Continuum spin system as an exactly solvable dynamical
  system,'' {\em Phys.\ Lett.\ A} \textbf{61} (1977), no.~1, 53 -- 54.}

\end{thebibliography}
\end{document}